\documentclass{aa}  

\usepackage{graphicx}
\usepackage{txfonts}
\usepackage{amssymb}
\usepackage{lineno}
\usepackage{txfonts}
\usepackage{natbib}   % the natbib package allows both number and author-year (Harvard)
\usepackage{amstext}
\usepackage{mathabx}
\usepackage{multirow}
\usepackage{ulem}
\usepackage{color}
\usepackage{pifont}
\usepackage{lscape}
\usepackage{array} %options pour les tableaux
\usepackage{tabularx}
\usepackage[version=3]{mhchem}

\usepackage{hyperref}
\newcommand{\fig}[1]{Fig.~\ref{#1}}
\newcommand{\figs}[2]{Figs.~\ref{#1} and~\ref{#2}}

\newcommand{\tab}[1]{Table~\ref{#1}}
\newcommand{\Kunit}{\,cm$^{2}.$s$^{-1}$}
\newcommand{\dix}[1]{$\times10^{#1}$}

\begin{document}

\title{A new chemical scheme for giant planet thermochemistry. Update of the methanol chemistry and new reduced chemical scheme.}
\author{O. Venot\inst{1}, T. Cavali\'e\inst{2,3}, R. Bounaceur\inst{4}, P. Tremblin\inst{5}, L. Brouillard\inst{6}, R. Lhoussaine Ben Brahim\inst{1}
}

\institute{Laboratoire Interuniversitaire des Syst\`{e}mes Atmosph\'{e}riques (LISA), UMR CNRS 7583, Universit\'{e} Paris-Est-Cr\'eteil, Universit\'e de Paris, Institut Pierre Simon Laplace, Cr\'{e}teil, France
\and Laboratoire d'Astrophysique de Bordeaux, Univ. Bordeaux, CNRS, B18N, all\'ee Geoffroy Saint-Hilaire, Pessac 33615, France
\and LESIA, Observatoire de Paris, Universit\'e PSL, CNRS, Sorbonne Universit\'e, Universit\'e Paris Diderot, Sorbonne Paris Cit\'e, 5 place Jules Janssen, 92195 Meudon, France
\and Laboratoire R\'{e}actions et G\'{e}nie des Proc\'{e}d\'{e}s, LRGP UMP 7274 CNRS, Universit\'{e} de Lorraine, 1 rue Grandville, BP 20401, F-54001 Nancy, France
\and Maison de la Simulation, CEA, CNRS, Univ. Paris-Sud, UVSQ, Universit\'e Paris-Saclay, 91191 Gif-sur-Yvette, France
\and Universit\'e de Bordeaux, 351 Cours de la Lib\'eration, 33400 Talence, France
}

\titlerunning{New Methanol}
\date{Received <date> /
Accepted <date>}

\abstract{
%Context
Several chemical networks have been developed to study warm (exo)planetary atmospheres. The kinetics of the reactions related to the methanol chemistry included in these schemes have been questioned.}
{
%Aims
The goal of this paper is to update the methanol chemistry for such chemical networks thanks to recent publications in the combustion literature.
We aim also at studying the consequences of this update on the atmospheric compositions of (exo)planetary atmospheres and brown dwarfs.}
{%Methods
We have performed an extensive review of combustion experimental studies and revisited the sub-mechanism describing methanol combustion in the scheme of Venot et al. (2012, A\&A 624, A58). The updated scheme involves 108 species linked by a total of 1906 reactions.
We have then applied our 1D kinetic model with this new scheme to several case studies (HD 209458b, HD 189733b, GJ 436b, GJ 1214b, ULAS J1335+11, Uranus, Neptune), and compared the results obtained with those obtained with the former scheme.}
{%Results
The update of the scheme has a negligible impact on hot Jupiters atmospheres. However, the atmospheric composition of warm Neptunes and brown dwarfs is modified sufficiently to impact observational spectra in the wavelength range JWST will operate. Concerning Uranus and Neptune, the update of the chemical scheme modifies the abundance of CO and thus impacts the deep oxygen abundance required to reproduce the observational data. For future 3D kinetics models, we also derived a reduced scheme containing 44 species and 582 reactions.}
{%Conclusion
Chemical schemes should be regularly updated in order to maintain a high level of reliability on the results of kinetic models and be able to improve our knowledge on planetary formation. 
}


\maketitle

\section{Introduction}

The knowledge of the deep composition of Solar System Giant Planets is essential to constrain their formation models \citep{Pollack1996,Boss1997,Owen1999,Gautier2005}. While only in situ measurements can provide ground truth measurements, their deep composition remains generally inaccessible to remote sensing techniques or the interpretation of the data has to relies on assumptions on temperature \citep[e.g.][]{dePater1989,dePater1991,Luszcz-Cook2013,Cavalie2014,Li2018}. Even if plans for future in situ exploration exist \citep{Arridge2012,Arridge2014,Mousis2014,Mousis2016,Mousis2018} there is only one such experiment that has been carried out, with the Galileo probe in Jupiter \citep{Atreya1999,Wong2004}. Therefore, thermochemical modelling remains a tool complementary to remote observations to infer the deep composition of the Solar System Giant Planets \citep{Lodders1994,Visscher2005,Visscher2010,Cavalie2017}

For H-dominated exoplanets, thermo- and photo-chemistry is used to predict the atmospheric composition \citep{Moses2011,Venot2012}. The atmosphere of hot exoplanets is schematically divided in three parts: 1) the deepest one, very hot, has a chemical composition governed by thermochemical equilibrium; 2) the middle one has a lower temperature and a composition controlled by transport-induced quenching; 3) the upper layers are subject to photochemistry \citep[][their Fig.1]{madhu2016}. Brown dwarfs are also subject to a transition between thermochemical equilibrium (part 1) and a quenching zone (part 2) but photochemistry can be ignored in this case because the object is isolated \citep[e.g.][]{griffith2000}. 
To interpret observations of brown dwarf/exoplanet atmospheres probing the regions governed by quenching (and eventually also influenced by photochemistry for exoplanets), it is mandatory to evaluate correctly the quenching level and the abundances of species at this point. It is of particular interest to explain which species are the reservoir of carbon (CO/CH$_4$) and nitrogen (NH$_3$/N$_2$). Indeed, the relative abundances of these species may vary depending on the pressure and temperature of the quenching level: at high temperatures and/or low pressures CO and N$_2$ are the main carbon and nitrogen bearing-species, respectively, while at low temperatures and/or high pressures CH$_4$ and NH$_3$ dominate. 

One of the main parameters in thermochemical modeling is the chemical scheme. \citet{Venot2012} have proposed a chemical scheme built with input data from the combustion industry for H, C, O, and N species, relevant for temperature and pressure ranges found in solar system giant planet deep tropospheres, in hot Jupiters, warm Neptunes, and brown dwarfs. However, \citet{Moses2014} has found differences between the latter model and hers in the chemistry of oxygen species that results in significant discrepancies in the abundances of some key (and observable) species, like CO in solar system giant planets. This was later confirmed by \citet{Wang2016}. \citet{Moses2014} identified the chemistry of methanol (CH$_3$OH) as being at the root of the differences. This has motivated the present study, in which we re-evaluate the chemistry of CH$_3$OH of \citet{Venot2012} and produce a new chemical scheme that accounts for these updates. We also produce a new reduced chemical scheme based on this new one, following \cite{Venot2019}, for future 3D kinetic modeling.

In this paper, we present a short review of methanol combustion studies (Section \ref{methanol_chemistry}), our new CH$_3$OH sub-scheme and its validation (Section \ref{validation}). We then apply it to typical cases (section \ref{applications}), analyse the differences with the previous model results (Section \ref{interpretation}), and discuss their implication for atmospheres (Section \ref{implication}). We give our conclusion (Section \ref{conclusion}). We present in Appendix \ref{reduced_scheme} a reduced chemical scheme extracted from the update for future 3D models.

\section{A short review of methanol combustion experimental studies \label{methanol_chemistry}}
The aim of \citet{Venot2012} was to propose a full and robust mechanism in order to model the combustion of compounds such as hydrogen, methane and ethane. Their chemical scheme, hereafter V12, which consisted in 105 species involved in 957 reversible and 6 irreversible reactions has been questioned by \citet{Moses2014}, pointing more specifically the reaction between methanol and hydrogen radical yielding to methyl radical and water (CH$_3$OH$+$H$\rightleftharpoons$CH$_3+$H$_2$O). This reaction was initially proposed by \citet{Hidaka1989}, with kinetic data for this reaction evaluated by analogy and optimised on a set of experimental data. More generally, the sub-mechanism for methanol combustion in V12 had been extracted from the work of \citet{Barbe1995}.

Many teams have studied the pyrolysis and combustion of methanol at different concentrations, pressures, temperatures, and with several kinds of reactors. Several studies were performed to measure ignition delay times, for example by \citet{Cooke1971}, \citet{Bowman1975}, \citet{Tsuboi1981}, and \citet{Natarajan1981}. These auto-ignition studies have used the shock tube apparatus and employed the reflected shock technique to study the auto-ignition characteristics at high temperatures (greater than 1300\,K) and moderate pressures (5\,bar). \citet{Kumar2011} and more recently \citet{Burke2016} studied the auto-ignition of methanol in a rapid compression machine for temperatures ranging from 800 to 1700\,K and pressures between 1 to 50\,bar. They have shown that under these experimental conditions, the ignition delay times of methanol are comparable to the other alcohols.

Several studies have attempted to measure laminar burning velocities for mixtures of methanol and many techniques were used such as counter-flow double flames, burner stabilised flames, heat flux method and closed bomb technique. Due to the high number of studies found in the literature, only the very large study of \citet{Liao2006} is summarised here. They have studied the influence of the initial temperature and equivalence ratio on the speed flame for an air/methanol mixture, at atmospheric pressure. They have used a closed bomb apparatus and they compared their experimental results to data obtained by other authors with the same experimental set up. The influence of the initial temperature on the laminar flame speed was also studied by \citet{Liao2006}. Different equivalence ratios have been considered, and an influence of the initial temperature on the laminar flame speed has been observed whatever the equivalence ratio. Thus, the laminar burning velocity is almost doubled when the initial temperature increases from 350 to 550\,K.

Finally, many other authors have studied the oxidation or pyrolysis of methanol using different reactors (such as static reactor or plug flow reactor) covering a large range of concentration, temperature and pressure and reported species profiles for products and intermediates. \tab{tab:combustion_CH3OH} gives an overview of the main studies published over the 50 past years on this topic and from which \citet{Burke2016} have built their CH$_3$OH sub-network.

\section{Validation of a new chemical scheme \label{validation}}
\citet{Burke2016} have recently published new experimental data on methanol combustion and proposed a revisited chemical model for this species. This kinetic model has been validated against those data and a set of previously published experimental data. The sub-mechanism of methanol is included in a more complete kinetic model to represent the combustion of mixtures with methane, ethane, propane, and butane. Their full model is composed of 1011 reactions and 160 species. 

We have first extracted the sub-mechanism of methanol combustion and the relevant thermodynamic data from the model of \citet{Burke2016} and updated the original model of \citet{Venot2012} with this new sub-network (see \tab{tab:new_scheme1}). The main difference with the previous methanol sub-scheme is that some reaction rates have an explicit logarithmic dependence in pressure (see Appendix \ref{Annex_PLOG}). These reactions are presented in \tab{tab:new_scheme2}. Another difference that can be highlighted is that the controversial reaction of \citet{Hidaka1989},  CH$_3$OH$+$H$\rightleftharpoons$CH$_3+$H$_2$O, is no longer explicitly present in the scheme. The removal of CH$_3$OH still exists and can eventually lead to the formation of CH$_3$ and H$_2$O, but through other destruction pathways (see Section \ref{interpretation}). 

The full and updated chemical scheme that we present in this paper, hereafter called V20, contains 108 species, 948 reversible reactions and 10 irreversible reactions (i.e. 1906 reactions in total). It can be downloaded from the KInetic Database for Astrochemistry (KIDA) \citep{Wakelam2012}\footnote{\href{http://kida.obs.u-bordeaux1.fr/}{kida.obs.u-bordeaux1.fr.}}. 

To validate the inclusion of the \citet{Burke2016} methanol sub-mechanism within our chemical scheme, we have compared simulation results with experimental results from various sources \citep{Aronowitz1979,Cathonnet1982,Norton1989,Held1994,Ren2013,Burke2016}. The model performance over a wide array of experimental conditions was found to be in better agreement than that of the original mechanism V12 (see figures in Appendix \ref{appendix_validation}).

\begin{figure}[!h]
  \includegraphics[width=\columnwidth]{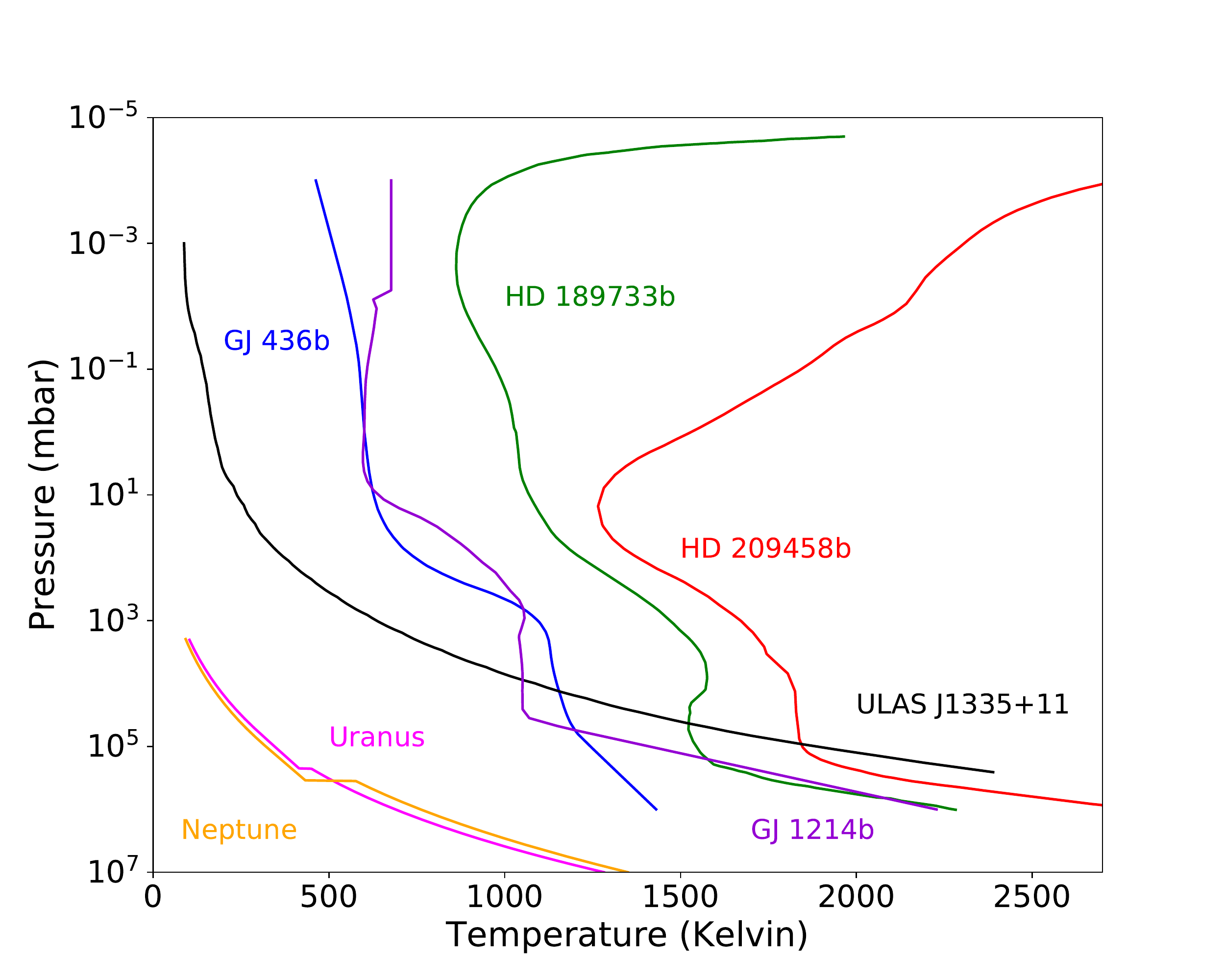}
  \caption{Adopted thermal profiles of the planets studied in this paper. }
  \label{fig:PTprofiles}
  \end{figure}

With the updated chemical scheme V20, we revisit in the next section the 1D thermo-photochemical model results for emblematic cases published in previous papers: HD 209458b and HD 189733b for hot Jupiters, GJ 436b and GJ 1214b for warm Neptunes, and Uranus and Neptune. We also model for the first time the T Dwarf ULAS J1335+11. Thermal profiles of these planets are shown in Fig.~\ref{fig:PTprofiles}.

\section{Applications \label{applications}}
  \subsection{Hot Jupiters}
  
\begin{figure}[!h]
  \includegraphics[width=\columnwidth]{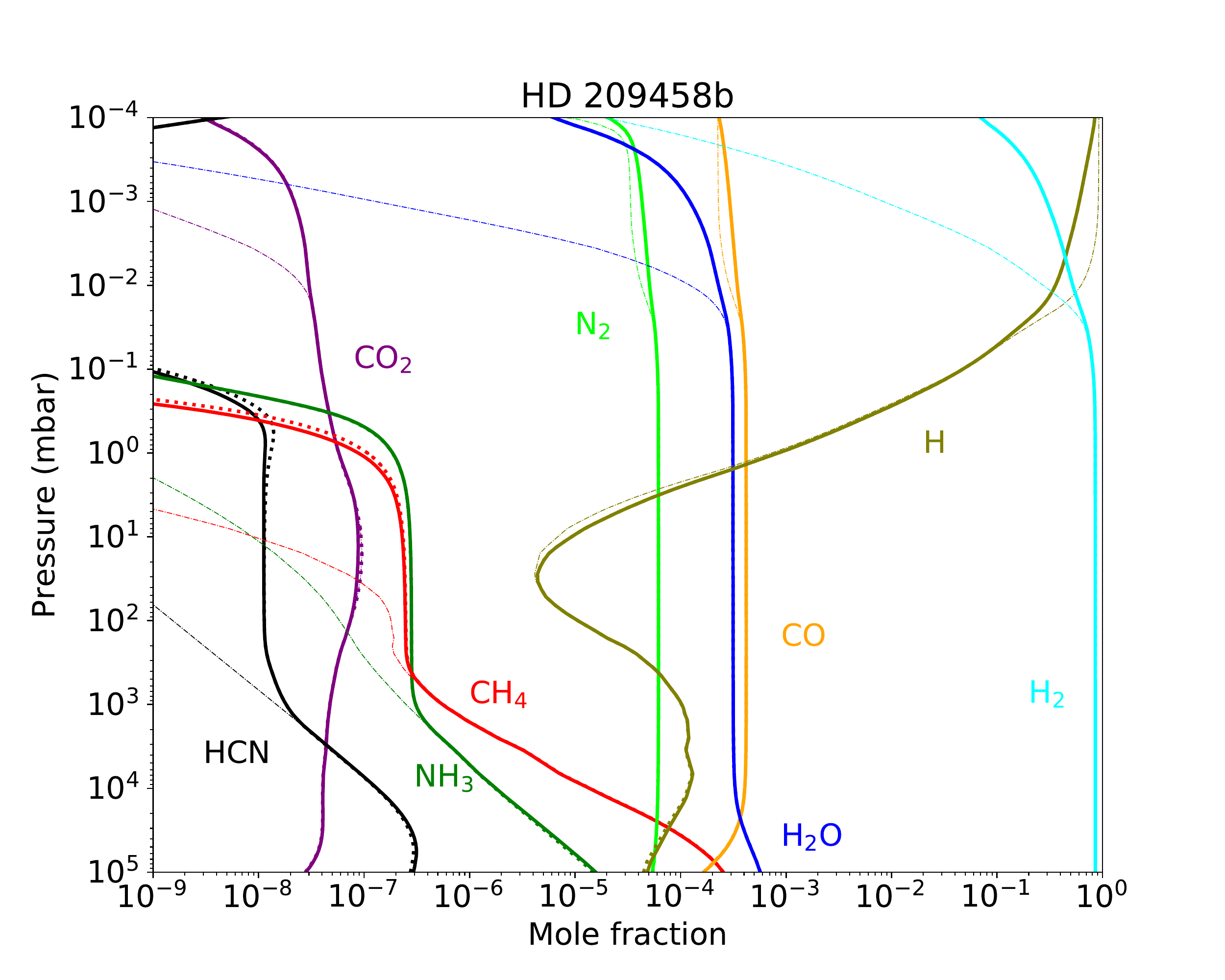}
  \includegraphics[width=\columnwidth]{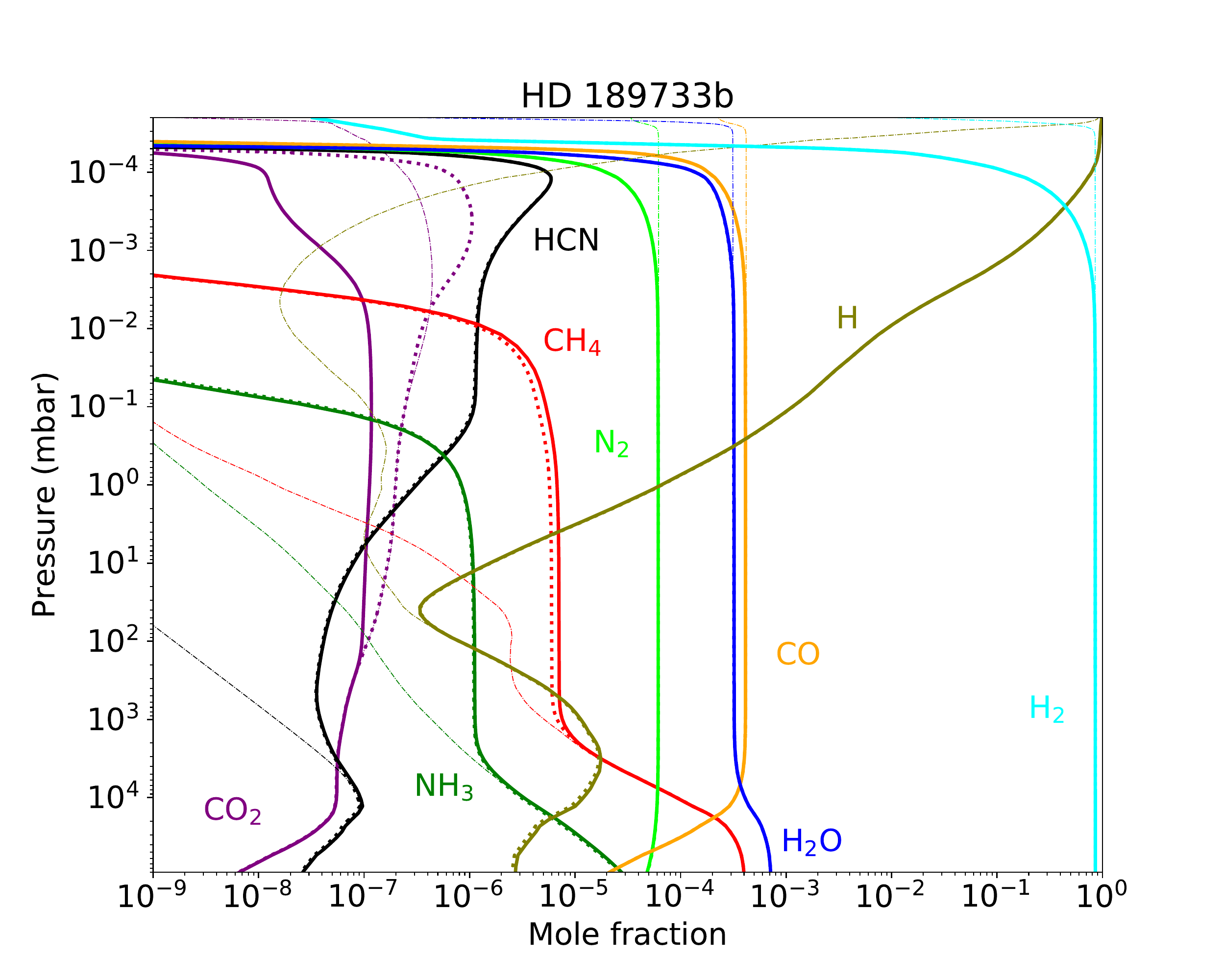}
  \caption{Vertical abundances profiles of the main atmospheric constituents in two Hot Jupiters: HD 209458b (top) and HD 189733b (bottom). The abundances obtained with the updated chemical scheme (solid lines) are compared to the ones obtained with the former one of \cite{Venot2012} (dotted lines). Thermochemical equilibrium is shown with thin dashed-dotted lines.}
  \label{fig:HD209_HD189}
  \end{figure}

  We have first applied our 1D kinetic model to the atmospheres of HD 209458b and HD 189733b. 
  We have used the same thermal (Fig.~\ref{fig:PTprofiles}) and eddy diffusion coefficient profiles as \cite{Moses2011}, that were used in \cite{Venot2012} with the original chemical scheme. The stellar and planetary characteristics are the same as in \cite{Venot2012}. We have used solar elemental abundances \citep{lodders2010}, but to account for sequestration of oxygen in refractory elements of the deep atmospheric layers, we have removed 20 \% of oxygen. As can be seen in \fig{fig:HD209_HD189}, the update of the chemical scheme has a very moderate effect on the chemical composition of these two planets. 
  Whereas quenching levels of all species remain the same in HD~209458b, one can notice variations in HD~189733b. With V20, CO$_2$ is quenched about 100 mbar whereas it was not with V12 and quenching of CH$_4$ happens slightly deeper than with V12, indicating that the chemical lifetime of these species is longer with V20. Although still different, this deeper quenching level of CH$_4$ goes in the direction of the results found by \cite{Moses2014} for this species. However, important differences are still present for the other species presented in this latter paper (i.e. C$_2$H$_2$, NH$_3$, HCN).

  \subsection{Warm Neptunes}
\begin{figure}[!h]
  \includegraphics[width=\columnwidth]{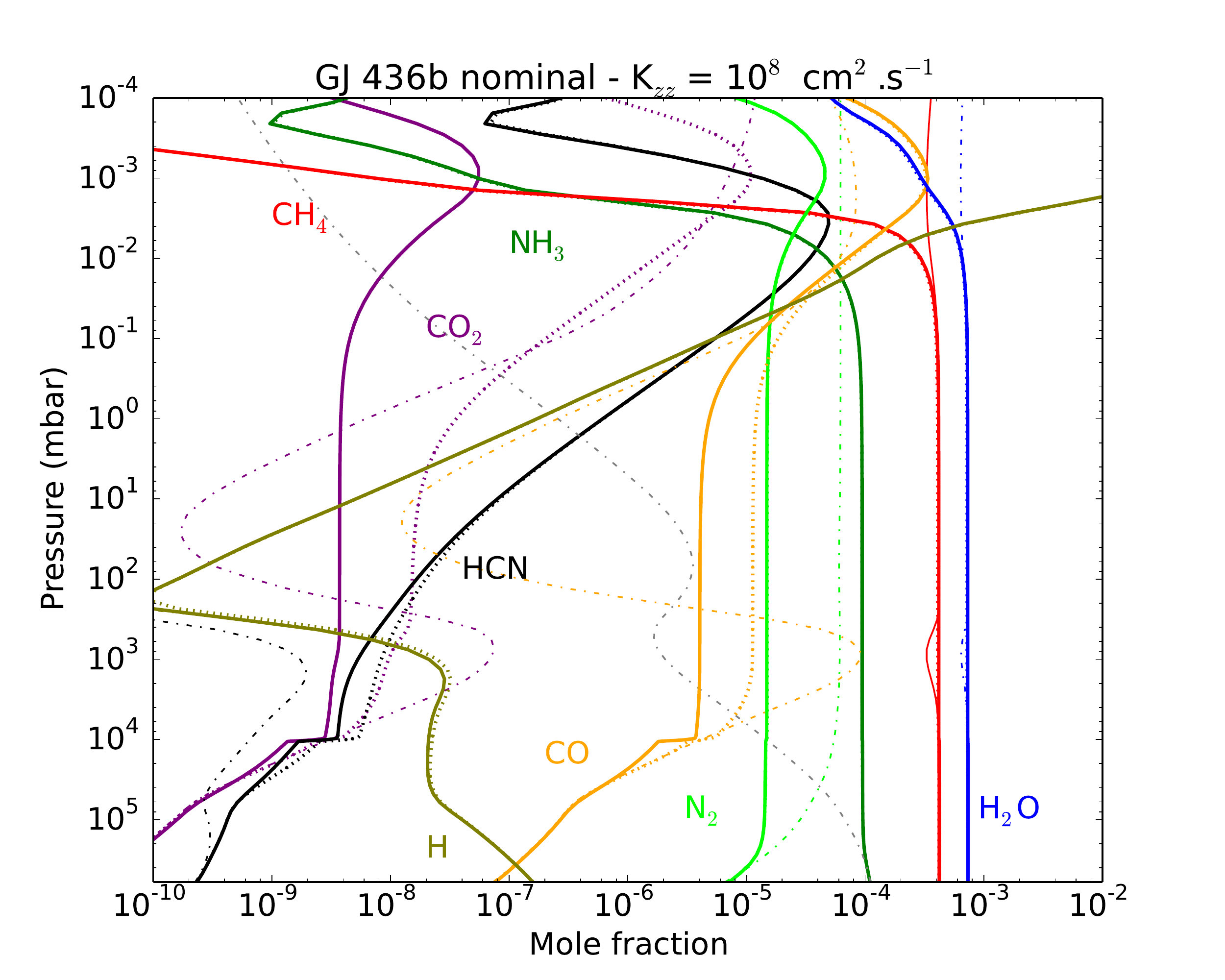}
  \includegraphics[width=\columnwidth]{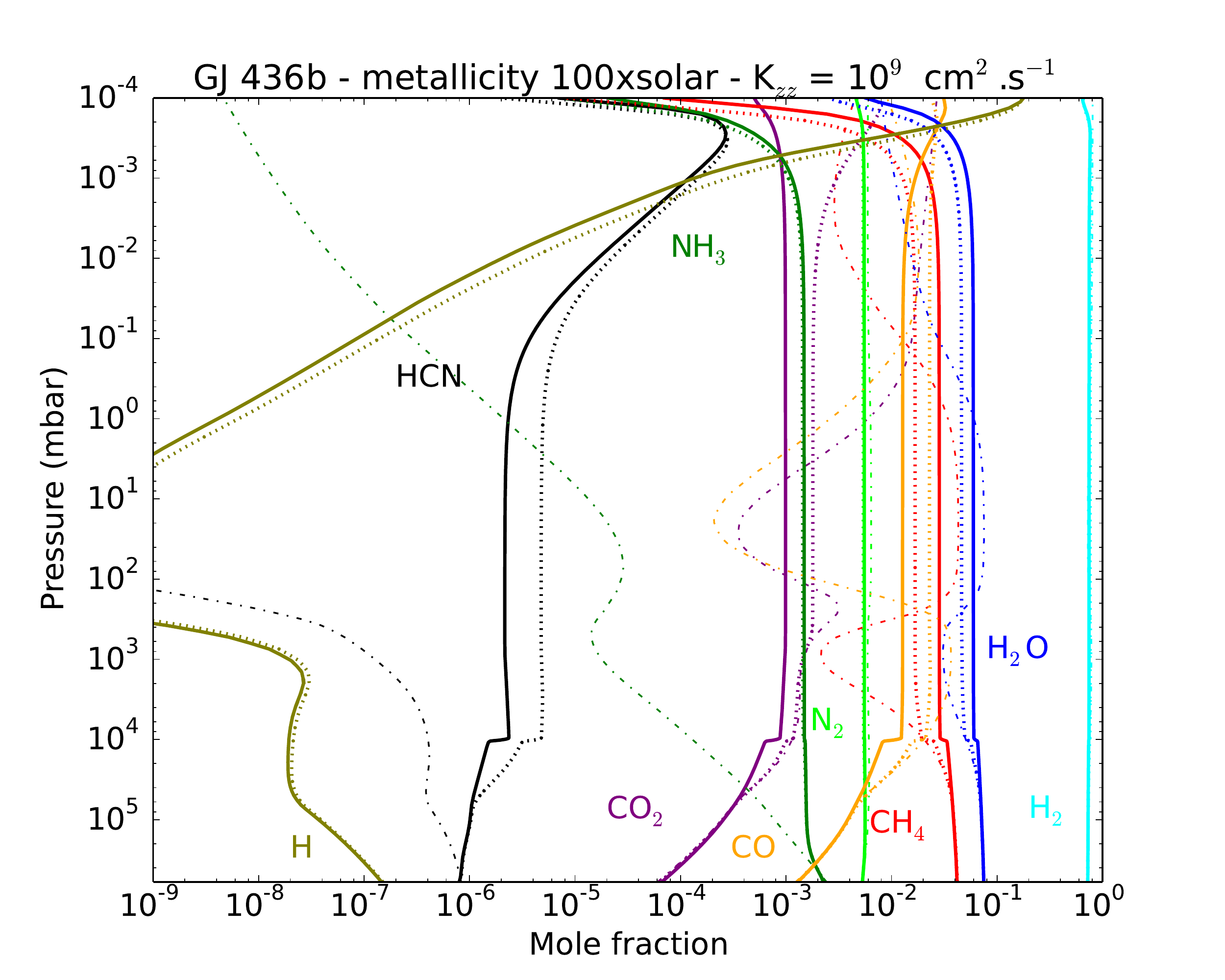}
  \caption{Same as Fig.~\ref{fig:HD209_HD189} for GJ 436b with a solar metallicity and K$_{zz}$=10$^8$cm$^2$s$^{-1}$ (top) and with a 100$\Sun$ metallicity and K$_{zz}$=10$^9$cm$^2$s$^{-1}$ (bottom). With the updated chemical scheme, CO sees its abundance decrease.}
  \label{fig:GJs}
  \end{figure}

\begin{figure}[!h]
 \includegraphics[width=\columnwidth]{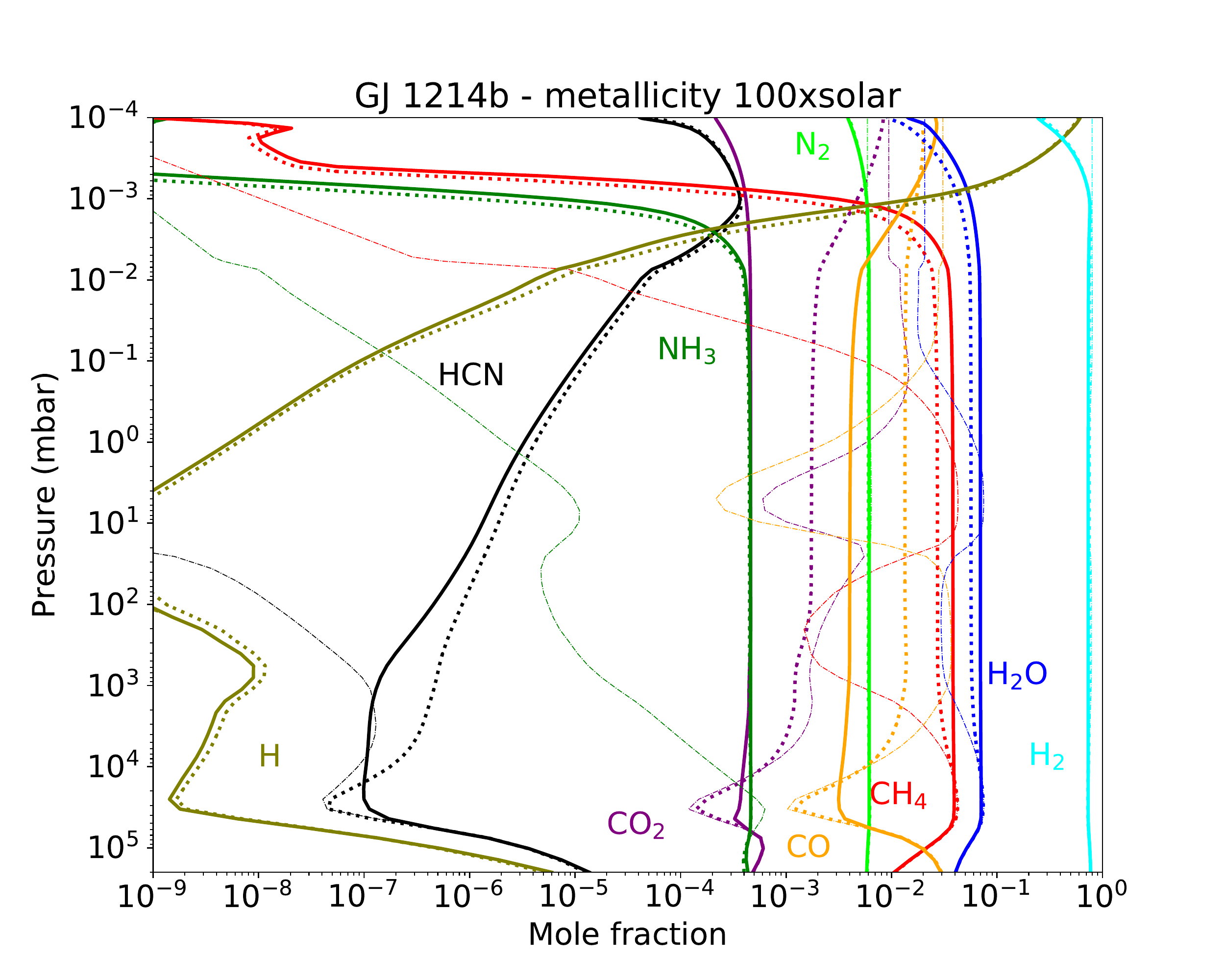}
  \caption{Same as Fig.~\ref{fig:HD209_HD189} for GJ 1214b with a 100$\Sun$ metallicity. As for GJ 436b, with the updated chemical scheme, CO sees its abundance decrease.}
  \label{fig:GJ1214}
  \end{figure}

  We have studied the effect of the methanol chemistry update on warm Neptunes, which are more temperate planets than hot Jupiters. We have applied our 1D kinetic model using alternatively the two chemical schemes to GJ 436b (see Fig.~\ref{fig:GJs}), assuming two different metallicities: solar and 100$\times$solar (100$\Sun$), as well as to GJ 1214b (see Fig.~\ref{fig:GJ1214}), assuming a metallicity 100$\Sun$.
  For both planets, the thermal profiles used are the same than in \cite{Venot2019}, i.e. determined with \texttt{ATMO} \citep{tremblin:2015aa} for GJ 436b and the \texttt{Generic LMDZ GCM} \citep{charnay2015} for GJ 1214b  (Fig.~\ref{fig:PTprofiles}). For GJ 436b, we have assumed a constant eddy diffusion coefficient with altitude, and used two values (10$^8$ and 10$^9$ cm$^2$s$^{-1}$). For GJ 1214b, we have used the formula determined by \cite{charnay2015}: $K_{zz}=3 \times 10^7 \times P^{-0.4}$cm$^2$s$^{-1}$, with $P$ in bar.
  For all the above cases, we observe the same trends: the update of the chemical scheme leads to deeper quenching level, and thus lower abundances for CO, CO$_2$, and HCN. On the contrary, but for the same reason, CH$_4$ and H$_2$O are found to be more abundant (Figs.~\ref{fig:GJs} and ~\ref{fig:GJ1214}).
  
  For the model of GJ 436b with a high metallicity, CO and CH$_4$ have abundances that are very close in the quenching area. With a K$_{zz}$ of 10$^8$s cm$^2$s$^{-1}$, CO is the main C-bearing species whatever the chemical scheme used, but with a stronger vertical mixing as the one presented in \fig{fig:GJs} (i.e. K$_{zz}$= 10$^9$s cm$^2$s$^{-1}$), the main C-bearing species depends on the chemical scheme: CO with V12 and CH$_4$ with V20.

  \subsection{T Dwarfs: ULAS J1335+11}
\begin{figure}[!h]
  \includegraphics[width=\columnwidth]{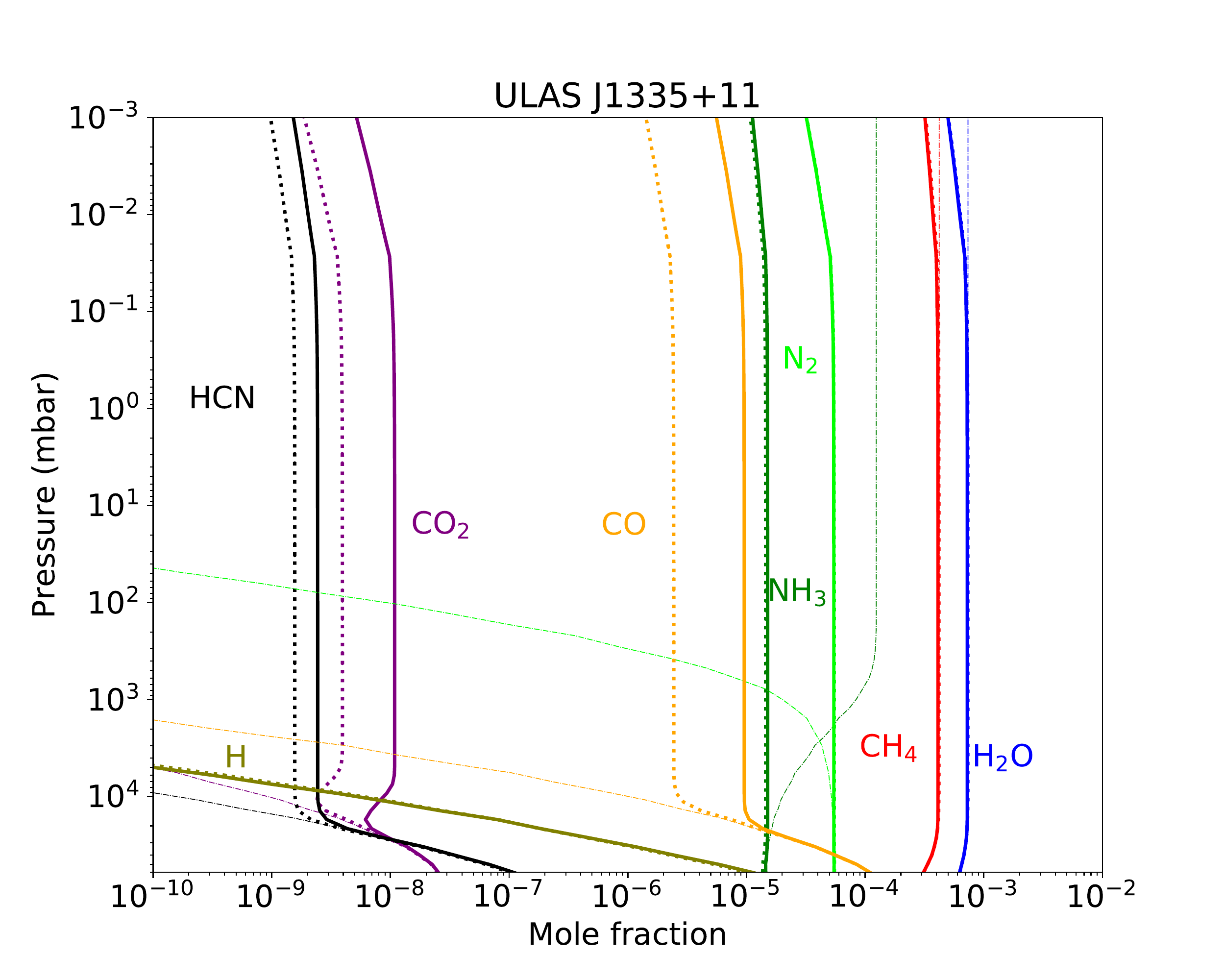}
  \caption{Same as Fig.~\ref{fig:HD209_HD189} for the brown dwarf ULAS J1335+11.}
  \label{fig:ULAS}
  \end{figure}

  We have modelled a typical T Dwarf ULAS J1335+11 \citep{leggett2009} with a thermal profile calculated with \texttt{ATMO} assuming an effective temperature of 500\,K and a surface gravity of $\log$(g)=4 (Fig.~\ref{fig:PTprofiles}). For the vertical mixing, we have assumed a constant eddy diffusion coefficient of 10$^6$ cm$^2$s$^{-1}$. Contrary to warm Neptunes, we observe that with V20, we obtain more CO and CO$_2$ in the atmosphere than with the former scheme (see Fig.~\ref{fig:ULAS}), because of the deeper quenching level. The increase in CO abundance is typically of a factor 3 at the effective temperature of late T dwarfs and can impact the CO absorption feature at 4.5 $\mu$m (see Sect.~\ref{implication}). At higher effective temperatures closer to the L/T transition, we did not observe any important differences between the updated and former scheme, similarly to the hot Jupiter cases.

  \subsection{Uranus and Neptune}
  For Uranus and Neptune, the update of the chemical scheme, coupled to the effect of composition on the thermal profile, has a significant effect on the oxygen chemistry. Taking the nominal cases of \citet{Cavalie2017} for both planets, i.e. O/H$<$160$\Sun$ (Uranus) and $=$480$\Sun$ (Neptune), a deep $K_{zz}$$=$$10^8$\Kunit, an upper tropospheric CH$_4$ mole fraction of 4\%, and a ``3-layer'' thermal profile, the model results in upper tropospheric mole fractions of CO of 7.8\dix{-8} and 3.8\dix{-6}, i.e. 34 and 19 times (respectively) above model results using the former chemical scheme and above the observed abundances.
  
\begin{figure}[!h]
  \includegraphics[width=\columnwidth]{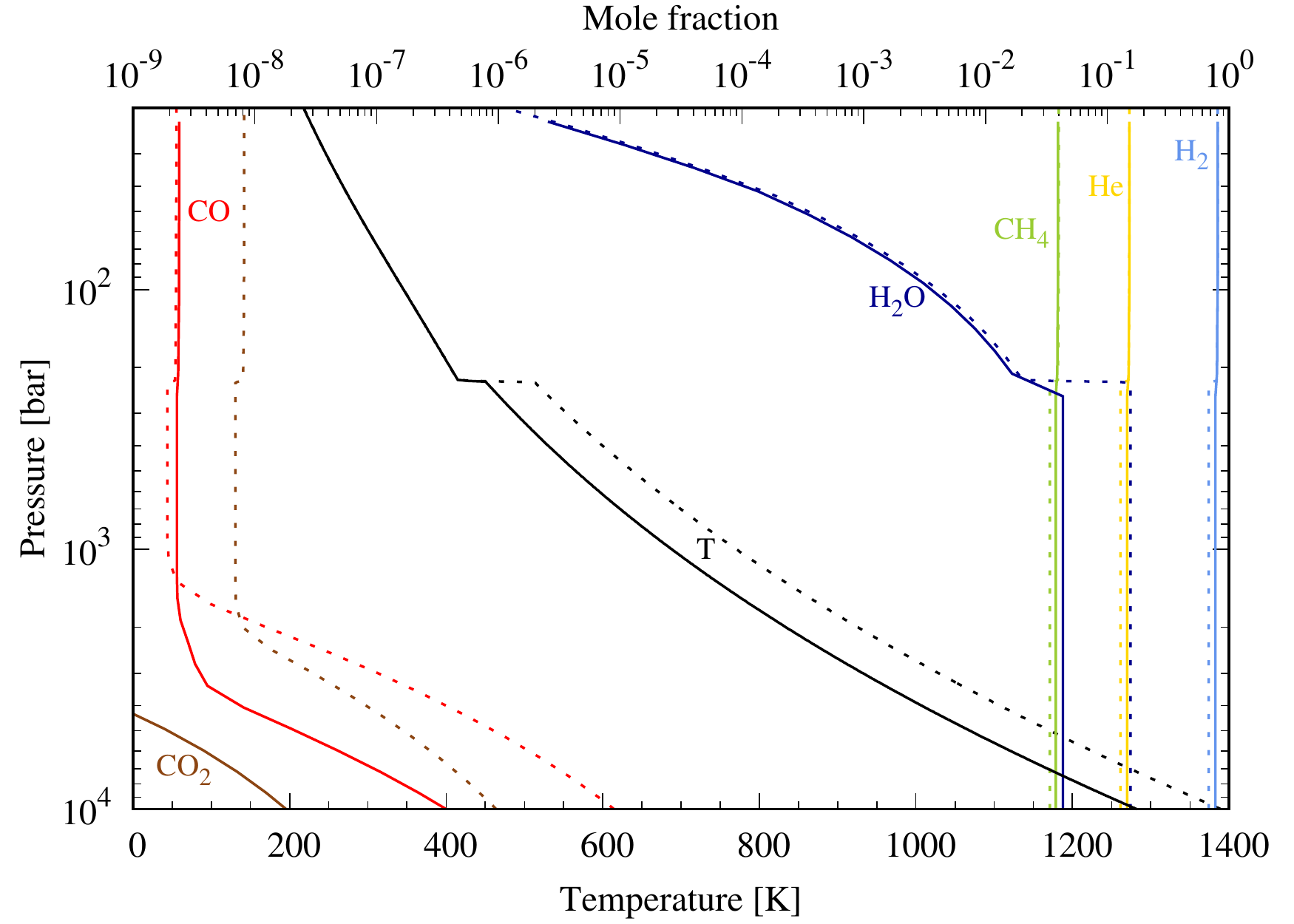}
  \includegraphics[width=\columnwidth]{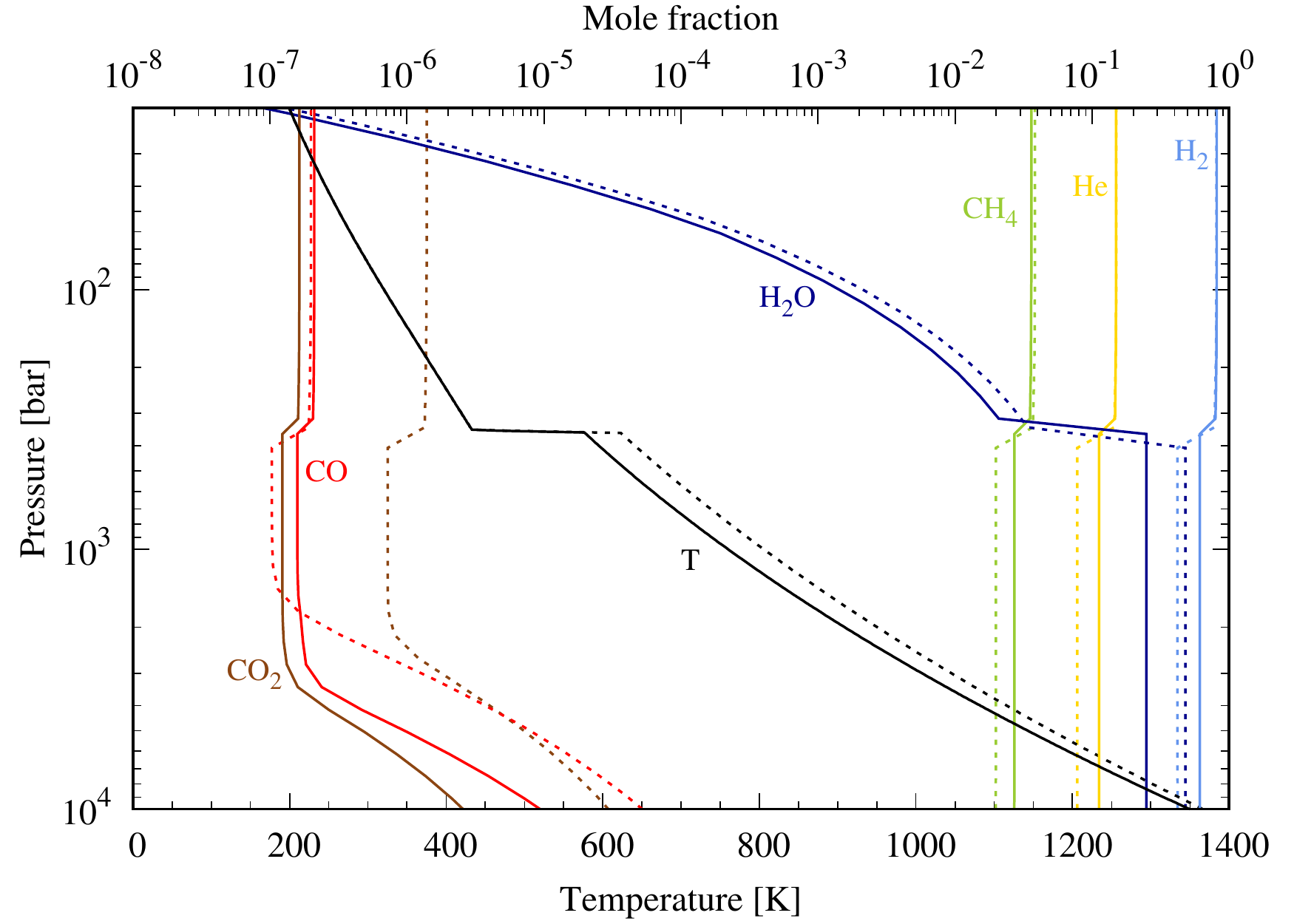}
  \caption{Vertical abundances profiles of the main atmospheric constituents and oxygen species in Uranus (top) and Neptune (bottom). The abundances obtained with the updated chemical scheme (solid lines), and O/H $<$45 and 250$\Sun$ for Uranus and Neptune (resp.), are compared to the ones obtained with the former one of \cite{Venot2012} (dotted lines), with $<$160 and 480$\Sun$ for Uranus and Neptune (resp.). The thermal and abundance profiles are thus not obtained with the same boundary conditions (see text for more details).}
  \label{fig:Uranus_Neptune}
  \end{figure}

  This implies that less H$_2$O is required in the layers where thermochemical equilibrium prevails to fit the observations of CO. As a consequence the ``3-layer'' temperature profiles are colder than in the nominal cases of \citet{Cavalie2017}, because the mean molecular weight gradient at the H$_2$O condensation level is smaller and produces therefore a less sharp temperature increase in this altitude region. The quench level then occurs deeper, enabling more CO to be transported towards the observable levels. We find that the upper tropospheric CO can be reproduced in Uranus and Neptune with an O/H of $<$45$\Sun$ and 250$\Sun$. The corresponding model results are displayed in \fig{fig:Uranus_Neptune}. The changes induced by the new chemical scheme are slightly more significant for Uranus than for Neptune. \\
  
  \subsection{Summary}
  The effect of the update depends on the temperature of the quenching level, as well as on the shape of abundance profiles. On one hand, if quenching happens at a temperature higher than $\sim$ 1500 K and at quite low pressure (0.1--1 bar, typically what happens in hot Jupiters atmospheres tested here), no substantial changes occur.
  On another hand, if quenching happens at lower temperature but higher pressure ($>$10 bars), then the quenching level is modified, consequently affecting the molecular abundances in upper layers. In all the cases we tested, we observe a deeper quenching level with the updated scheme V20.
  Molecular abundances are affected by the update depending on their slope at the now deeper quench level: if the abundance increases with altitude, the abundance will be lowered (e.g. CO and CO$_2$ in GJ 436b); and if the abundance decreases with altitude, the abundance will be enhanced (CO in Uranus, Neptune, and ULAS J1335+11).

\section{Interpretation of the results \label{interpretation}}

\subsection{0D model}

To understand the changes of kinetics, and thus of abundances, observed in the atmospheres modeled in this paper, we run our 0D model at the pressure and temperature where CO is quenched in GJ 436b (i.e. 10 bars and 1150 K) and where CH$_4$ is quenched in HD 209458b (i.e. 0.4 bar and 1500 K), and in HD 189733b (i.e. 1.5 bar and 1500 K) with our chemical schemes.

\begin{figure*}[!h]
  \includegraphics[width=\columnwidth]{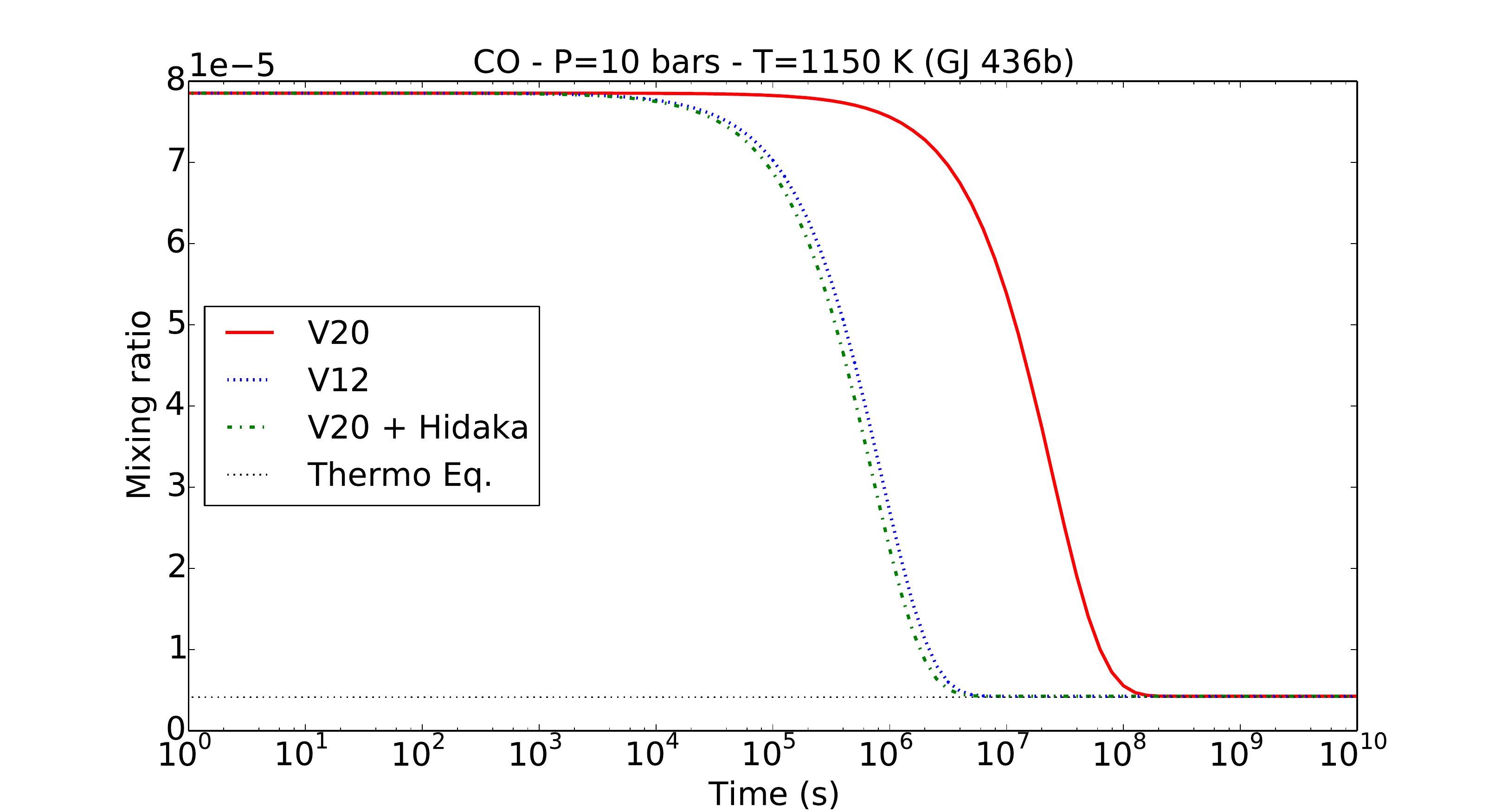}
  \includegraphics[width=\columnwidth]{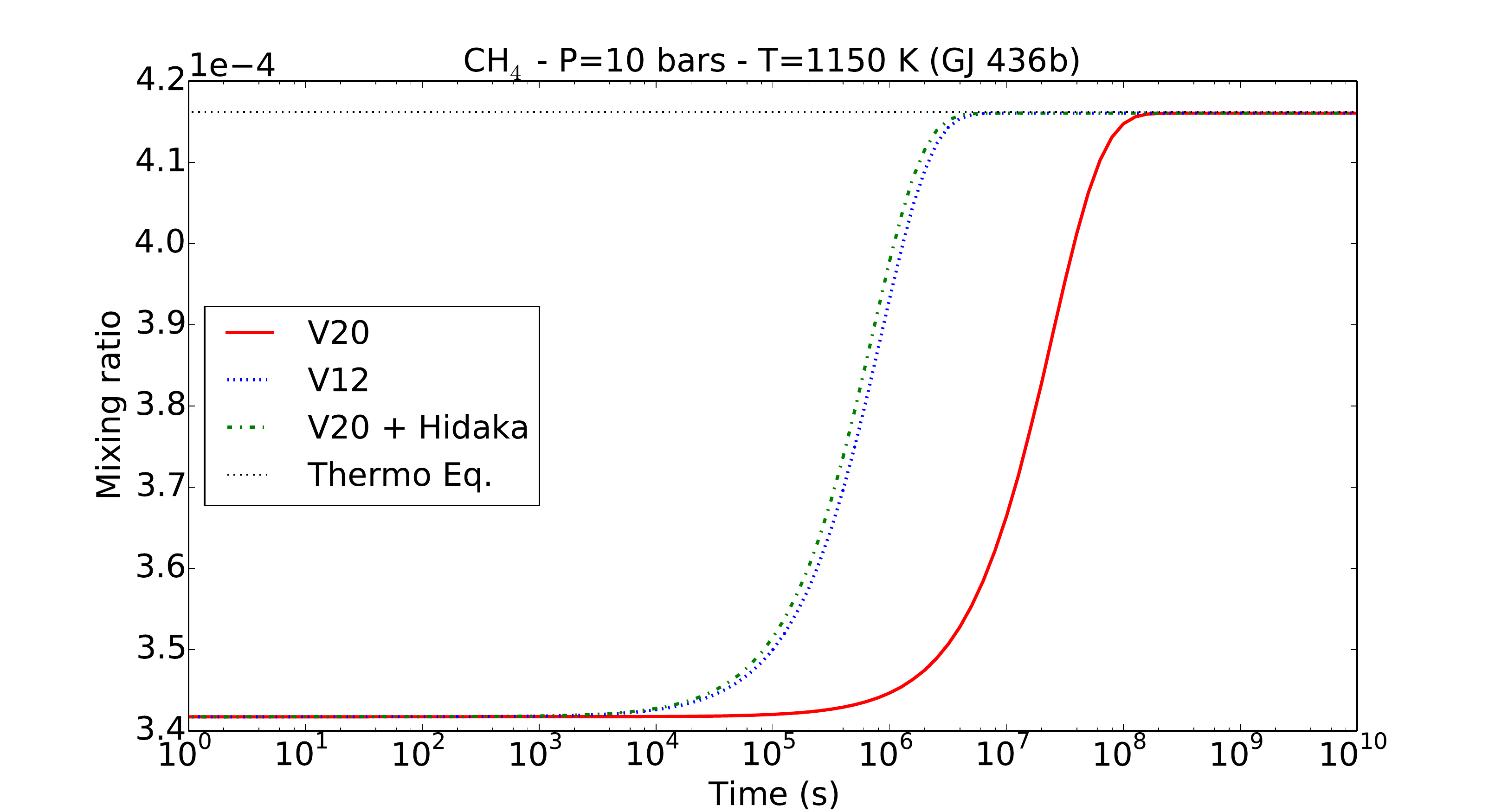}
  \includegraphics[width=\columnwidth]{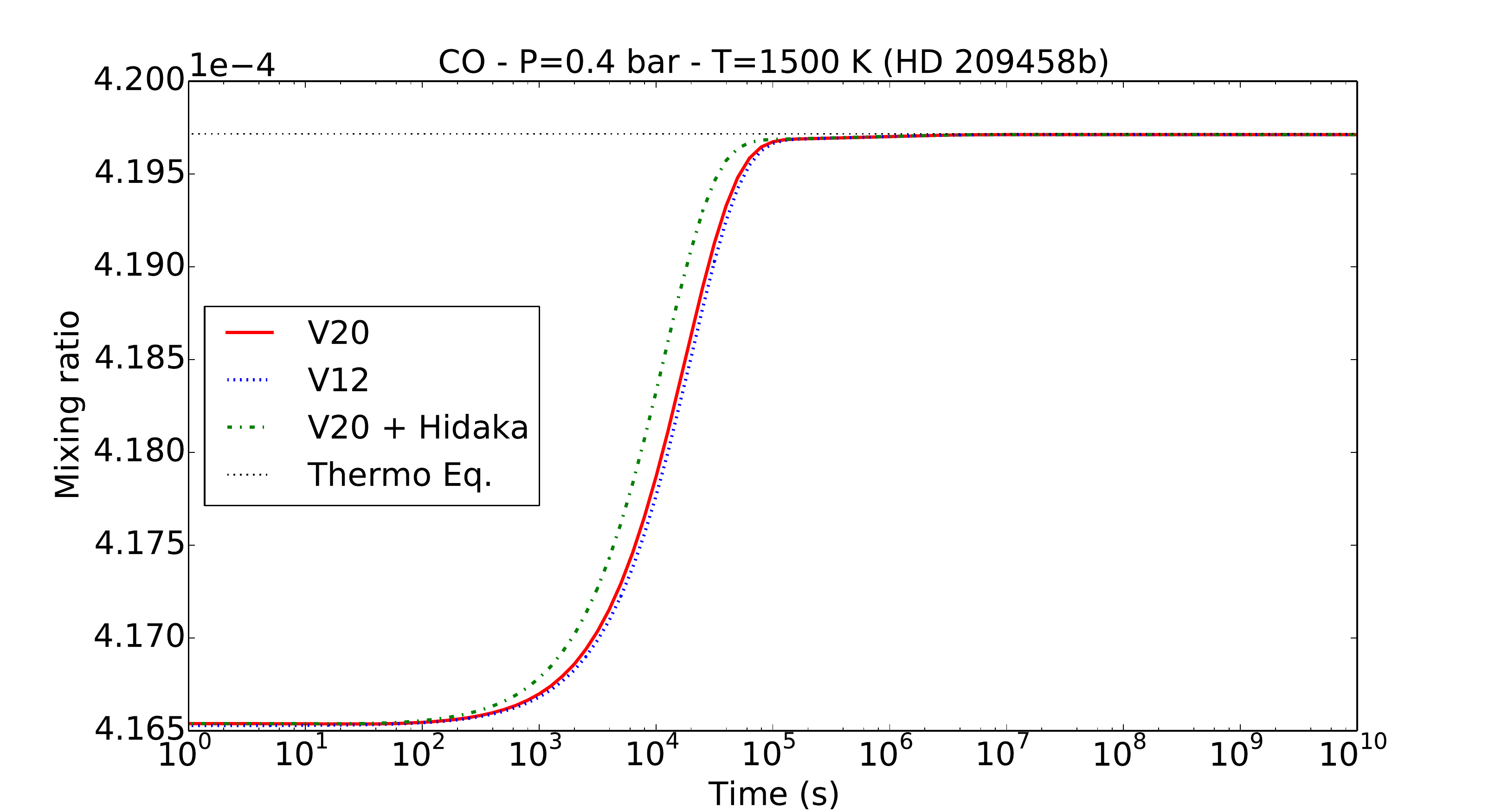}
  \includegraphics[width=\columnwidth]{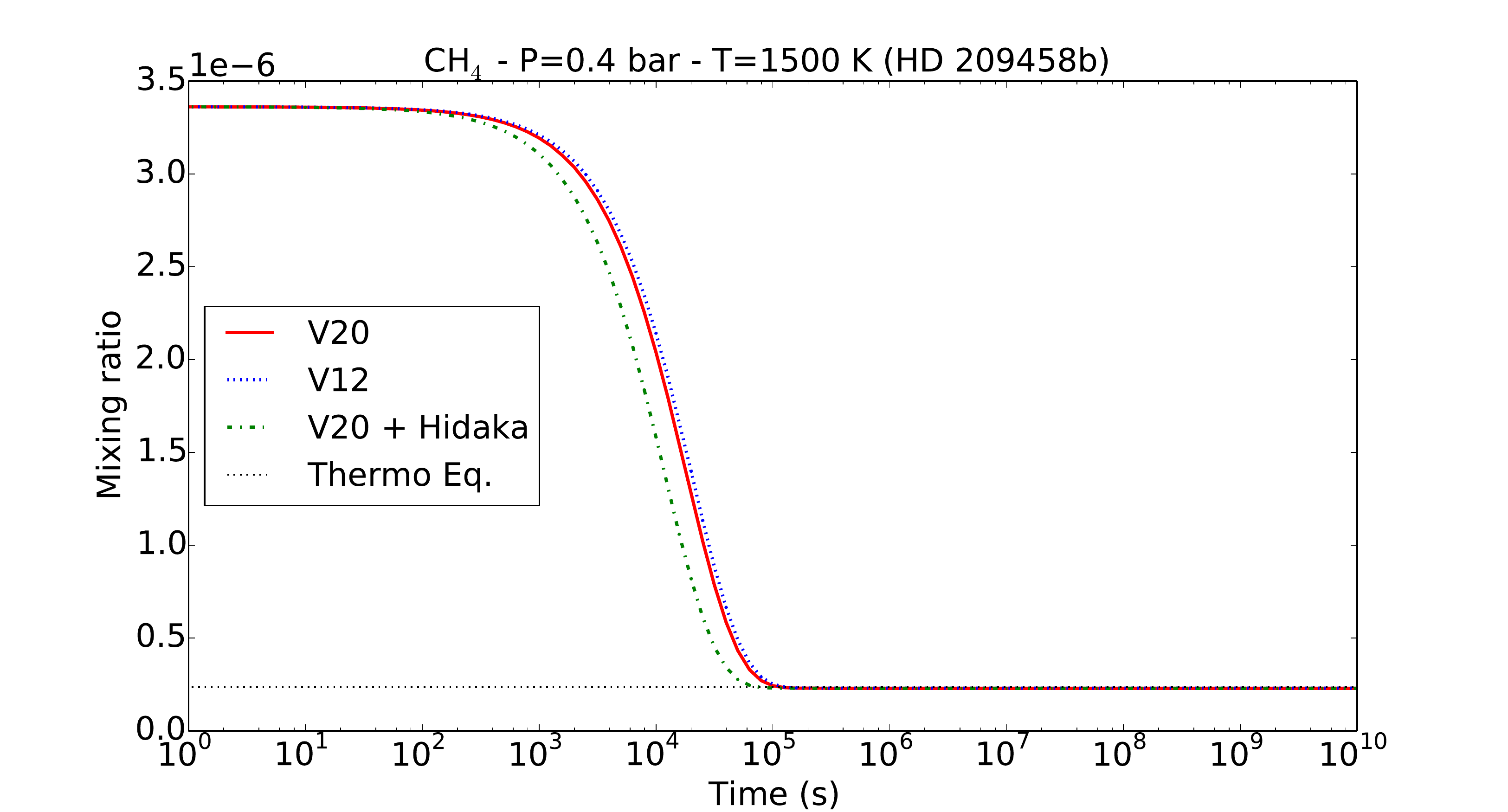}
  \includegraphics[width=\columnwidth]{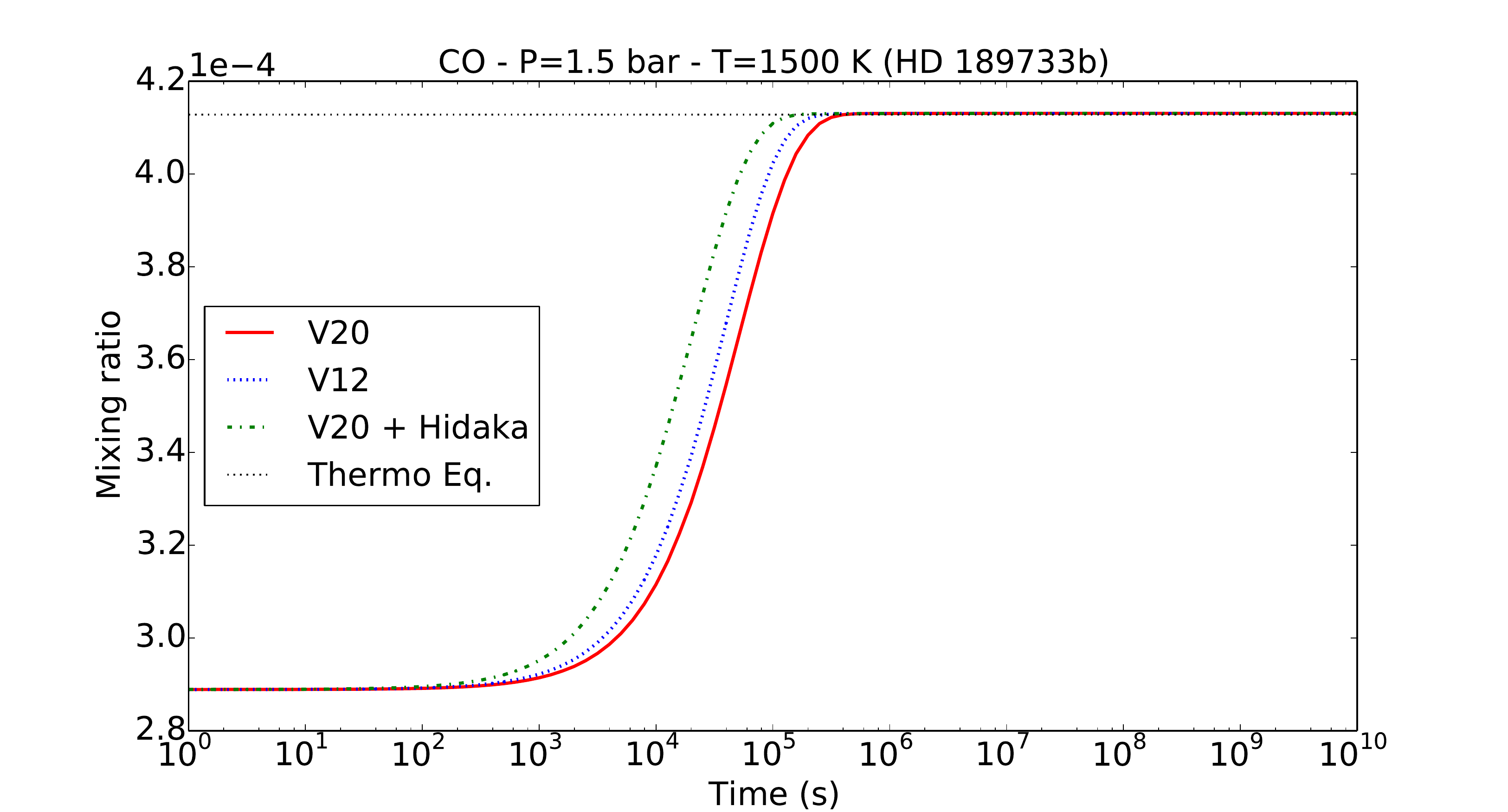}
  \includegraphics[width=\columnwidth]{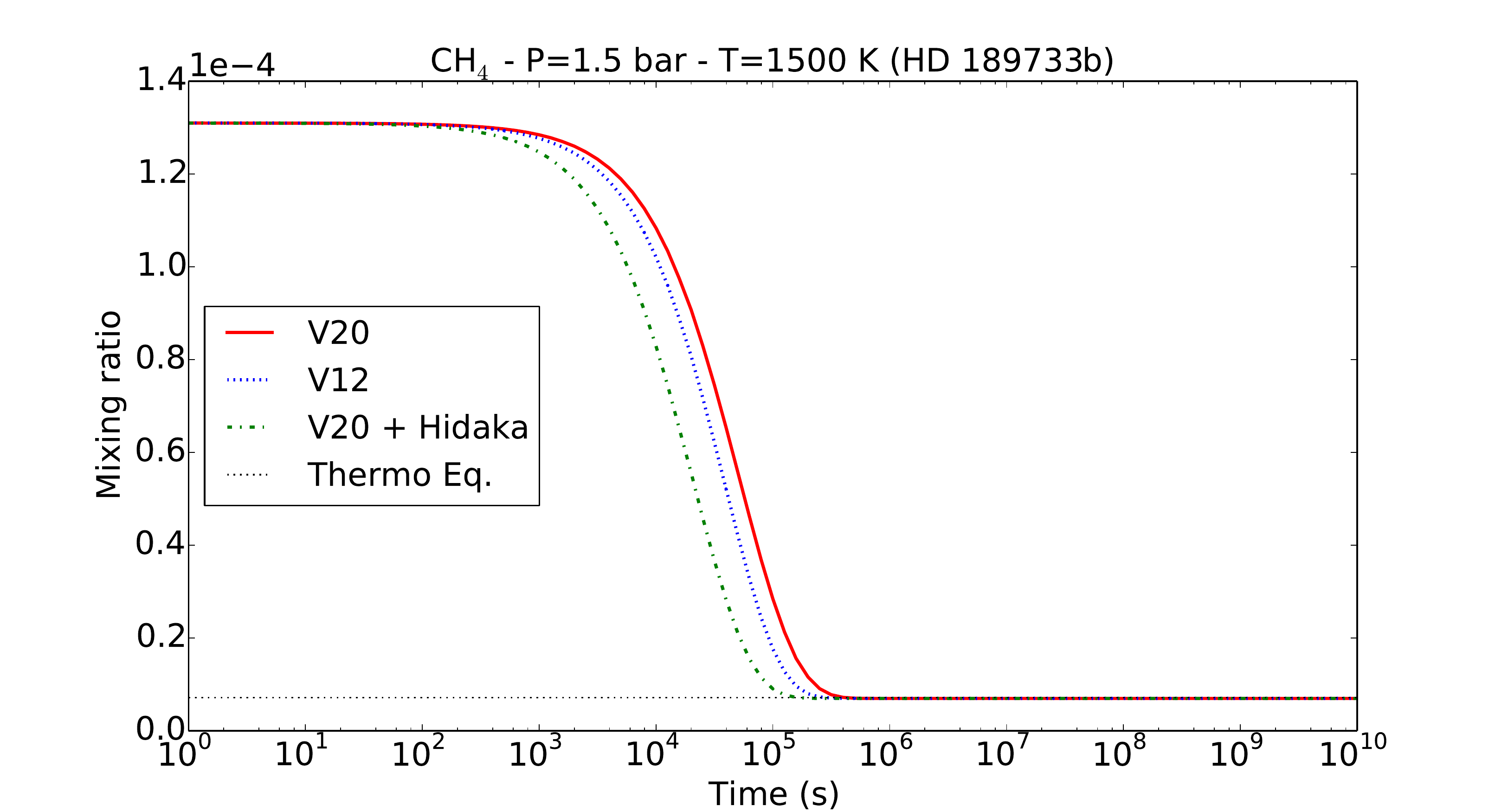}
  \caption{Temporal evolution of the abundance of CO (left) and CH$_4$ (right) at their quenching levels in GJ 436b (top), HD 209458b (middle), and HD 189733b (bottom). The corresponding pressures and temperatures are indicated on each panel. The abundances obtained with the updated chemical scheme (solid red lines) are compared to the ones obtained with the former one of \cite{Venot2012} (dotted blue lines), to the updated chemical scheme to which has been added Hidaka's reaction (dotted-dashed green lines) and to the thermochemical equilibrium at these conditions of P and T (dotted black lines). The initial conditions are the thermochemical abundances (assuming solar elemental abundances) at 1 bar and 1100 K (for GJ 436b), at 3.5 bars and 1750 K (for HD 209458b), and at 13 bars and 1570 K (for HD 189733b)}
  \label{fig:0D}
  \end{figure*}

\begin{figure}[!h]
  \includegraphics[width=\columnwidth]{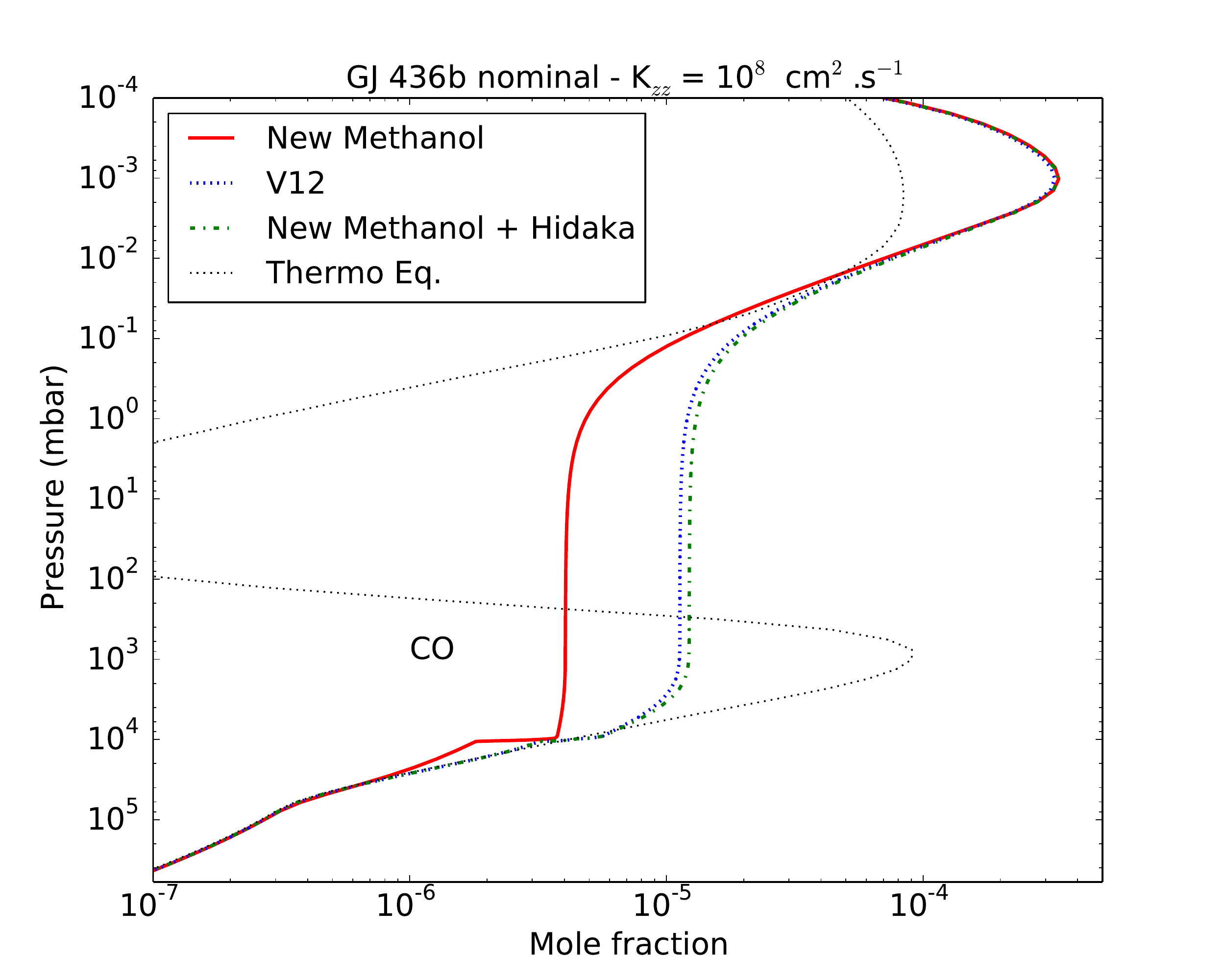}
  \includegraphics[width=\columnwidth]{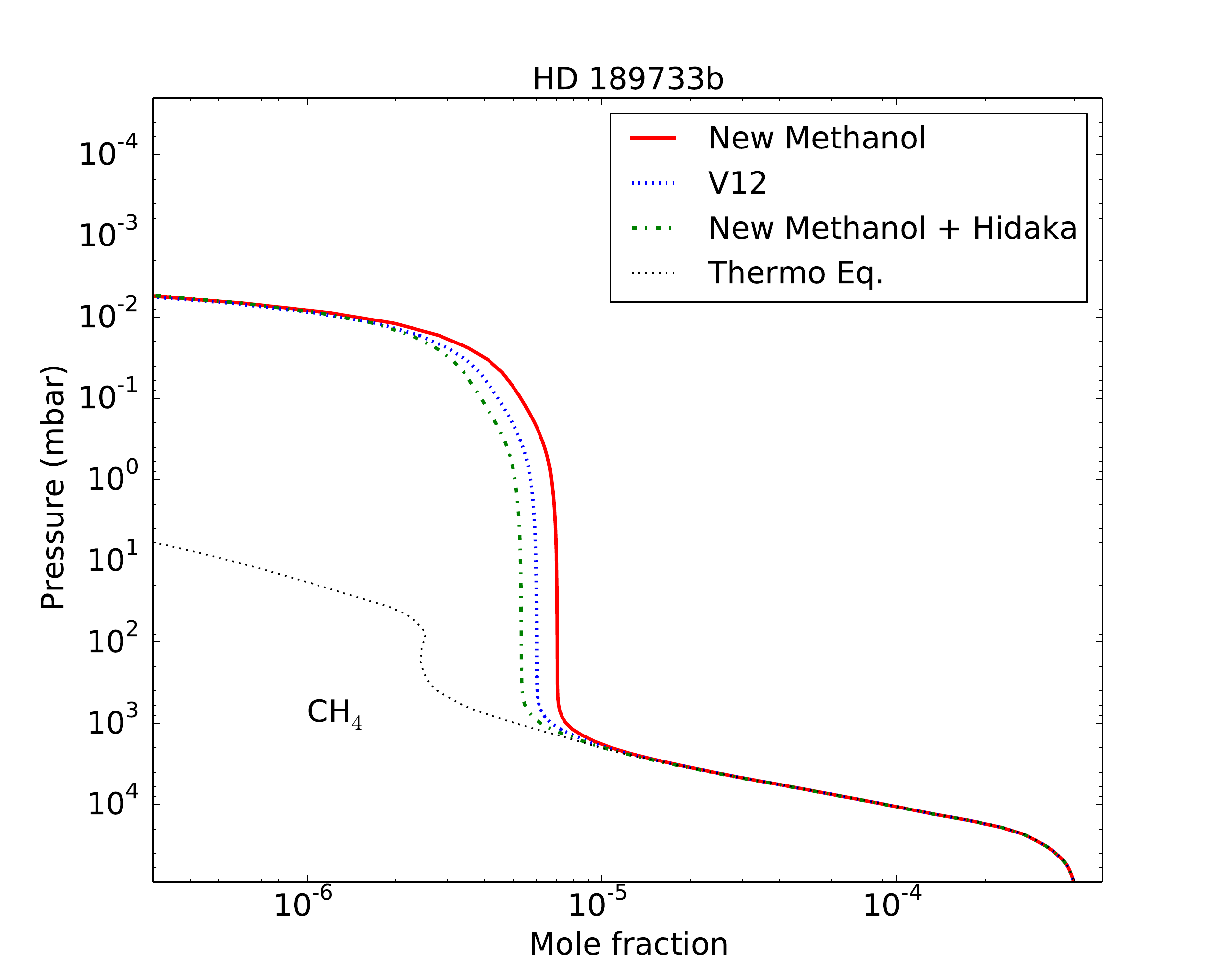}  \includegraphics[width=\columnwidth]{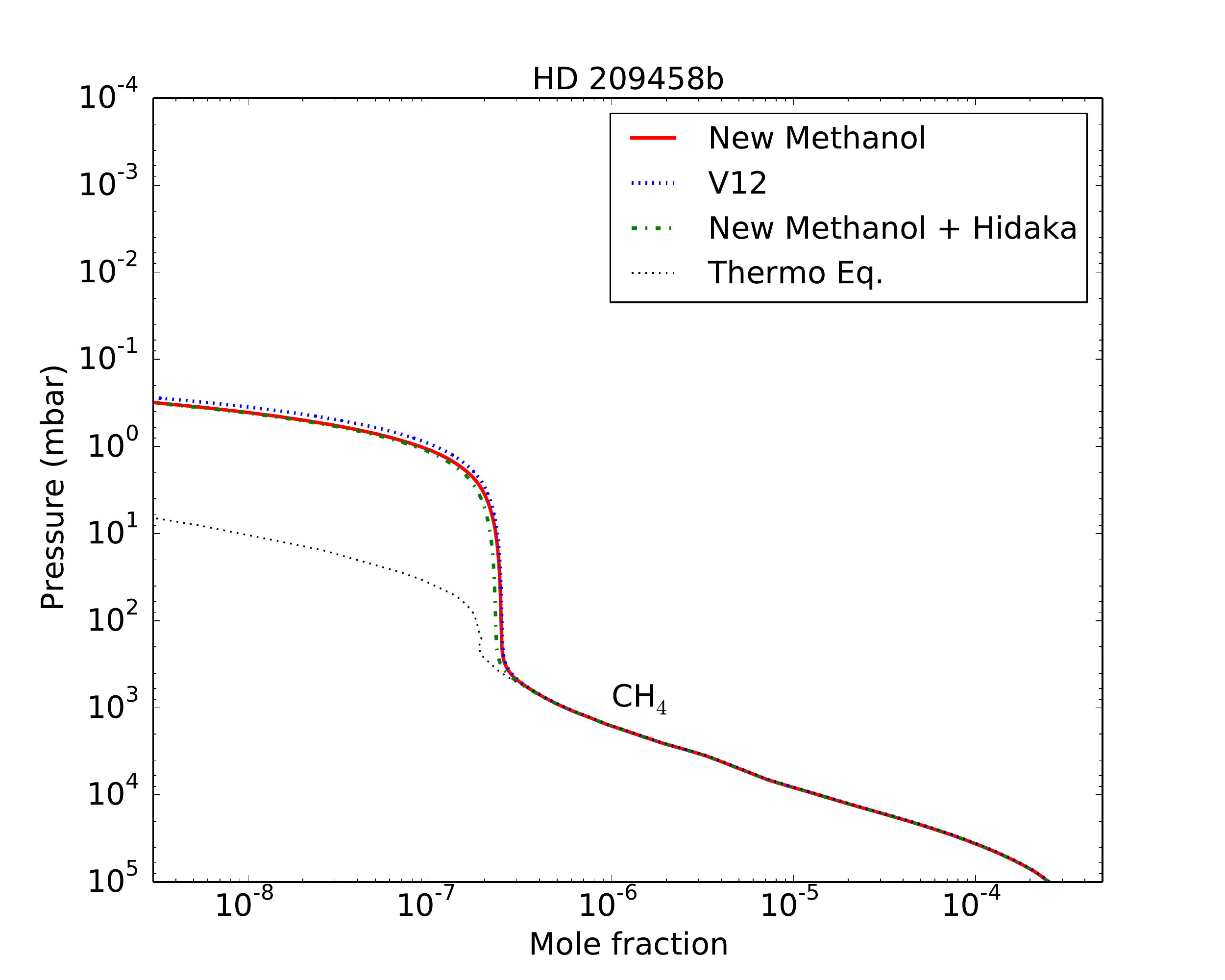}
  \caption{Vertical abundances profiles of CO in GJ 436b (top) and of CH$_4$ in HD 189733b (middle) and in HD 209458b (bottom) using different chemical schemes, as labelled.}
  \label{fig:1D_modif10}
  \end{figure}

On one side, at the level of CO quenching in GJ 436b (Fig.~\ref{fig:0D}), we observed that the kinetics of CO and CH$_4$ are very much slower with V20 than with V12. The difference is of two orders of magnitude.
We identify that this slow-down in the updated scheme is due to the \textit{non-presence} of the reaction CH$_3$OH$+$H$\rightleftharpoons$CH$_3+$H$_2$O, which is included in the scheme of V12 with the reaction rate proposed by \citet{Hidaka1989}.
The addition of this unique reaction to our new chemical scheme (scheme called hereafter ``V20+Hidaka'') accelerates the kinetics of CH$_4$ and CO (see Fig.~\ref{fig:0D}, top) and brings the abundances of CO (as well as CO$_2$ and HCN) in the 1D model very close to that found with V12 (Fig.\ref{fig:1D_modif10}). The difference of CO abundance at 100 mbar is reduced from 7 ppm to 1 ppm (i.e. a factor 2.8 and 1.1 respectively). Note that the change in CO$_2$ abundance is due to the Hidaka reaction for pressures greater than 1 bar, but also to the reaction CO$+$OH$\rightleftharpoons$CO$_2+$H for lower pressures.

On the other side, at the levels of CH$_4$ quenching in HD~209458b and in HD~189733b (Fig.~\ref{fig:0D}, middle and bottom), there is only a minor difference (less than a factor 2) concerning the kinetics of CO and CH$_4$ in V12 and V20. This explains why we obtain (almost) the same chemical composition for these planets with both chemical schemes. Here also, adding Hidaka's reaction to V20 accelerates slightly the kinetics of CO and CH$_4$, but the variation remains small, about a factor $\sim$2. One can note also that the kinetics of ``V20+Hidaka'' is in reality further away from V12 than V20 is. This excessive acceleration explains the 1D abundance profiles of methane determined for these planets (Fig.~\ref{fig:1D_modif10}). CH$_4$ quenches at (slightly) higher altitude when Hidaka's reaction is included than with the original V20, even higher than what is obtained with V12. For HD 209458b, the deviation of CH$_4$ abundance at 100 mbar between V12 and V20 is of 4.5 ppb, whereas the gap between V12 and ``V20+Hidaka'' is about 20 ppb. These differences are really small, a factor 1.02 and 1.09 respectively.
In the case of CH$_4$ in HD 189733b (at 10 mbar), the difference between V12 and V20 is a little more important (1 ppm, i.e. a factor 1.2) than the gap between V12 and ``V20+Hidaka'' (0.6 ppm, i.e. a factor 1.1). However, compared to the factor 2.8 of deviation observed for CO in GJ 436b, all the differences of CH$_4$ abundances in hot Jupiters remain really minor. In this case of HD 189733b, it is interesting to compare the methane abundances obtained with those found in \cite{Moses2014}. This paper focuses in HD 189733b and compares the atmospheric abundances of several species, including CH$_4$, obtained using V12 and \cite{Moses2011}'s chemical scheme. At 10 mbar, CH$_4$ has an abundance of $\sim$10$^{-5}$ with \cite{Moses2011}'s scheme, and $\sim$6$\times$10$^{-6}$ with V12 (like in this study). The update of the scheme we perform here leads indeed to an increase of CH$_4$ abundance (to 7$\times$10$^{-6}$), so towards the result obtained with \cite{Moses2011}'s scheme, but the new value we derive remains lower, and still closer to the previous value obtained with V12.
\\
Finally, we can say that the reaction CH$_3$OH$+$H$\rightleftharpoons$CH$_3+$H$_2$O, with the reaction rate of \cite{Hidaka1989}, do have a role on the chemical composition of hot Jupiters, but the amplitude of variation generated by the addition of this single reaction in the new V20 scheme remains very small and is not crucial for the kinetics of conversion of CO/CH$_4$.

\subsection{Chemical pathways}

To understand the differences between the different panels of Figs. \ref{fig:0D} and \ref{fig:1D_modif10}, and thus why the update of the chemical scheme modifies significantly the atmospheric composition of warm Neptunes, T dwarfs, Giant Planets, but not hot Jupiters, we analysed the chemical pathways occurring in the different P-T conditions. We found that the behaviour of the hydrogen radical is the key to explain the differences. 

\begin{figure}[!h]
  \includegraphics[width=\columnwidth]{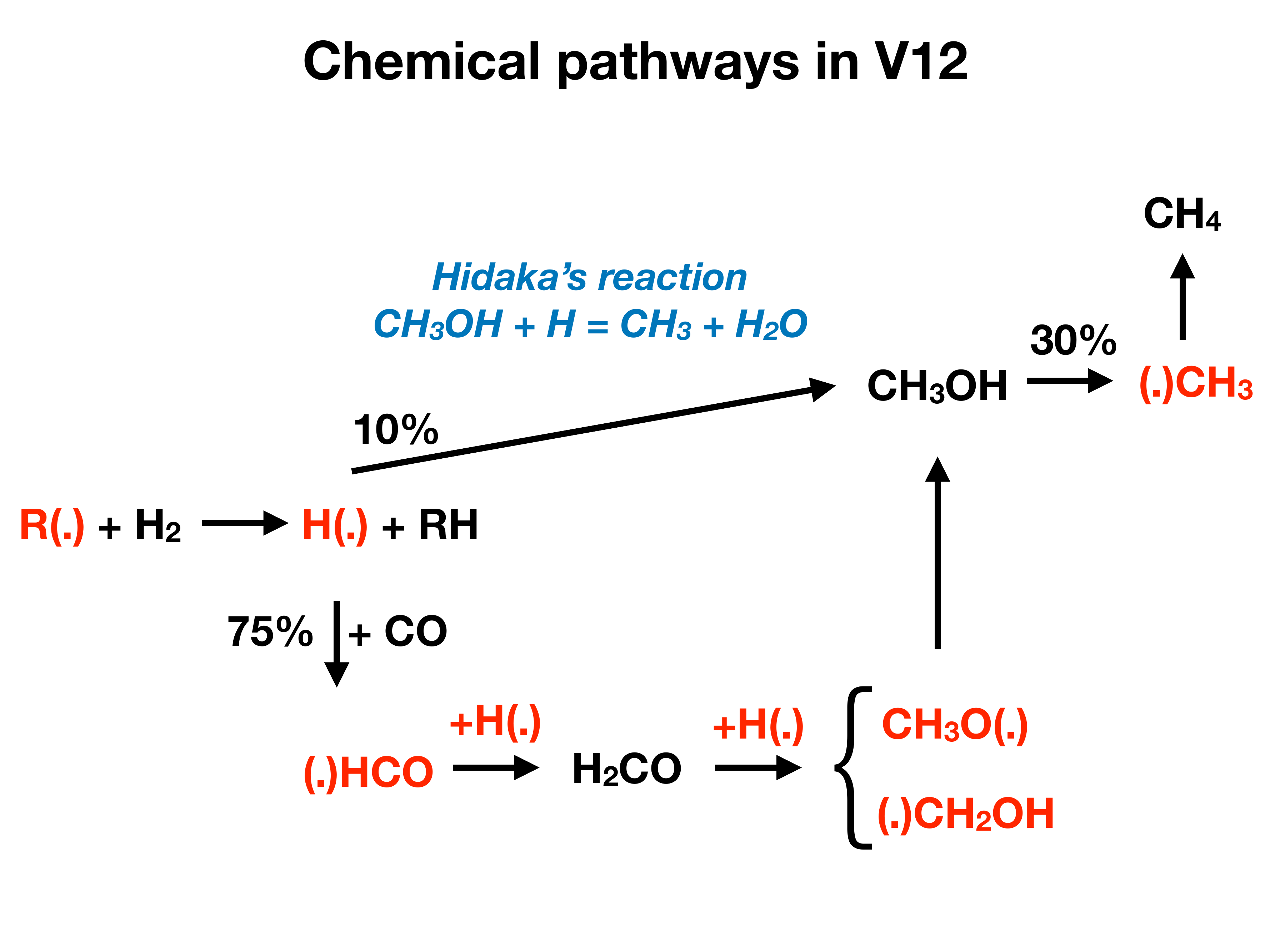}
    \caption{Chemical pathways controlling H<->CH$_4$ conversion in the chemical scheme of \cite{Venot2012} at 10 bars and 1150K.}
  \label{fig:pathV12}
  \end{figure}

At 10 bars and 1150 K (i.e. CO quenching level in GJ 436b), whatever the chemical scheme, the net production rate of H is positive. 
The kinetic analysis of V12 scheme is represented in Fig.~\ref{fig:pathV12}.
Hydrogen radical comes mainly from metathesis (H-transfer reactions) between H$_2$ and another radical (R(.)). 75\% of H react with CO to form HCO, which then reacts mainly with H to give formaldehyde (H$_2$CO). Then, by addition of H again, H$_2$CO forms either the CH$_2$OH or CH$_3$O radical. These two species, by metathesis, are transformed into methanol. 10\% of the hydrogen present in the atmosphere react with the formed methanol, through Hidaka's reaction CH$_3$OH + H $\longrightarrow$ CH$_3$ + H$_2$O, to form the methyl radical. CH$_3$ then reacts with H or H$_2$ to create CH$_4$. In this P-T condition, with this chemical scheme, 30\% of CH$_3$ comes from Hidaka's reaction. This reaction is thus very important in this context.

  \begin{figure}[!h]
  \includegraphics[width=\columnwidth]{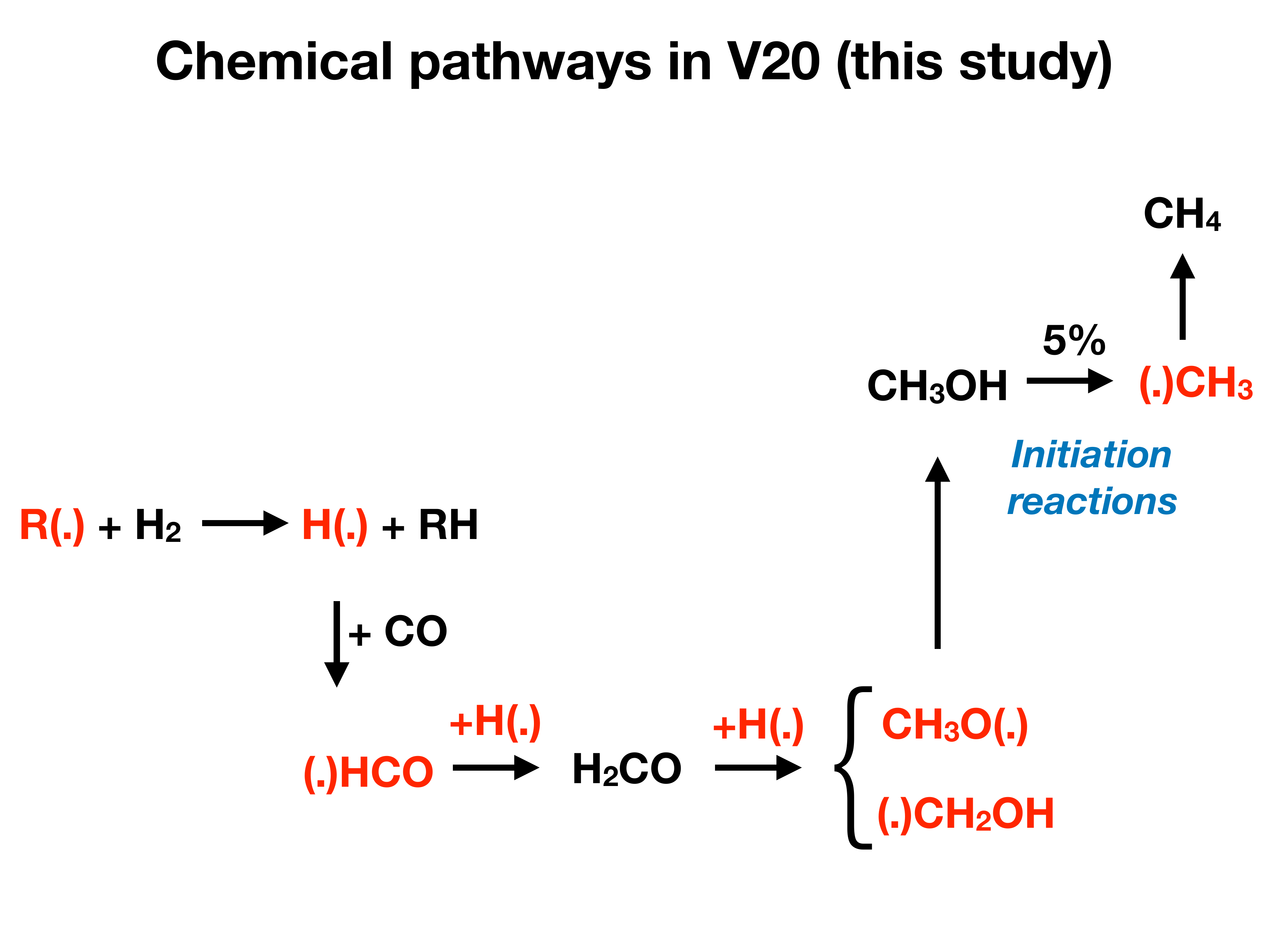}
  \caption{Chemical pathways controlling H$\leftrightarrow$CH$_4$ conversion in the updated chemical scheme at 10 bars and 1150K.}
  \label{fig:pathNM}
  \end{figure}
  
We performed the same analysis with the updated scheme (Fig.~\ref{fig:pathNM}). The production of methanol from H$_2$ is identical to that of V12. Then, Hidaka's reaction not being included in this scheme, H cannot react with CH$_3$OH to form CH$_3$. In V20 scheme, only 5\% of CH$_3$ comes from methanol, through the priming reaction CH$_3$OH (+M) $\longrightarrow$ CH$_3$ + OH (+M). The majority of methyl radical comes from the initiation reactions of methane (CH$_4$ (+M) $\longrightarrow$ CH$_3$ + H) and ethane (C$_2$H$_6$ (+M) $\longrightarrow$ CH$_3$ (+M)). Here also, CH$_3$ then reacts with H or H$_2$ to create CH$_4$. We see that the main difference between the two chemical schemes is due to the chemical pathways between CH$_3$OH and CH$_3$.

We performed the same analysis at 0.4 bar and 1500 K, i.e. CH$_4$ quenching level in HD 209458b. We found that the main chemical pathways are the same with the two chemical schemes. Contrary to the previous case, at this lower pressure, the net production rate of H is negative (i.e. positive loss rate). The majority of hydrogen (90 \%) is equally consumed to give H$_2$, CH$_3$, and CH$_4$. The remaining 10\% are involved in the following loop:
\begin{center}
\begin{tabular}{rcl}
H $+$ CO & $\longrightarrow$ & HCO\\
HCO $+$ H & $\longrightarrow$ & CO  $+$ H$_2$\\
\end{tabular}
\end{center}
Our analysis shows that at these pressure and temperature, Hidaka's reaction does not step in in the overall production/destruction of CH$_4$, CH$_3$, and CO, which leads to identical results between the two schemes.\\
The same analysis has been performed for the quenching level in HD~189733b and leads to the same global conclusion than in HD~209458b. However at these pressure and temperature (1.5 bar and 1500 K), Hidaka's reaction plays a minor role in V12: 0.1\% of CH$_3$ is produced through this reaction (vs 0\% and 30\% in the cases of HD~209458b and GJ436b, respectively) which explains why there is a larger difference between V12 and V20 for HD 189733b than for HD~209458b.\\

To summarise, the key to explain our results is the production rate of hydrogen. On one side, in a P-T domain where the production rate of H is positive, Hidaka's reaction will play a major role in V12 and thus there will be differences between the two schemes. On the other side, in a P-T domain where the loss rate of H is positive, then Hidaka's reaction does not play its rate-accelerating effect and results obtained with the two schemes will be very similar.

\section{Discussion}\label{implication}

\subsection{Implications for hot Jupiters}

The update of the chemical scheme does not impact fundamentally the predicted atmospheric composition of HD 209458b and HD 189733b, which can be considered as typical hot Jupiters, with a solar elemental composition. The main variation of abundance is the decrease of CO$_2$ in the upper atmosphere of HD~189733b. We calculated the synthetic transmission spectra of this planet with the forward model \texttt{TauRex} \citep{waldmann2015, waldmann2015b} and observed only a slight variation in the CO$_2$ absorption band at 4-5 $\mu$m (50 ppm). This difference would hardly be distinguishable with future observations performed with JWST/NIRSpec or ARIEL, at least with one single observation. Stacking together several transits data will reduce the error bars, making the distinction eventually possible \citep{mugnaiEPSC2019}. The abundance of CO$_2$ being dependent of the quenching level in HD 189733b, an accurate estimation of its abundance could help to constrain and better understand the mixing occurring in hot jupiters atmospheres.

We confirm the abundances of NH$_3$, HCN, CH$_4$, and C$_2$H$_2$ obtained in \cite{Venot2012} with the previous chemical scheme. Although in the atmosphere of HD 189733b quenching of CH$_4$ happens deeper than with V12 (leading to a very small increase of the abundance of this species), the other aforementioned species are not affected by the update of the scheme. Thus, our global results are not modified in a way that would bring them closer to the results obtained by \cite{Moses2014}. As we explained in Sect.~\ref{interpretation}, in the atmosphere of hot Jupiters, the differences between our results and that of \cite{Moses2014} are thus not due only to the choice of the reaction rate of CH$_3$OH$+$H$\rightleftharpoons$CH$_3+$H$_2$O. This result comforts us with the idea that a global validation of a scheme prevails compared to individual reaction calculations.

\subsection{Implications for warm Neptunes}
The update of the chemical scheme has important consequences on the molecular composition of warm Neptunes, especially for atmospheres with high metallicities. The quenching level of CO$_2$, CO, and CH$_4$ being modified, the abundances of these species vary and even a change of the main C-bearing species can occur (Fig.~\ref{fig:GJs}). The change of chemical composition found for warm Neptunes has observational consequences. 

\begin{figure}[!h]
  \includegraphics[width=\columnwidth]{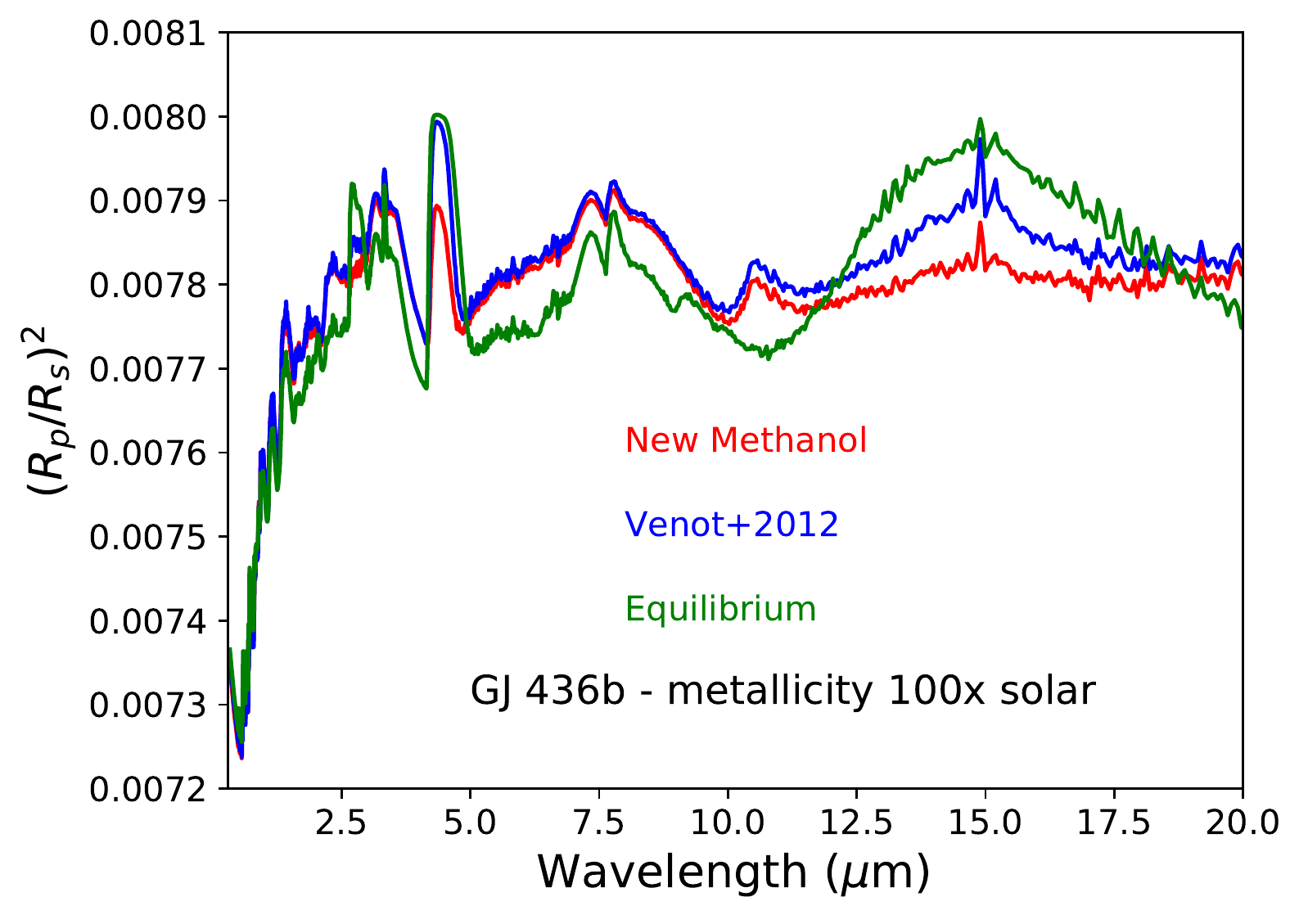}
  \caption{Synthetic transmission spectra of GJ 436b's atmosphere with a metallicity 100$\Sun$. The different spectra corresponds to the compositions at chemical equilibrium (green), and with disequilibrium compositions (K$_{zz}$=10$^9$cm$^2$s$^{-1}$) calculated with the \cite{Venot2012}'s scheme (blue), and with the updated scheme (red). The spectral resolving power is 300.}
  \label{fig:spectraGJ436b}
  \end{figure}

With the forward model \texttt{TauRex}, we have computed the synthetic transmission spectra for our models of GJ 436b with a high metallicity. We have calculated the spectra corresponding to the compositions at equilibrium, determined with V12 and the updated scheme (Fig. \ref{fig:spectraGJ436b}). 
First, we can note the important variations between the disequilibrium spectra and the one at equilibrium between 1--10 $\mu$m and in NH$_3$ band (11 $\mu$m), which are due to the high NH$_3$ abundance in disequilibrium models. The important departures in CO$_2$ band (15 $\mu$m) is due to the high abundance of CO$_2$ at low pressure in the equilibrium model. We can expect that future high-resolution observations of warm Neptunes such as GJ 436b could be able to detect the possible disequilibrium composition of these planets, even if cloudy \citep{kawashima2019}, and would certainly help to constrain the vertical mixing responsible of quenched abundances. 
Then, between the two disequilibrium spectra, important variations are visible in CO$_2$ absorption bands (4-5 $\mu$m, 15 $\mu$m). As this species is less abundant with the updated scheme, the absorption is lower in these bands, resulting in a lower $(R_p/R_s)^2$. Such a departure (up to 100 ppm) will be easily observable with future instruments such as JWST/MIRI. Thus, the choice of the chemical scheme is critical for an accurate constraint on vertical mixing.

\subsection{Implications for Brown Dwarfs}

\begin{figure}[!h]
  \includegraphics[width=\columnwidth]{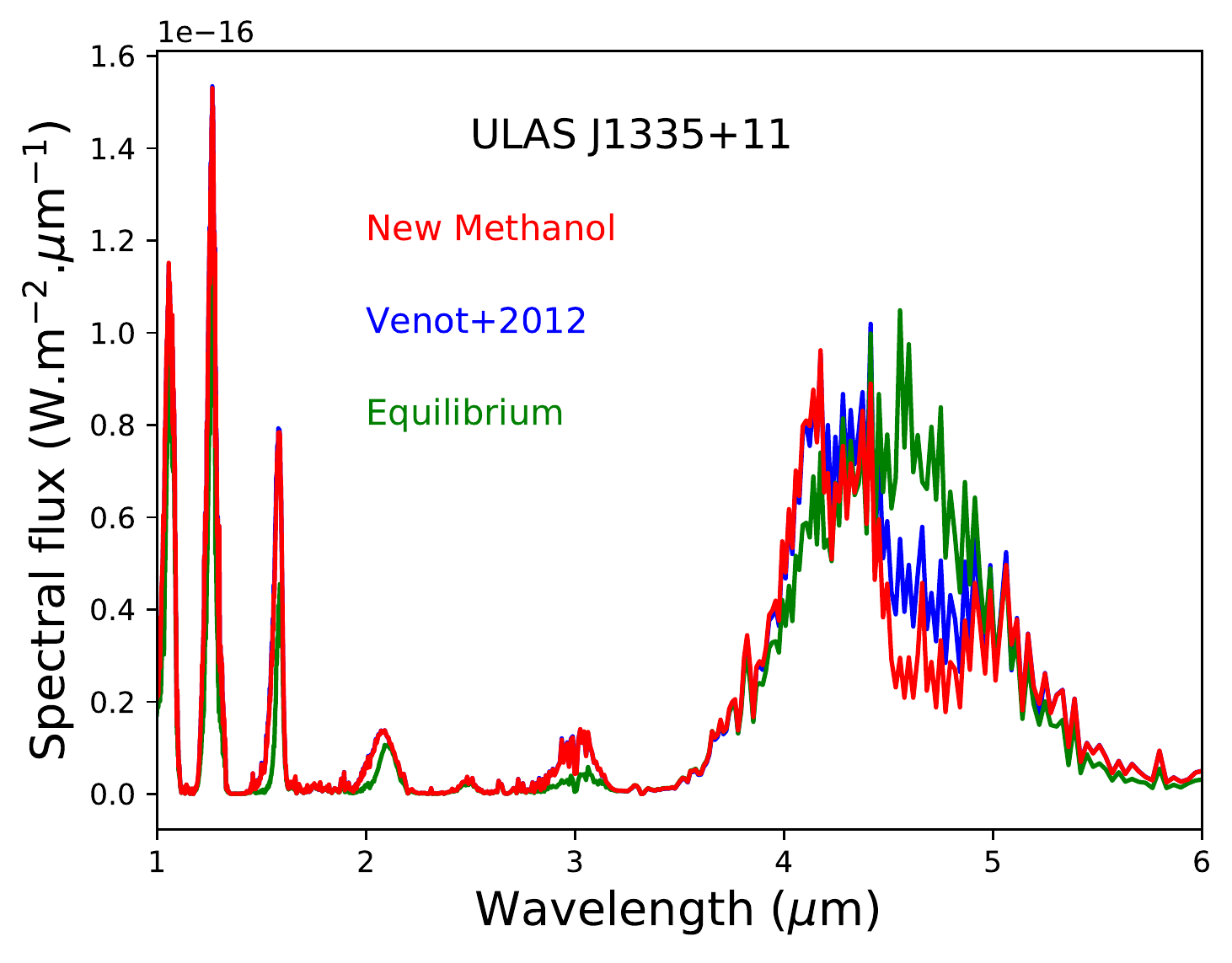}
  \caption{Emission spectra for ULAS J1335+11 (for R=0.1R$_{\Sun}$ at 10 pc) obtained with the updated chemical scheme (red), compared with the ones obtained with the former one of \cite{Venot2012} (blue) and with equilibrium chemistry (green).}
  \label{fig:spec_ULAS}
  \end{figure}

The updated scheme has a significant impact on the abundance of CO in late T dwarfs. This has a direct impact on the planet spectrum in the 4.7-$\mu$m window, because CO is a strong absorber at these wavelengths. We show in Fig.~\ref{fig:spec_ULAS} the emission spectrum at equilibrium, with the former and the updated scheme. The new scheme can lead up to a factor 2 decrease in the flux in this window because of the increase of the CO abundance. Such a difference will be easily constrained by JWST/NIRSpec measurements. The updated scheme combined with JWST measurements will therefore allow to better characterise the strength of vertical mixing that is necessary to reproduce the out-of-equilibrium abundance of CO in cold brown dwarfs.

\subsection{Implications for the formation of Uranus and Neptune}

The results obtained for Uranus and Neptune in this paper with the thermochemical model of \citet{Venot2012} and the updated chemical scheme for methanol do not waive the difference found since more than two decades between the two planets in terms of deep oxygen abundance. This difference primarily results from their different tropospheric CO abundances. And while \citet{Teanby2019} recently proposed from their \textit{Herschel}-SPIRE data a model without any tropospheric CO in Neptune, i.e. quite similar to Uranus, they probably lacked sensitivity in the upper troposphere to make this result robust. \citet{Moreno2011} showed in a preliminary combined analysis of \textit{Herschel}-SPIRE and IRAM-30m data, including the CO(1-0) line that is most sensitive to the tropospheric CO, that the tropospheric CO in Neptune was 0.20$\pm$0.05\,ppm. %New wideband observations of the CO(1-0) line are required to settle this question.

Assuming the CO abundance difference between the two planets is representative of their respective deep oxygen abundances, and according to our new results, only Neptune could in principle have formed from ices condensed in clathrates (C/O$\sim$0.12). On the other hand, the low upper limit on O/H for Uranus is in contradiction with such a process (C/O$\sim$1). Interestingly though, this upper limit is close to the C/H required to fit CH$_4$ (10.383 dex vs. 10.331 dex, respectively), which is one of the conditions under which Uranus planetesimal ices could have formed on the CO snow-line and be mainly composed of CO rather than H$_2$O \citep{Ali-Dib2014}. Such a low upper limit may also derive from inhibited convection in the deep layers of Uranus precluding any tropospheric abundance measurements to be representative of the bulk composition of the planet. One should however not forget that several model parameters remain uncertain, like the deep $K_{zz}$. A lower $K_{zz}$ than that assumed in our nominal models would result in higher O/H \citep{Cavalie2017} and therefore change our interpretation.

\section{Conclusion \label{conclusion}}
We present in this paper an update of the chemical scheme of \cite{Venot2012}. The analysis of \cite{Moses2014} denotes that discrepancies between her results and \cite{Venot2012} could be due to differences in chemical rates involving methanol. This has motivated us to update \cite{Venot2012}'s chemical network by replacing their methanol sub-network by the one put together by \citet{Burke2016}, following a comprehensive study on methanol combustion. We have validated this new network against experimental measurements. We emphasise that one change, among others, in our new chemical network is that the controversial reaction CH$_3$OH$+$H$\rightleftharpoons$CH$_3+$H$_2$O has been removed.

The new updated scheme V20 gives quite similar results as the former one for hot Jupiters. A variation of CO$_2$ abundance is observed in HD 189733b atmosphere, but only modifies the synthetic spectra to a lower extent (50 ppm at 4-5$\mu$m). A very small change of CH$_4$ quenching level, which modifies in return slightly the abundance of this species, is also observed in HD 189733b, without any impact on the observable. 

For warm Neptunes and T Dwarfs, the update has more significant implications because the reaction CH$_3$OH$+$H$\rightleftharpoons$CH$_3+$H$_2$ played an important role in the former scheme of V12. The quenching of CO, CO$_2$ (and eventually H$_2$O and CH$_4$ in high metallicity atmospheres) happening deeper with the new scheme, the abundances of these species are modified compared to the results obtained with \cite{Venot2012}'s chemical scheme. The change is important enough to affect the synthetic spectra. The differences with the former scheme (up to 100 ppm in transmission for warm Neptune and a factor 2 in emission for the T Dwarf) will certainly be detectable with future instruments, such as JWST. Using an accurate and updated chemical scheme is thus paramount for a correct interpretation of future observations, and for a better comprehension of mixing processes at play in these atmospheres.

The consequence of the update is also fundamental for our understanding of the formation Uranus and Neptune. For a given O/H ratio, the abundance of CO is higher with the updated scheme than with the former one. Consequently, the O/H ratios necessary to reproduce the tropospheric observations of CO is lower than what had been found previously. The updated scheme indicates O/H of $<$45 and 250$\Sun$ for Uranus and Neptune, respectively. 

Finally, we have derived a reduced chemical scheme from this update, for future 3D kinetic models that are crucial (and the next step) for our understanding of (exo)planetary atmospheres.

The next steps on the improvement of our chemical scheme will imply adding new species, like sulphur species, following the recent detection of H$_2$S in Uranus and Neptune \citep{Irwin2018,Irwin2019}. Phosphorus species could also be of interest to extend the scope of our work, as PH$_3$ can provide additional constraints on the deep oxygen abundance \citep{Visscher2005}. The use of this species as a tracer for O abundance will be possible only if PH$_3$ is quenched in giant planets atmosphere, which is an expected behaviour of this molecule \cite{FegleyLodders1994,Visscher2006}. However, PH$_3$ remains undetected in Uranus and Neptune \citep{Moreno2009}. Although these species have not been detected yet in exoplanet atmospheres, their presence is expected and it has been shown that they should be observable with JWST \citep{baudino2017,wang2017}.

We show with this study that collaborations between astrophysicists and combustion specialists are really fruitful to accurately study high-temperature atmospheres. The intensive work performed in the field of combustion is paramount to perform reliable atmospheric modeling, leading to a correct interpretation of observations.

\begin{acknowledgements}
The author thanks the anonymous referee for his/her careful review that helps improving the manuscript. O.V. and T.C. thank the CNRS/INSU Programme National de Plan\'etologie (PNP) and CNES for funding support. P.T. acknowledges support from the European Research Council (grant no. 757858 -- ATMO). The authors thank B. Edwards, Q. Changeat, I. Waldmann for useful discussions on JWST and ARIEL observations.
\end{acknowledgements}

\bibliographystyle{aa}
\bibliography{NewMethanol}

%%%%%%%%%%%%%%%%%%%%%%%%%%%%%%%%%%%
\begin{appendix} %First appendix
\section{A short review of methanol combustion experimental studies (continued)}
\tab{tab:combustion_CH3OH} gives an overview of the main studies published over the 50 past year on the pyrolysis of methanol. The CH$_3$OH sub-network from \citet{Burke2016} that we have implemented in our model results from these studies.

\begin{table*}
  \caption{Overview of the main studies published over the 50 past year on the pyrolysis of methanol. Reactor types are: shock tube (A), plug flow reactor (B), Rapid Compression Machine (C), stirred reactor (D), static reactor (E), premixed flame (F).}  
  \label{tab:combustion_CH3OH}      
  \begin{center}          
  \begin{tabular}{lllll}
    \hline
    Reference & Reactor & Temperature range (K) & Pressure & Equivalence ratio \\
    \hline
\citet{Cooke1971}         & A & 1570–1879      & 1\,atm           & 1.00 \\
\citet{Bowman1975}        & A & 1545-2180      & 0.18-0.46\,MPa   & 0.375-6.0 \\
\citet{Akrich1978}        & F & 298            & 0.11\,atm        & 0.77–1.53 \\
\citet{Aronowitz1979}     & B & 1070-1225      & 0.1\,MPa         & 0.03-3.16 \\
\citet{Westbrook1979}     & A-B & 1000-2180    & 0.1-0.5\,MPa     & 0.05-3.0 \\
\citet{Tsuboi1981}        & A & 1200-1800      &                  & 0.2-2.0 \\
\citet{Natarajan1981}     & A & 1300-1700      & 0.25-0.45\,MPa   & 0.5-1.5 \\
\citet{Cathonnet1982}     & D & 700-900        & 0.02-0.05\,MPa   & 0.5-4.0 \\
\citet{Metghalchi1982}    & F & 300-500        & 0.1\,MPa         & 0.5-1.4 \\
\citet{Yano1983}          & C & 700-1000       &                  & \\
\citet{Cribb1984}         & A & 2000           & 0.04\,MPa        & \\
\citet{Hidaka1989}        & A & 1372-1842      &                  & \\
\citet{Norton1989}        & B & 1025-1090      & 0.1\,MPa         & 0.6-1.6 \\
\citet{Chen1991}          & D & & & \\
\citet{Egolfopoulos1992}  & A-B-E-F & 820-2180 & 0.005-0.47\,MPa  & 0.05 \\
\citet{Grotheer1992}      & C-F & & & \\
\citet{Held1994}          & B & 810–1043       & 1–10\,atm        & 0.60–1.60 \\
\citet{Aniolek1995}       & D & 650–700        & 0.92\,atm        & 0.50–1.50 \\
\citet{Fieweger1997}      & A & 800–1200       & 12.83–39.48\,atm & 1.00 \\
\citet{Held1998}          & A-B-E-F & 633-2050 & 0.026-2\,MPa     & 0.05-2.6 \\
\citet{Alzueta2001}       & B & 700–1500       & 1\,atm           & 0.07–2.70 \\
\citet{Lindstedt2002}     & A-B-F & & & \\
\citet{Ing2003}           & B & 873–1073       & 1–5\,atm         & 0.75–1.00 \\
\citet{Ing2003}           & B & 1073           & 1–10\,atm        &  \\
\citet{Rasmussen2008}     & B & 650–1350       & 1.00\,atm        & 0.004–0.08 \\
\citet{Liao2006}          & D & 300-550        & 0.1\,MPa         & 0.6-1.4 \\
\citet{Dayma2007}         & D & 700–1090       & 10\,atm          & 0.30–1.00 \\
\citet{Li2007}            & A-B-F & 300-2200   & 0.1-2\,MPa       & 0.05-6.0 \\
\citet{Noorani2010}       & A & 1068–1776      & 2–12\,atm        & 0.50–2.00 \\
\citet{Veloo2010}         & C & 343            & 1\,atm           & 0.70–1.50 \\
\citet{Kumar2011}         & C & 850–1100       & 6.91–29.61\,atm  & 0.25–1.00 \\
\citet{Aranda2013}        & B & 600–900        & 20–100\,atm      & 4.35–0.06 \\
\citet{Ren2013}           & A & 1200-1650      & 0.1-0.3\,MPa	  & \\
\citet{Burke2016}         & A-C & 820-1650     & 0.2-5\,MPa       & 0.5-2.0 \\
    \hline
  \end{tabular}
  \end{center}
\end{table*}

\section{New CH$_3$OH sub-scheme and reactions with a logarithmic dependence in pressure \label{Annex_PLOG}}
Under certain conditions, some reaction rate expressions depend on pressure as well as temperature. Generally speaking, the rate for unimolecular/recombination fall-off reactions increases with increasing pressure, while the rate for chemically activated bimolecular reactions decreases with increasing pressure. Several expressions are available in the literature to express the variation of the kinetic data between high- and low- pressure limit. The Lindemann approach \citep{Lindemann1922}, the Troe form \citep{Gilbert1983} or the approach taken at SRI International by \citet{Stewart1989} are the main expressions commonly used for the pressure-dependent reactions.
The sub-mechanism of methanol combustion uses another kind of expression for the pressure dependence using logarithmic interpolation with the key word PLOG. Miller and Lutz (2003, pers. comm.) developed a generalised method for describing the pressure dependence of a reaction rate based on direct interpolation of reaction rates specified at individual pressures. In this formulation, the reaction rate is described in terms of the standard modified Arrhenius rate parameters. Different rate parameters are given for discrete pressures within the pressure range of interest. When the actual reaction rate is computed, the rate parameters will be determined through logarithmic interpolation of the specified rate constants, at the current pressure from the simulation. This approach provides a very straightforward way for users to include rate data from more than one pressure regime.

\tab{tab:new_scheme1} lists the reactions of the new methanol sub-scheme we include in our kinetic model. We list in \tab{tab:new_scheme2} the reactions of the new CH$_3$OH sub-scheme that present an explicit logarithmic dependence with pressure. The chemical rate of such a reaction is computed by interpolating over pressure at the considered temperature.

\begin{table*}
  \caption{Reactions of the new chemical sub-network of CH$_3$OH not involving a logarithmic dependence with pressure, extracted from \citet{Burke2016}. These reactions are either totally new compared to the former scheme, or the reaction rate has been modified. The corresponding reaction rates are expressed with a modified Arrhenius law $k(T) = A \times T^n exp^{-\frac{E_a}{RT}}$, with $T$ in Kelvin, $E_a/R$ in Kelvin, and $n$ dimensionless. $k_0$ and $k_{\infty}$ are the reaction rates in the low and high pressures regimes, respectively For $k_0$, unit of $A$ is: s$^{-1}$.K$^{-n}$ for thermal dissociations, cm$^3$.molecule$^{-1}$.s$^{-1}$.K$^{-n}$ for bimolecular reactions or decomposition reaction with a second-body M, and cm$^6$.molecule$^{-2}$.s$^{-1}$.K$^{-n}$ for combination reactions with a third-body M. For $k_{\infty}$, unit of $A$ is: s$^{-1}$.K$^{-n}$ for decomposition reactions (behaviour of a thermal dissociation), and cm$^3$.molecule$^{-1}$.s$^{-1}$.K$^{-n}$ for combination reactions (behaviour of bimolecular reactions). }
  \label{tab:new_scheme1}      
  \begin{center}          
  \begin{tabular}{ll}
    \hline
    Reaction & Rate \\ 
    \hline
    %new reaction compared to C0C2
     HCOH $+$ O$_2$ $\longrightarrow$ CO$_2$ $+$ H $+$ OH & $k_0=8.306$\dix{-12}\\
     HCOH $+$ O$^3$P $\longrightarrow$ CO$_2$ $+$ H $+$ H & $k_0=8.306$\dix{-11}\\
     HCOH $+$ O$^3$P $\longrightarrow$ CO $+$ OH $+$ H & $k_0=4.983$\dix{-11}\\
     HCOH $+$ O$_2$ $\rightleftharpoons$  CO$_2$ $+$ H$_2$O & $k_0=4.983$\dix{-11}\\
     HCOH $+$ H $\rightleftharpoons$  H$_2$CO $+$ H & $k_0=3.322$\dix{-10}\\
     HCOH $+$ OH $\rightleftharpoons$  HCO $+$ H$_2$O & $k_0=3.322$\dix{-11}\\
     HOCHO $\rightleftharpoons$ CO $+$ H$_2$O & $k_0=2.45$\dix{12}$e^{-30400/T}$\\
     HOCHO $\rightleftharpoons$ CO$_2$ $+$ H$_2$ & $k_0=2.95$\dix{9}$e^{-24390/T}$\\
     HOCHO $+$ H $\longrightarrow$ H$_2$ $+$ CO$_2$ $+$ H & $k_0=7.043$\dix{-18}$T^{2.1}e^{-2447/T}$\\
     HOCHO $+$ H $\longrightarrow$ H$_2$ $+$ CO $+$ OH & $k_0=1.002$\dix{-10}$T^{-0.35}e^{-1502/T}$\\
     HOCHO $+$ O$^3$P $\longrightarrow$ CO $+$ 2 OH & $k_0=2.94$\dix{-6}$T^{-1.9}e^{-1496/T}$\\
     HOCHO $+$ OH $\longrightarrow$ H$_2$O $+$ CO$_2$ $+$ H & $k_0=4.352$\dix{-18}$T^{2.06}e^{-460.5/T}$\\
     HOCHO $+$ OH $\longrightarrow$ H$_2$O $+$ CO $+$ OH & $k_0=3.073$\dix{-17}$T^{1.51}e^{483.7/T}$\\
     HOCHO $+$ CH$_3$ $\longrightarrow$ CH$_4$ $+$ CO $+$ OH & $k_0=6.478$\dix{-31}$T^{5.8}e^{-1106/T}$\\
     HOCHO $+$ OOH $\longrightarrow$ H$_2$O$_2$ $+$ CO $+$ OH & $k_0=1.661$\dix{-12}$e^{-5993/T}$\\
     H$_2$CO $+$ H (+M) $\rightleftharpoons$ CH$_2$OH (+M)&
     $\left\{\begin{array}{l} 
     k_0=3.504\times10^{-16}T^{-4.82}e^{-3283/T}\\
     k_{\infty}=8.970\times10^{-13}T^{0.454}e^{-1810/T}\\
   \end{array}\right.$ \\
     H$_2$CO $+$ OH $\rightleftharpoons$  HOCH$_2$O & $k_0=7.475$\dix{-9}$T^{-1.1}$\\
     HOCH$_2$O $\rightleftharpoons$ HOCHO $+$ H & $k_0=1.0$\dix{14}$e^{-7491/T}$\\
     CH$_3$OH (+M) $\rightleftharpoons$ CH$_3$ $+$ OH (+M) & $\left\{\begin{array}{l}
     k_0=2.492\times10^{19}T^{-6.995}e^{-49270/T}\\
     k_{\infty}=2.084\times10^{18}T^{-0.615}e^{-46530/T}\\
     \end{array}\right.$ \\
     CH$_3$OH (+M) $\rightleftharpoons$ $^3$CH$_2$ $+$ H$_2$O (+M) & $\left\{\begin{array}{l}
     k_0=2.375\times10^{23}T^{-8.227}e^{-49980/T}\\
     k_{\infty}=3.121\times10^{18}T^{-1.017}e^{-46110/T}\\
     \end{array}\right.$ \\
     CH$_3$OH (+M) $\rightleftharpoons$ CH$_2$OH $+$ H (+M) & $\left\{\begin{array}{l} 
     k_0=5.631\times10^{18}T^{-7.244}e^{-52910/T} \\
     k_{\infty}=7.896\times10^{-3}T^{5.038}e^{-42470/T}\\
   \end{array}\right.$ \\
   CH$_3$OH $+$ H $\rightleftharpoons$ CH$_2$OH $+$ H$_2$ & $k_0=5.1$\dix{-19}$T^{2.55}e^{-2735/T}$\\
   CH$_3$OH $+$ O$_2$ $\rightleftharpoons$ CH$_3$O $+$ OOH & $k_0=5.947$\dix{-20}$T^{2.27}e^{-21500/T}$\\
   CH$_3$OH $+$ OOH $\rightleftharpoons$ CH$_3$O $+$ H$_2$O$_2$ & $k_0=2.027$\dix{-12}$e^{-10090/T}$\\
   CH$_3$OH $+$ CH$_3$OO $\rightleftharpoons$ CH$_2$OH $+$ CH$_3$OOH & $k_0=3.007$\dix{-12}$e^{-6893/T}$\\
   CH$_2$OH $+$ OOH $\rightleftharpoons$  HOCH$_2$O $+$ OH & $k_0=1.661$\dix{-11}\\
   CH$_2$OH $+$ O$_2$ $\rightleftharpoons$  H$_2$CO $+$ OOH & $k_0=2.508$\dix{-9}$T^{-1}$\\
   CH$_2$OH $+$ O$_2$ $\rightleftharpoons$  H$_2$CO $+$ OOH & $k_0=4.003$\dix{-10}$e^{-2522/T}$\\
   CH$_2$OH $+$ HCO $\rightleftharpoons$ CH$_3$OH $+$ CO & $k_0=1.661$\dix{-11}\\
   CH$_2$OH $+$ CH$_2$OH $\rightleftharpoons$  H$_2$CO $+$ CH$_3$OH & $k_0=4.983$\dix{-12}\\
   CH$_3$OH $+$ CH$_3$ $\rightleftharpoons$ CH$_2$OH $+$ CH$_4$ & $k_0=3.538$\dix{-25}$T^{3.953}e^{-3547/T}$\\
   CH$_3$OH $+$ CH$_3$ $\rightleftharpoons$ CH$_3$O $+$ CH$_4$ & $k_0=5.349$\dix{-21}$T^{2.425}e^{-4313/T}$\\
   CH$_3$OH $+$ HCO $\rightleftharpoons$ CH$_2$OH $+$ H$_2$CO & $k_0=1.6$\dix{-20}$T^{2.9}e^{-6591/T}$\\
   CH$_3$OH $+$ H $\rightleftharpoons$ CH$_3$O $+$ H$_2$ & $k_0=3.306$\dix{-19}$T^{2.56}e^{-5178/T}$\\
   CH$_3$OH $+$ O$^3$P $\rightleftharpoons$ CH$_3$O $+$ OH & $k_0=6.445$\dix{-20}$T^{2.5}e^{-1549/T}$\\
   CH$_3$OH $+$ O$^3$P $\rightleftharpoons$ CH$_2$OH $+$ OH & $k_0=6.445$\dix{-19}$T^{2.5}e^{-1549/T}$\\
   CH$_3$OH $+$ OH $\rightleftharpoons$ CH$_3$O $+$ H$_2$O & $k_0=2.492$\dix{-22}$T^{3.03}e^{383.6/T}$\\
   CH$_3$OH $+$ OH $\rightleftharpoons$ CH$_2$OH $+$ H$_2$O & $k_0=5.116$\dix{-20}$T^{2.65}e^{405.6/T}$\\
   CH$_3$OH $+$ O$_2$ $\rightleftharpoons$ CH$_2$OH $+$ OOH & $k_0=5.947$\dix{-19}$T^{2.27}e^{-21500/T}$\\
   CH$_3$OH $+$ OOH $\rightleftharpoons$ CH$_2$OH $+$ H$_2$O$_2$ & $k_0=5.415$\dix{-11}$e^{-9443/T}$\\
   CH$_3$ $+$ OOH $\rightleftharpoons$ CH$_3$O $+$ OH & $k_0=1.661$\dix{-12}$T^{0.269}e^{345.7/T}$\\
   CH$_3$O (+M) $\rightleftharpoons$ H$_2$CO $+$ H (+M) &
   $\left\{\begin{array}{l}
   k_0=3.101\times10^{1}T^{-3}e^{-12220/T}\\
   k_{\infty}=6.8\times10^{13}e^{-13160/T}\\
   \end{array}\right.$\\
   CH$_3$O $+$ O$_2$ $\rightleftharpoons$ H$_2$CO $+$ OOH & $k_0=7.276$\dix{-43}$T^{9.5}e^{2766/T}$\\
   CH$_3$O $+$ H $\rightleftharpoons$ H$_2$CO $+$ H$_2$ & $k_0=3.322$\dix{-11}\\
   CH$_3$O $+$ CH$_3$ $\rightleftharpoons$ H$_2$CO $+$ CH$_4$ & $k_0=1.993$\dix{-11}\\
   H$_2$CO $+$ CH$_3$O $\rightleftharpoons$ HCO $+$ CH$_3$OH & $k_0=1.1$\dix{-12}$e^{-1153/T}$\\
   C$_2$H$_4$ $+$ CH$_3$O $\rightleftharpoons$  C$_2$H$_3$ $+$ CH$_3$OH & $k_0=1.993$\dix{-13}$e^{-3394/T}$\\
   2 OH (+M) $\rightleftharpoons$ H$_2$O$_2$ (+M) &
   $\left\{\begin{array}{l} 
   k_0=1.526\times10^{-28}T^{-0.76}\\
   k_{\infty}=1.201\times10^{-10}T^{-0.37}\\
   \end{array}\right.$\\
   CO $+$ OH $\rightleftharpoons$ CO$_2$ $+$ H & $k_0=4.983$\dix{-20}$T^{1.5}e^{251.4/T}$\\
    \hline
  \end{tabular}
  \end{center}
\end{table*}

\begin{table*}
  \caption{Reactions of the new chemical sub-network of CH$_3$OH involving a logarithmic dependence with pressure, extracted from \citet{Burke2016}. The corresponding reaction rates are expressed with a modified Arrhenius law $k_0(T) = A \times T^n exp^{-\frac{E_a}{RT}}$, with $T$ in Kelvin, $E_a/R$ in Kelvin, and $n$ dimensionless. $A$ is in s$^{-1}$.K$^{-n}$ for the thermal dissociation and cm$^3$.molecule$^{-1}$.s$^{-1}$.K$^{-n}$ for bimolecular reactions.}            
  \label{tab:new_scheme2}      
  \begin{center}          
  \begin{tabular}{cc}
    \hline
    Reaction & Rate \\ 
    \hline
     CH$_3$ $+$ OH $\rightleftharpoons$ HCOH $+$ H$_2$ & $k_0=\left\{\begin{array}{l}
     1.441\times10^{-15}T^{0.787}e^{1531/T}, p=0.01\,\mathrm{atm} \\
     5.173\times10^{-15}T^{0.630}e^{1343/T}, p=0.1\,\mathrm{atm} \\
     2.585\times10^{-13}T^{0.156}e^{688/T}, p=1\,\mathrm{atm} \\
     2.830\times10^{-3}T^{-2.641}e^{3227/T}, p=10\,\mathrm{atm} \\
     1.204\times10^{-3}T^{-2.402}e^{4851/T}, p=100\,\mathrm{atm} \\
     \end{array}\right.$ \\

    CH$_3$ $+$ OH $\rightleftharpoons$ CH$_2$OH $+$ H & 
    $k_0=\left\{\begin{array}{l} 
     2.692\times10^{-14}T^{0.965}e^{-1617/T}, p=0.01\,\mathrm{atm} \\
     3.001\times10^{-14}T^{0.950}e^{-1632/T}, p=0.1\,\mathrm{atm} \\
     7.781\times10^{-14}T^{0.833}e^{-1794/T}, p=1\,\mathrm{atm} \\
     2.532\times10^{-11}T^{0.134}e^{-2839/T}, p=10\,\mathrm{atm} \\
     5.961\times10^{-10}T^{-0.186}e^{-4328/T}, p=100\,\mathrm{atm} \\
     \end{array}\right.$ \\

    CH$_3$ $+$ OH $\rightleftharpoons$ H $+$ CH$_3$O & 
    $k_0=\left\{\begin{array}{l} 
     1.969\times10^{-15}T^{1.016}e^{-6008/T}, p=0.01\,\mathrm{atm} \\
     1.973\times10^{-15}T^{1.016}e^{-6008/T}, p=0.1\,\mathrm{atm} \\
     2.042\times10^{-15}T^{1.011}e^{-6013/T}, p=1\,\mathrm{atm} \\
     2.986\times10^{-15}T^{0.965}e^{-6069/T}, p=10\,\mathrm{atm} \\
     8.705\times10^{-14}T^{0.551}e^{-6577/T}, p=100\,\mathrm{atm} \\
     \end{array}\right.$ \\

    C$_2$H$_3$ $+$ O$_2$ $\rightleftharpoons$ CO $+$ CH$_3$O & 
    $k_0=\left\{\begin{array}{l} 
     1.360\times10^{-5} T^{-2.66}e^{-1611/T}, p=0.01\,\mathrm{atm} \\
     6.742\times10^{-10}T^{-1.32}e^{-446/T}, p=0.1\,\mathrm{atm} \\
     7.207\times10^{-10}T^{-1.33}e^{-453/T}, p=0.316\,\mathrm{atm} \\
     1.710\times10^{-13}T^{-0.33}e^{376/T}, p=1\,\mathrm{atm} \\
     3.138\times10^{-12}T^{-3.00}e^{4526/T}, p=3.16\,\mathrm{atm} \\
     3.205         T^{-5.63}e^{-1/T}, p=10\,\mathrm{atm} \\
     1.827\times10^{-6} T^{-2.22}e^{-2606/T}, p=31.6\,\mathrm{atm} \\
     9.615\times10^{8}  T^{-6.45}e^{-8459/T}, p=100\,\mathrm{atm} \\
     \end{array}\right.$ \\

    C$_2$H$_3$ $+$ O$_2$ $\rightleftharpoons$ CO $+$ CH$_3$O & 
    $k_0=\left\{\begin{array}{l} 
     2.142\times10^{-15} T^{0.18}e^{864/T}, p=0.01\,\mathrm{atm} \\
     9.947\times10^{-13}T^{-2.93}e^{4813/T}, p=0.1\,\mathrm{atm} \\
     4.832\times10^{-13}T^{-2.93}e^{5093/T}, p=0.316\,\mathrm{atm} \\
     9.581\times10^{-3} T^{-3.54}e^{-2401/T}, p=1\,\mathrm{atm} \\
     8.286\times10^{-9} T^{-1.62}e^{-930/T}, p=3.16\,\mathrm{atm} \\
     1.549\times10^{-7} T^{-1.96}e^{-1673/T}, p=10\,\mathrm{atm} \\
     1.694\times10^{48} T^{-20.69}e^{-7981/T}, p=31.6\,\mathrm{atm} \\
     1.827\times10^{-15}T^{0.31}e^{-515/T}, p=100\,\mathrm{atm} \\
     \end{array}\right.$ \\

    C$_2$H$_5$OH $\rightleftharpoons$ CH$_3$ $+$ CH$_2$OH & 
    $k_0=\left\{\begin{array}{l} 
     1.20\times10^{54}T^{-1.29}e^{-50330/T}, p=0.001\,\mathrm{atm} \\
     5.18\times10^{59}T^{-14.0}e^{-50280/T}, p=0.01\,\mathrm{atm} \\
     1.62\times10^{66}T^{-15.3}e^{-53040/T}, p=0.1\,\mathrm{atm} \\
     5.55\times10^{64}T^{-14.5}e^{-53430/T}, p=1\,\mathrm{atm} \\
     1.55\times10^{58}T^{-12.3}e^{-53230/T}, p=10\,\mathrm{atm} \\
     1.78\times10^{47}T^{-8.96}e^{-50860/T}, p=100\,\mathrm{atm} \\
     \end{array}\right.$ \\
    \hline
  \end{tabular}
  \end{center}
\end{table*}

\section{Validation of the new chemical scheme \label{appendix_validation}}
In what follows, we present model comparisons with experimental data for the cases where the new CH$_3$OH sub-scheme improvement is most noticeable. 

\begin{figure}[!h]
\includegraphics[width=\columnwidth]{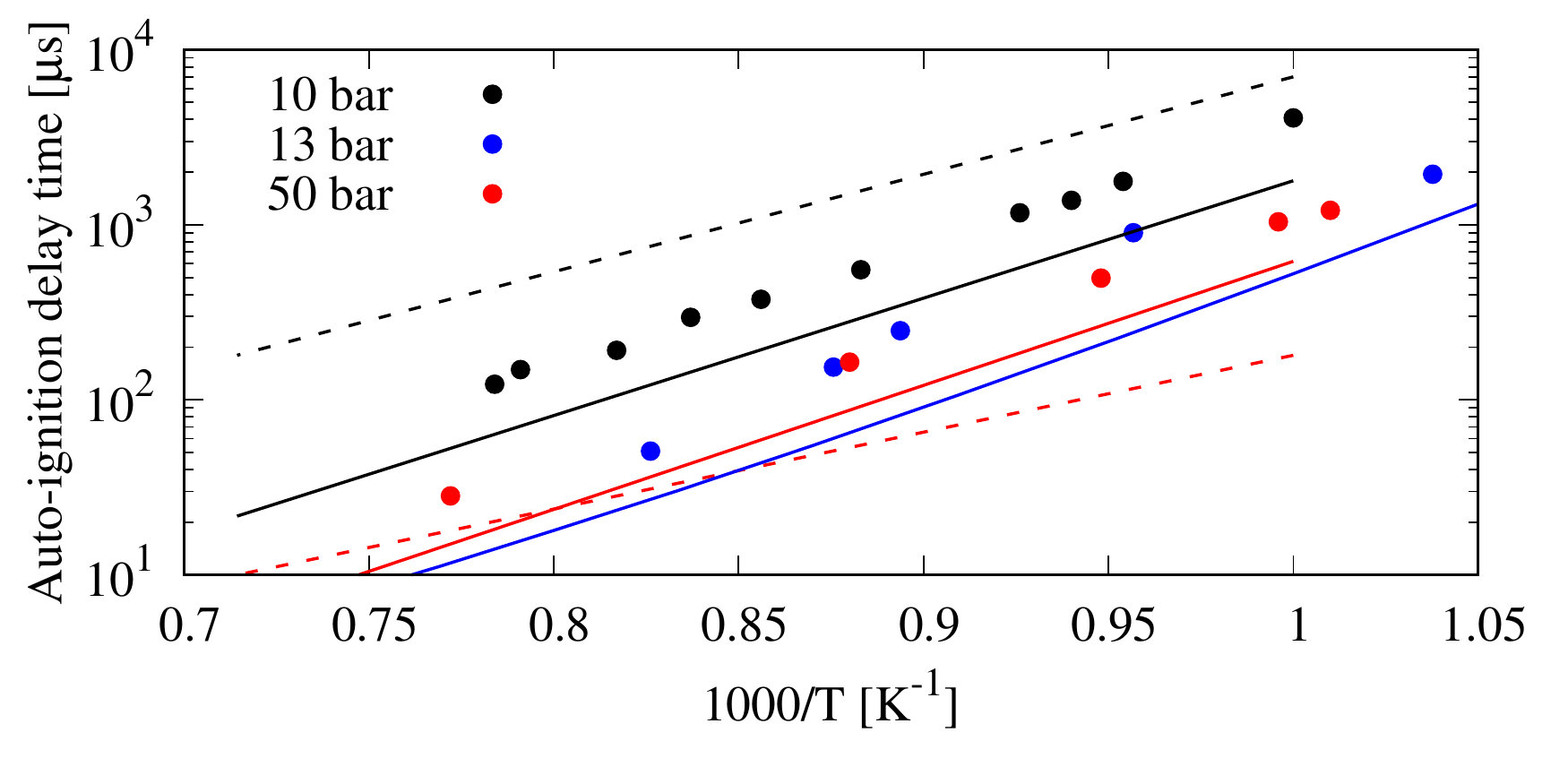}
\caption{Auto-ignition delay times of methanol in shock tube under high pressure. Points are experimental data from \citet{Burke2016} and \citet{Fieweger1997}, and lines are simulations with the chemical scheme of \citet{Venot2012} (dashed) and with the updated chemical scheme of this paper (solid). Composition: 5.7 mol\% CH$_3$OH$+$8.55 mol\% O$_2$ $+$ N$_2$.}
\label{fig:Tig}
\end{figure}

\citet{Burke2016} have studied the combustion of methanol in a shock tube at several pressures and temperatures. \fig{fig:Tig} shows, for the chemical scheme of \citet{Venot2012} and the new scheme of this paper, the variations of the auto-ignition delay times at two different pressures (10 and 50\,bar), for temperatures ranging from 1000 to 1500\,K and for an equivalence ratio of 1. We also include simulations with the new scheme compared with the data from \citet{Fieweger1997} at 13\,bar.

\begin{figure}[!h]
\includegraphics[width=\columnwidth]{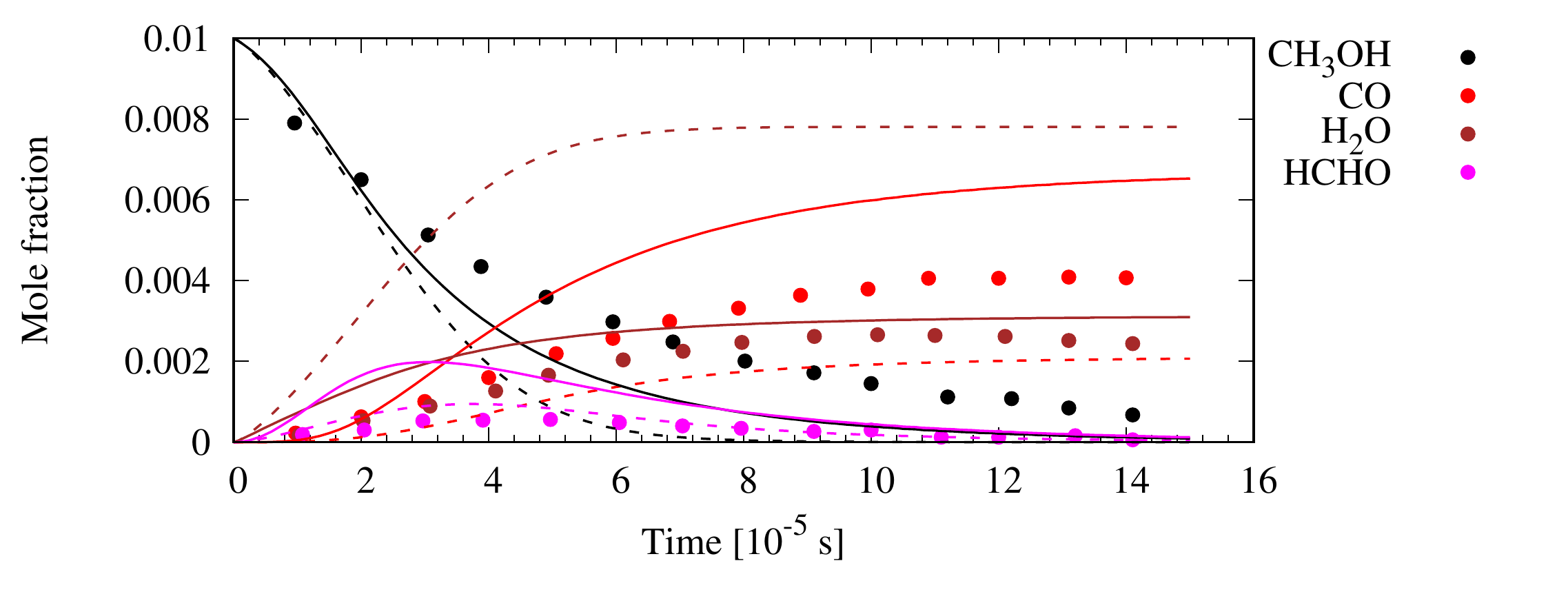}
\includegraphics[width=\columnwidth]{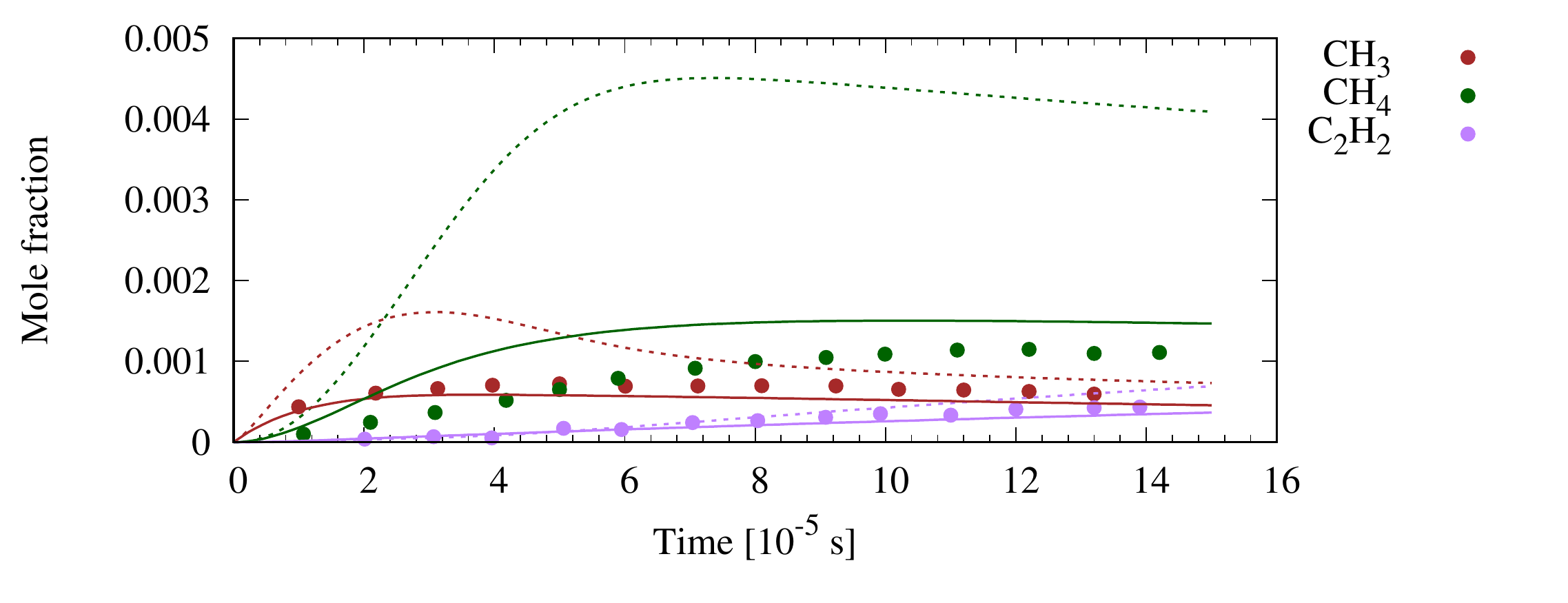}
\caption{Comparison of the predictions of the mechanism (solid: new ; dashed: original) with the experimental data from \citet{Cribb1984}, for which T$=$2000\,K, P$=$0.354\,atm, 1\% CH$_3$OH $+$ 6\% H$_2$ balanced with Ar.}
\label{fig:Cribb}
\end{figure}

The study of the pyrolysis of methanol at a very high temperature of about 2000\,K and low pressure, around 0.4\,atm, in a shock tube by \citet{Cribb1984} is displayed in \fig{fig:Cribb}. 

\begin{figure}[!h]
\includegraphics[width=\columnwidth]{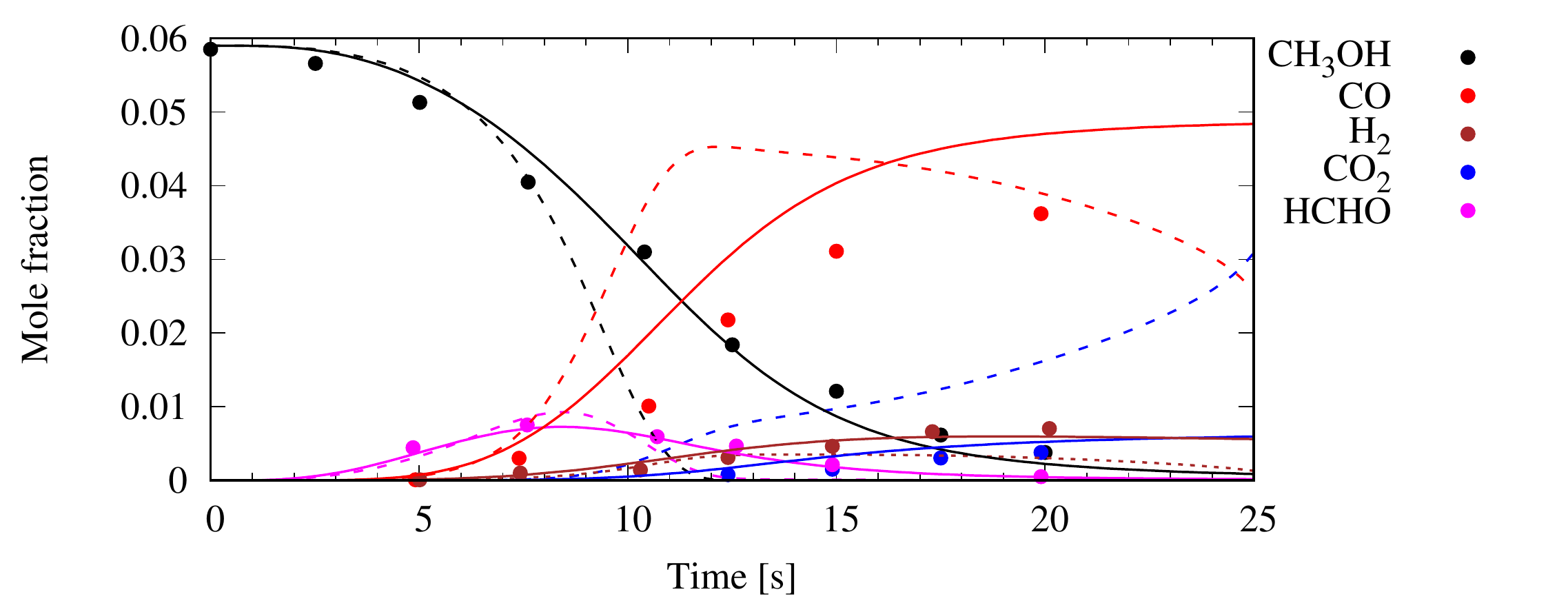}
\caption{Species profiles for static reactor experiments where symbols denote experimental measurements from \citet{Cathonnet1982} and curves modeling results using the chemical scheme of \citet{Venot2012} (dashed) and with the updated chemical scheme of this paper (solid). Experimental conditions: 5.89 mol\% CH$_3$ $+$ 8.84 mol\% O$_2$ $+$ N$_2$, T$=$823\,K, P$=$0.026 MPa.}
\label{fig:Batch}
\end{figure}

The variation of mole fraction of different compounds obtained in a batch reactor obtained by \citet{Cathonnet1982} at relatively low temperature, around 800\,K, is presented in \fig{fig:Batch}.

\begin{figure}[!h]
\includegraphics[width=\columnwidth]{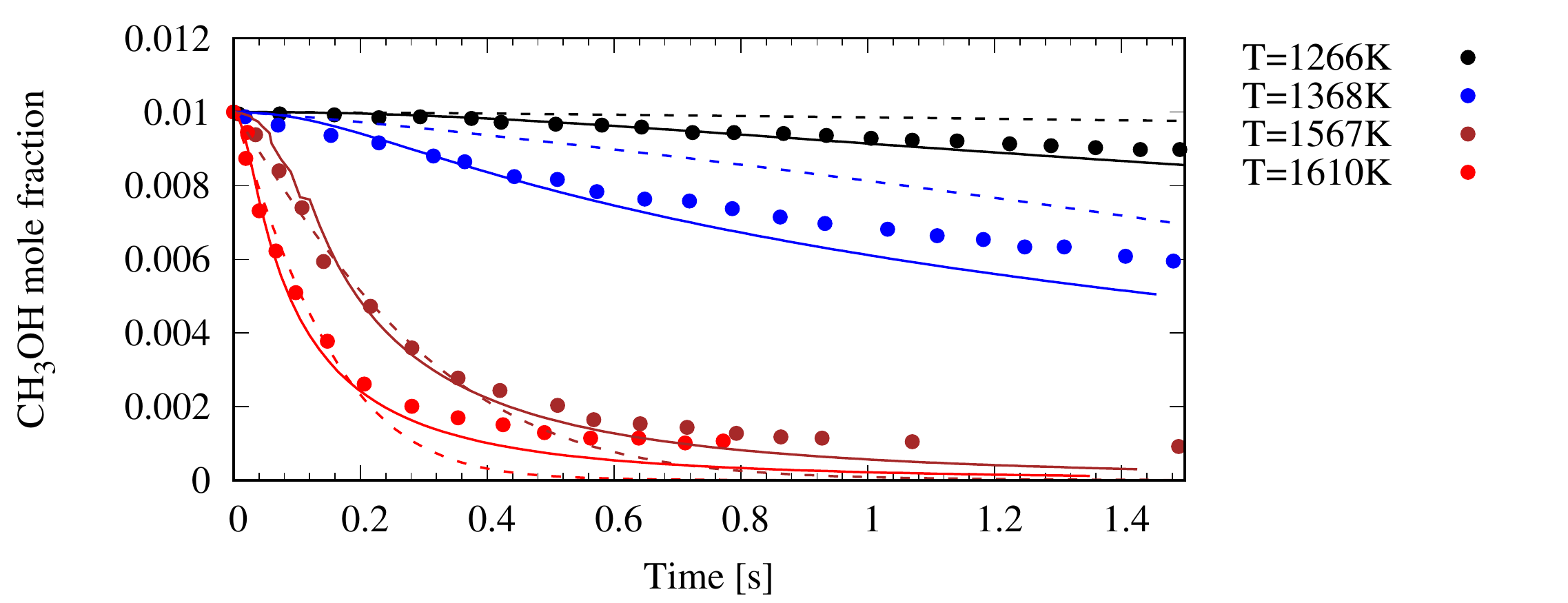}
\includegraphics[width=\columnwidth]{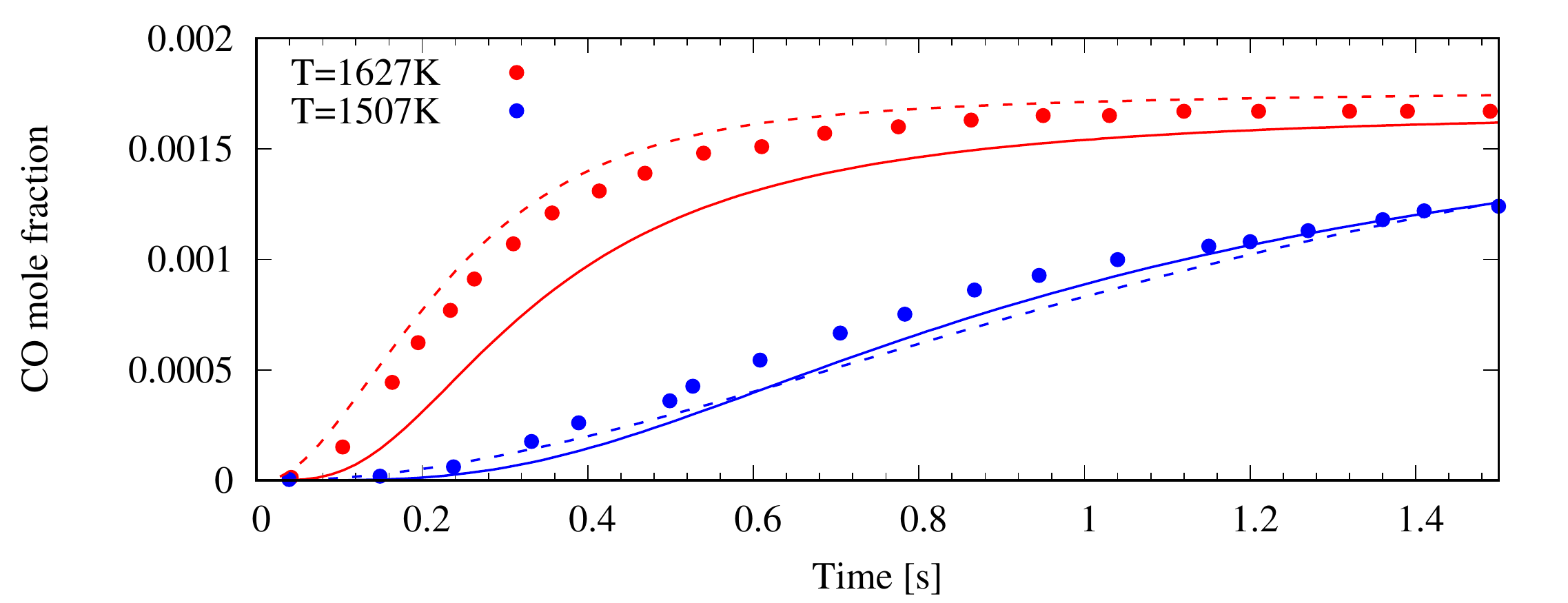}
\caption{Mole fraction of methanol (top) and carbon monoxide (bottom) versus time for the combustion of methanol in a shock-tube for a pressure around 2.5\,atm, and an equivalence ratio of 1. Points are experimental data from \citet{Ren2013}, and lines are modeling results using the chemical scheme of \citet{Venot2012} (dashed) and with the updated chemical scheme of this paper (solid).
P$=$2.2\,atm, (right) P$=$1.1\,atm.}
\label{fig:Ren}
\end{figure}

\fig{fig:Ren} shows the variation of mole fraction of methanol and carbon monoxide versus time obtained in a Shock-Tube during the pyrolysis of methanol diluted in Argon (1/99) at different temperature and for a pressure of 2.2 and 1.1\,atm by \citet{Ren2013}.
 
\begin{figure}[!h]
\includegraphics[width=\columnwidth]{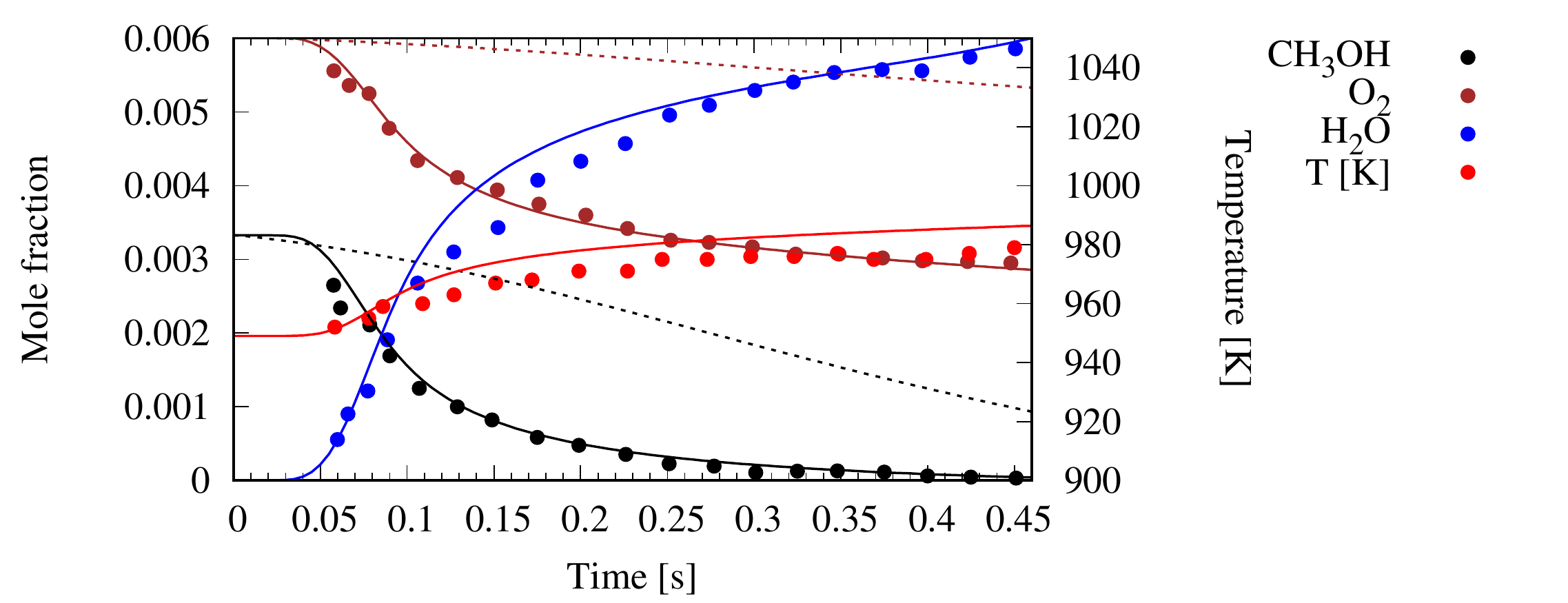}
\caption{Reaction profiles of CH$_3$OH/air mixtures in a flow reactor, where symbols represent the experimental data of \citet{Held1994} and curves modeling results using the chemical scheme of \citet{Venot2012} (dashed lines) and with the updated chemical scheme of this paper (solid lines). Experimental conditions: 0.33 mol\% CH$_3$OH $+$ 0.6 mol\% O$_2$ $+$ N$_2$, T around 1000\,K, P$=$0.25 MPa.}
\label{fig:Plug}
\end{figure}

In addition, we have compared the experimental data of \citet{Held1994} obtained in a plug flow reactor against simulated results with the updated chemical scheme of this paper, at a pressure of 0.26\,MPa and a temperature around 1000\,K (see \fig{fig:Plug}). 

\begin{figure}[!h]
\includegraphics[width=\columnwidth]{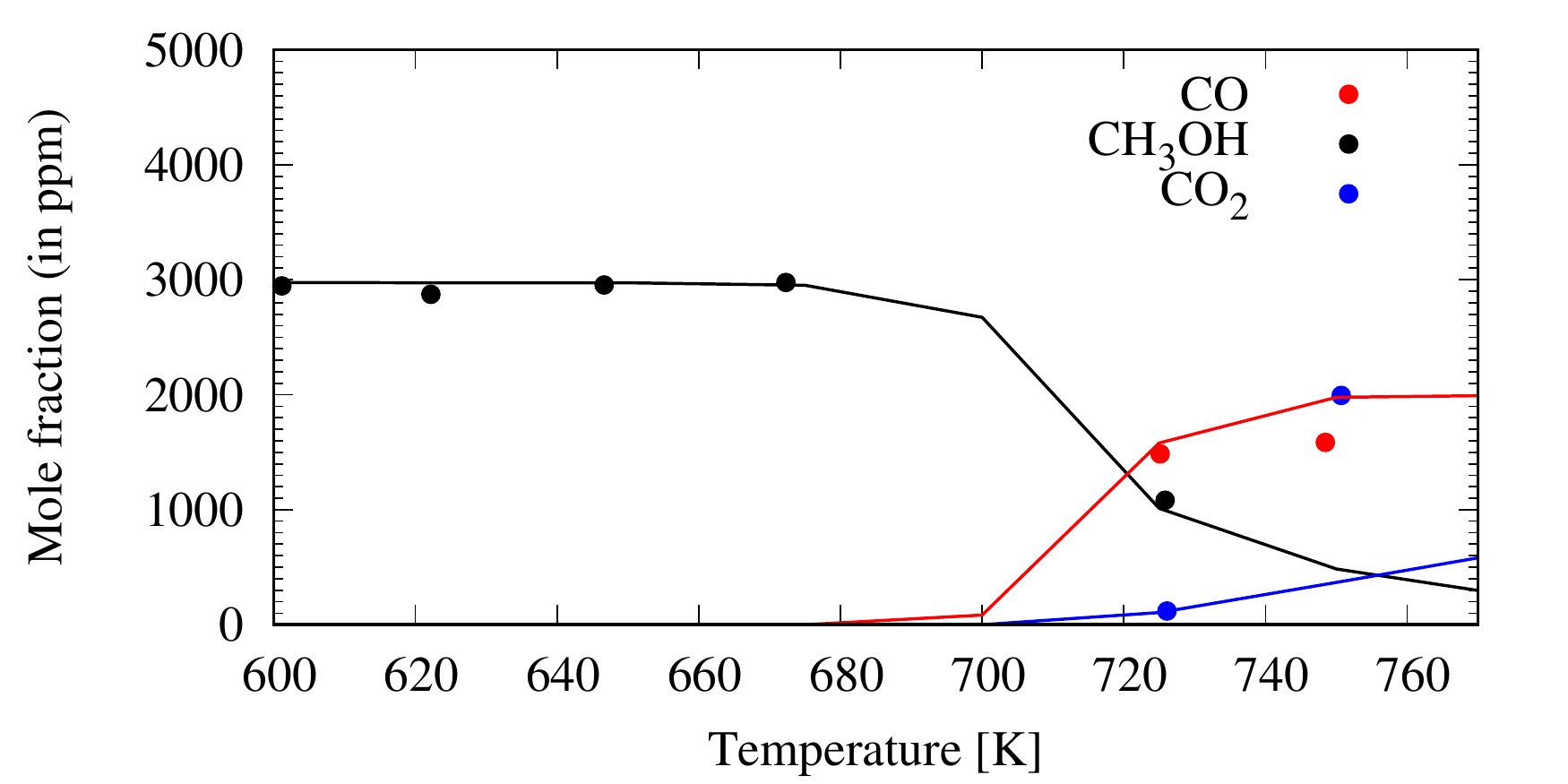}
\includegraphics[width=\columnwidth]{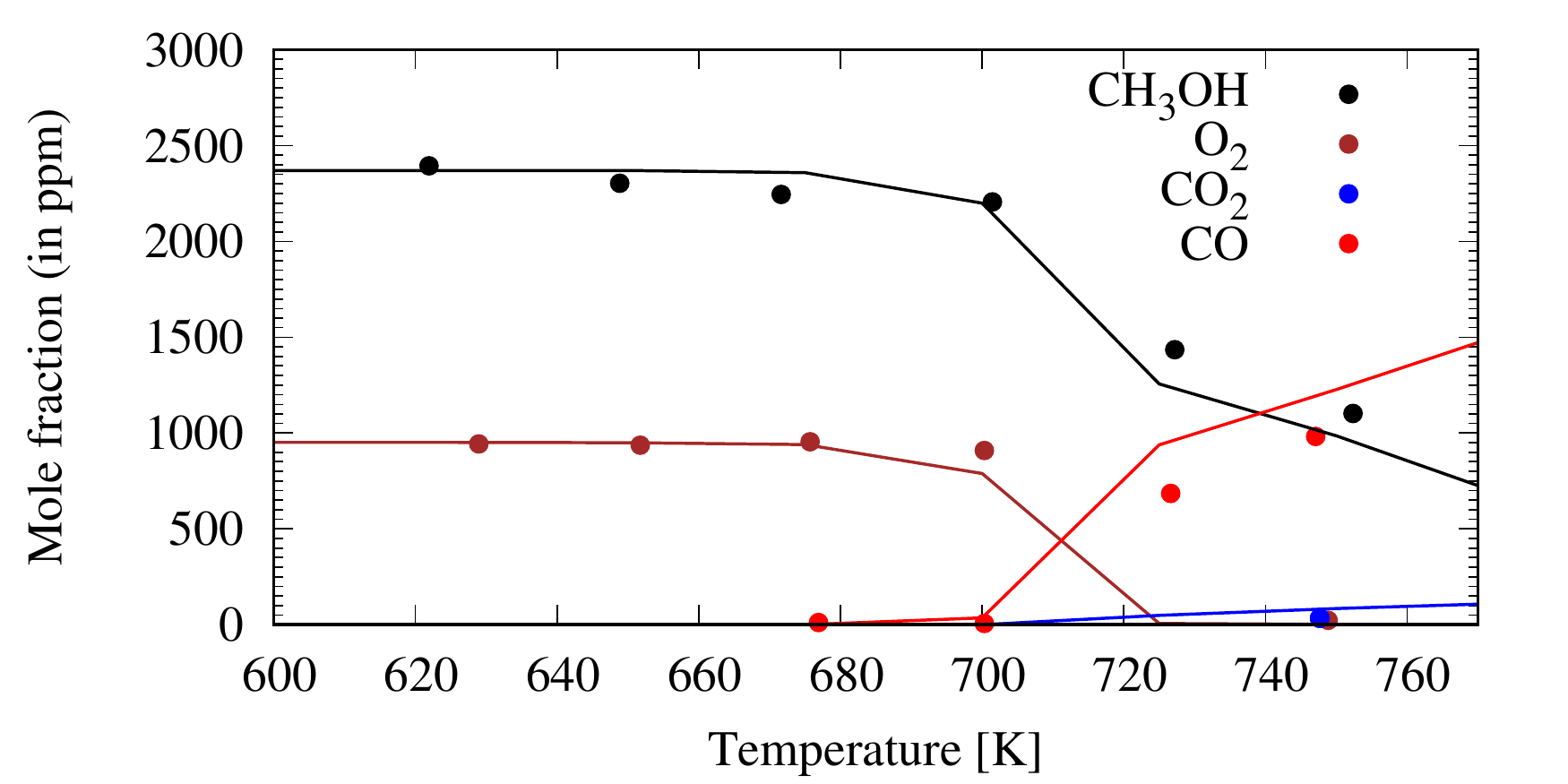}
\caption{High pressure (100\,bar) CH$_3$OH combustion experimental data (dots) of \citet{Aranda2013} in lean (top) and fuel-rich (bottom) conditions compared to our new chemical model simulations (solid lines).}
\label{fig:aramdi}
\end{figure}

We have also checked the high pressure regime, to test the PLOG formalism for some kinetic rates (see \tab{tab:new_scheme2}), and we find a good agreement for our new chemical scheme with the data from \citet{Aranda2013}, as shown in \fig{fig:aramdi}.

\begin{figure}[!h]
\includegraphics[width=\columnwidth]{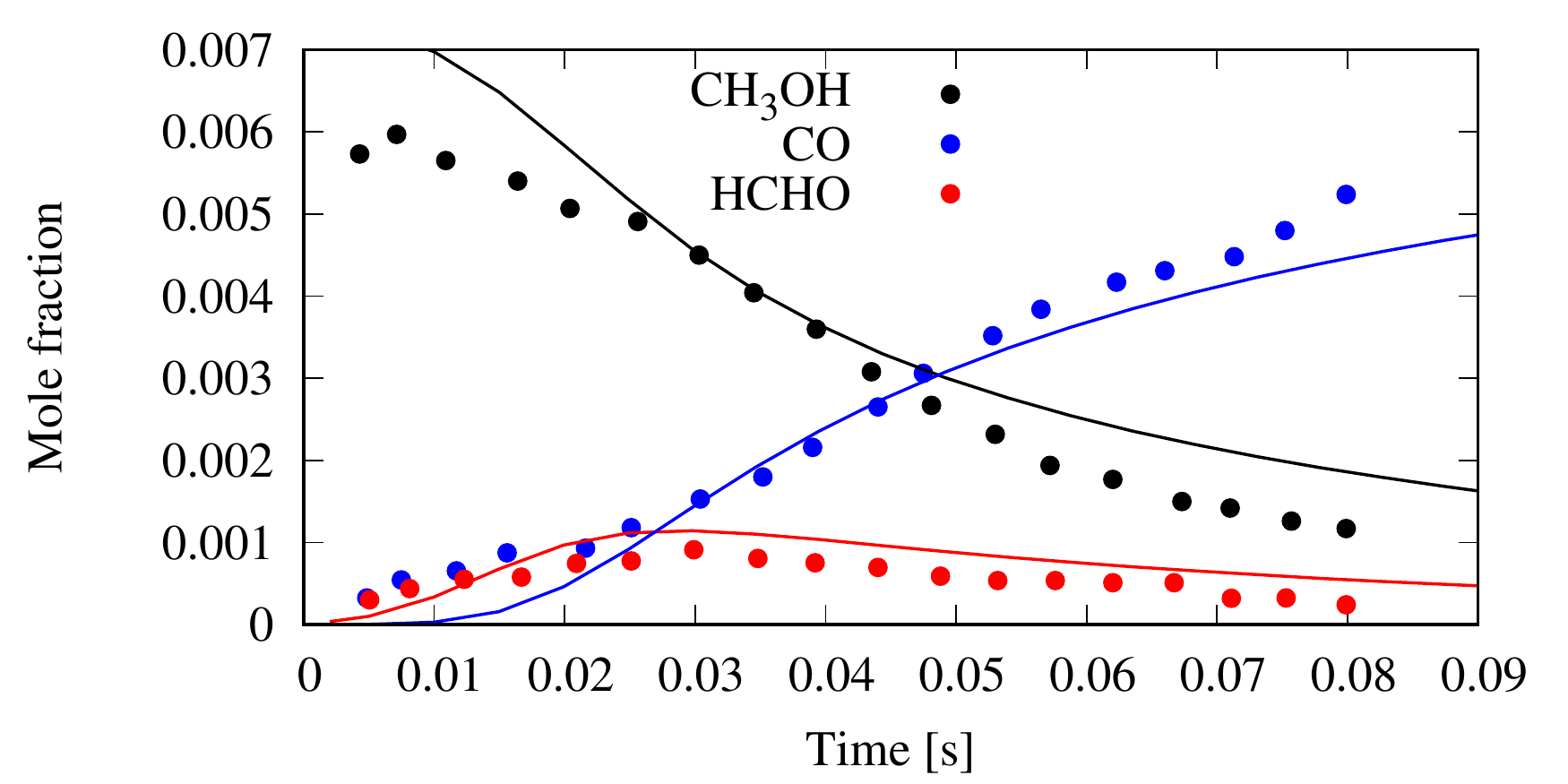}
\includegraphics[width=\columnwidth]{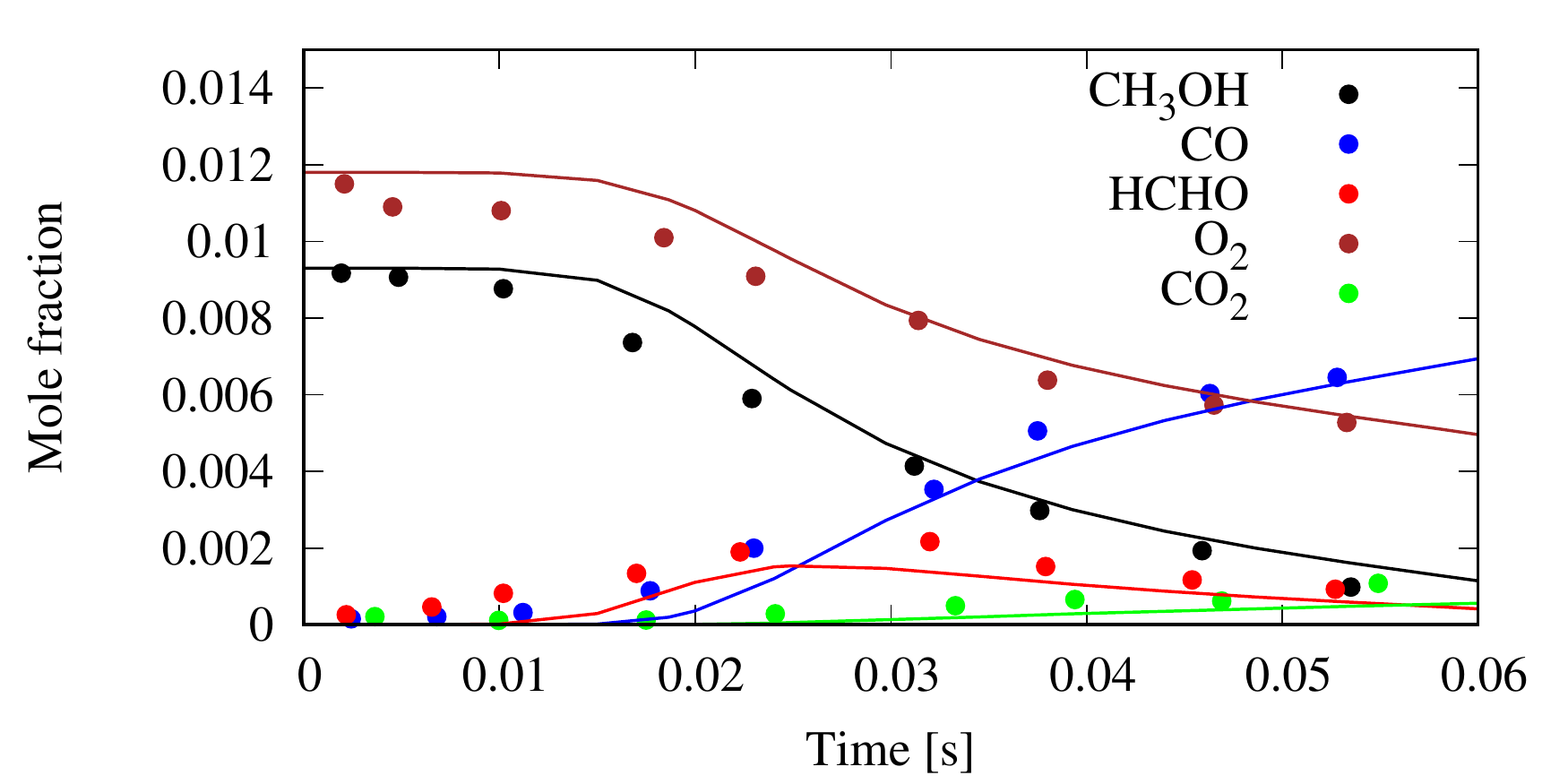}
\caption{Species mole fractions versus time in a plug Flow Reactor. solid lines are obtained from simulations using updated chemical scheme of this paper. Top: $p=1$\,atm, $T=1000$\,K, 6200ppm CH$_3$OH $+$ 6500ppm O$_2$ balanced with N$_2$, \citep{Aronowitz1979}. Bottom: $P=1$\,atm, $T=1031$\,K, 9300ppm CH$_3$OH $+$ 11800ppm O$_2$ balanced with N$_2$ \citep{Norton1989}.}
\label{fig:Plug_flow}
\end{figure}

Finally, comparisons in \figs{fig:Plug_flow}{fig:JSR} demonstrate that the predictions from the new sub-mechanism of methanol oxidation are in good agreement with the species time and temperature history measurements in plug flow or jet-stirred reactors at different pressures \citep{Aronowitz1979,Norton1989,Held1994,Burke2016}

\begin{figure}[!h]
\includegraphics[width=\columnwidth]{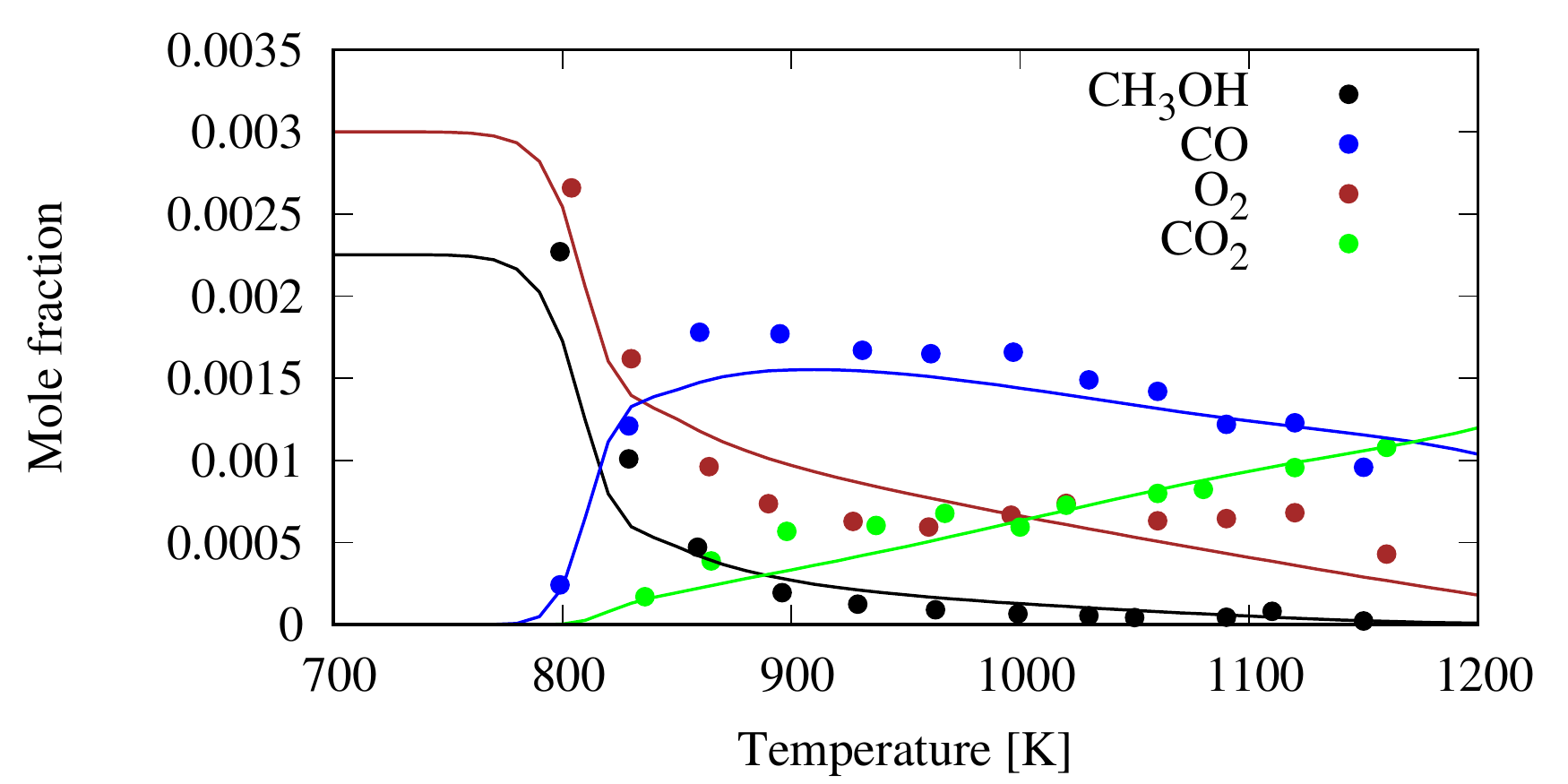}
\includegraphics[width=\columnwidth]{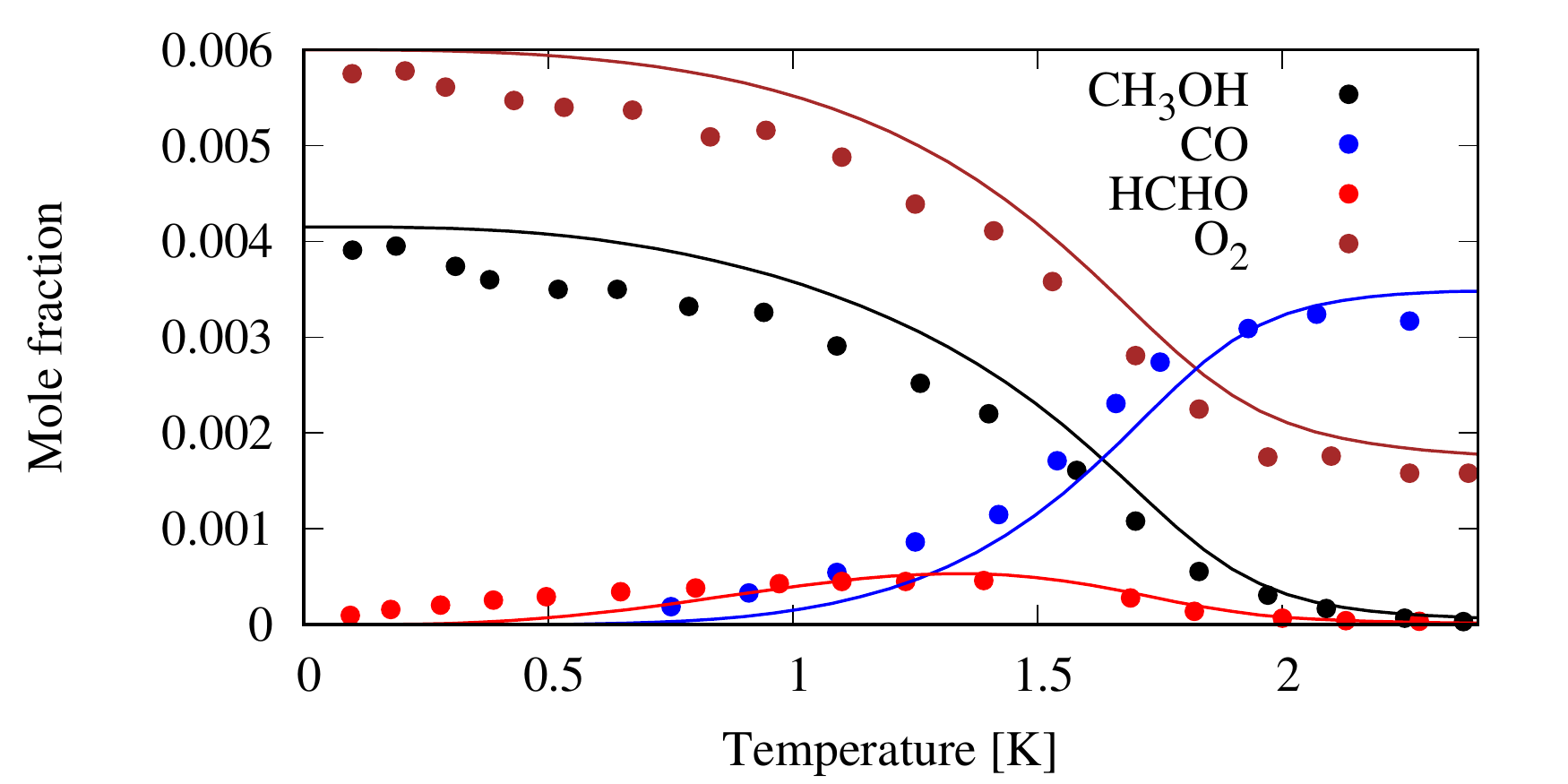}
\caption{Species mole fractions versus temperature and time. The top panel corresponds to an experiment conducted with a jet-stirred reactor.  Experimental conditions: $P=20$\,atm, equivalence ratio$=$1, 2400ppm CH$_3$OH $+$ 3000ppm O$_2$ balanced with N$_2$ \citep{Burke2016}. The bottom panel corresponds to an experiment conducted with a plug flow reactor. Experimental conditions are: $P=1$\,atm, 4150ppm  CH$_3$OH $+$ 6000ppm of O$_2$ balanced with N$_2$, $T=783$K, $P=15$\,atm \citep{Held1994}. Data are compared with simulations using the updated chemical scheme of this paper (solid lines).}
\label{fig:JSR}
\end{figure}

\section{New reduced chemical scheme \label{reduced_scheme}}

A reduced chemical scheme of V12 was recently developed by \cite{Venot2019} to reproduce the abundances of H$_2$O, CH$_4$, CO, CO$_2$, NH$_3$, and HCN, i.e. species already detected in (exo)planet atmospheres. Following our update of the former full scheme, we also provide an update for the reduced scheme. We have derived the new reduced scheme by following the same methodology as in \cite{Venot2019}. We have used the ANSYS Chemkin-Pro Reaction Workbench package \citeyearpar{chemkin}, with the method \textit{Directed Relation Graph with Error Propagation} (DRGEP), followed by a \textit{Sensitivity Analysis}. After several reduction attempts, we have ended with the reduced scheme presented here. It is the best compromise between number of species, number of reactions, applicability range, and abundances accuracy. As in \cite{Venot2019}, the scheme has been developed primarily for GJ 436b-like planets, in order to reproduce the abundances of the current observed neutral species (listed previously), as well as C$_2$H$_2$, but it can be applied to hot Jupiters, brown dwarfs and solar system giant planets as well. Acetylene was not included in the former reduced scheme, which prevented its use for modeling very hot C-rich atmospheres. Thus, this updated reduced network is sensibly larger than the previous one (i.e. 30 species, 181 reversible reactions) and contains 44 species, 288 reversible and 6 irreversible reactions, i.e. a total of 582 reactions. Like the updated full chemical scheme, it is available on KIDA \citep{Wakelam2012}. 

\begin{figure}[!h]
  \includegraphics[width=\columnwidth]{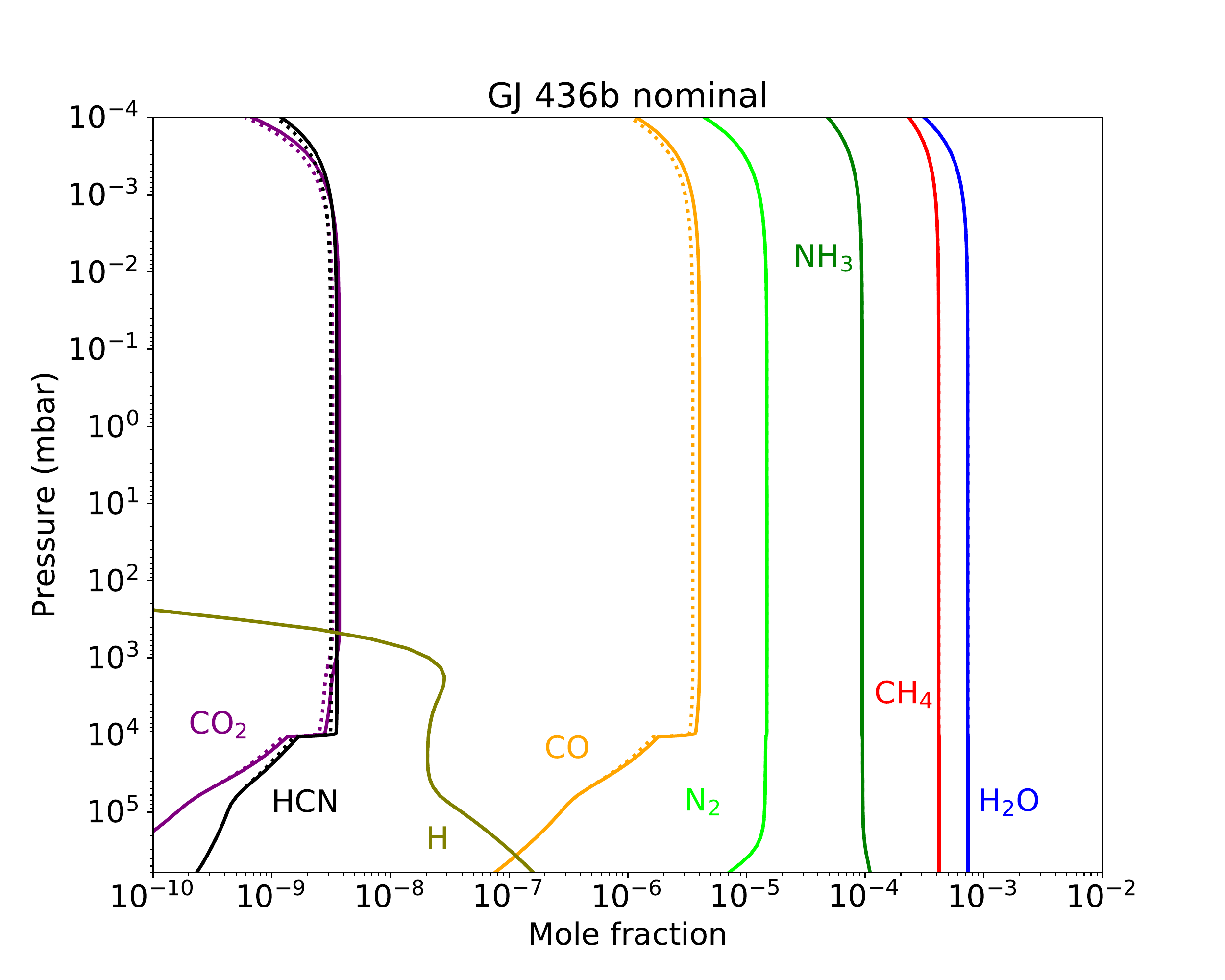}
   \includegraphics[width=\columnwidth]{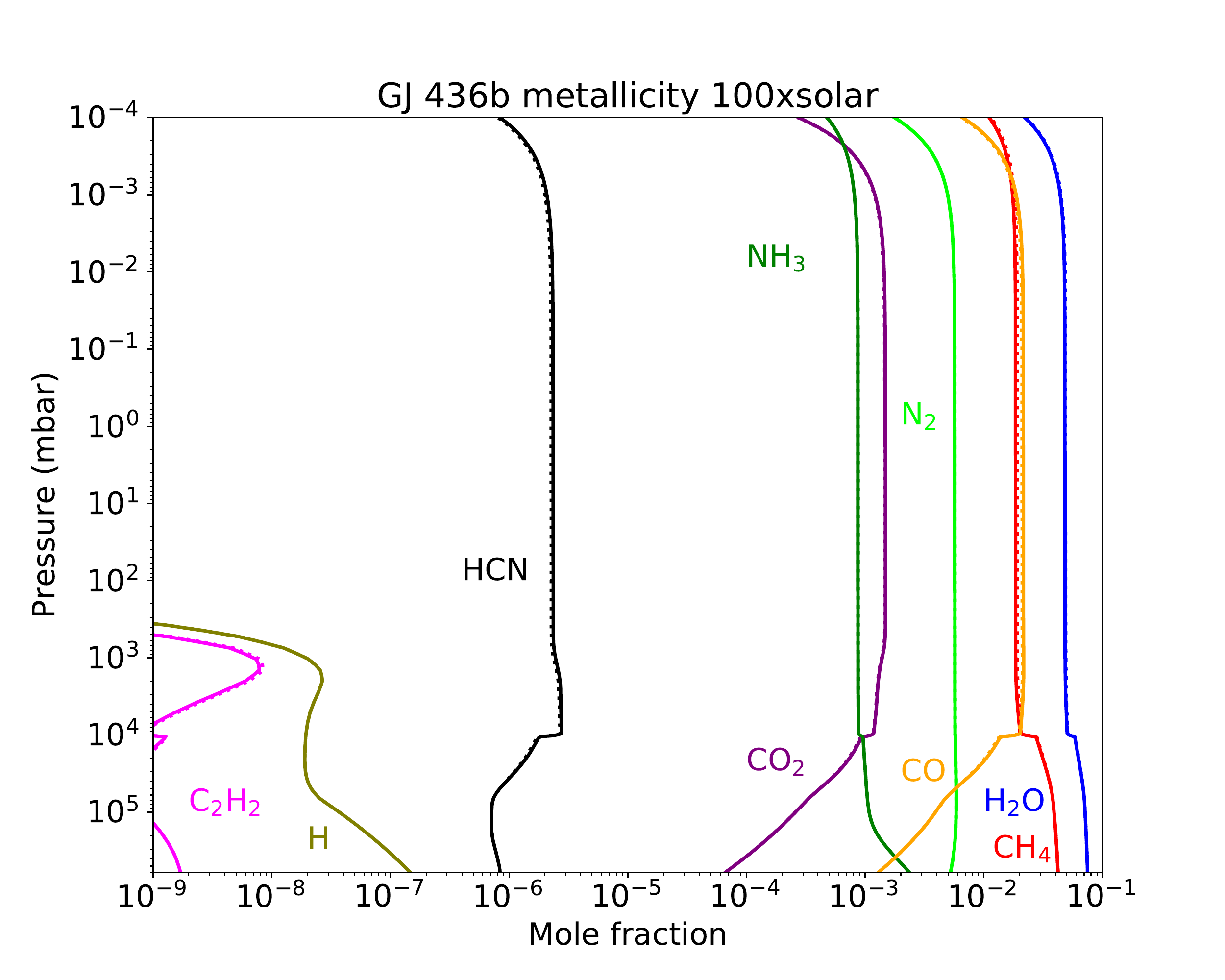}
  \includegraphics[width=\columnwidth]{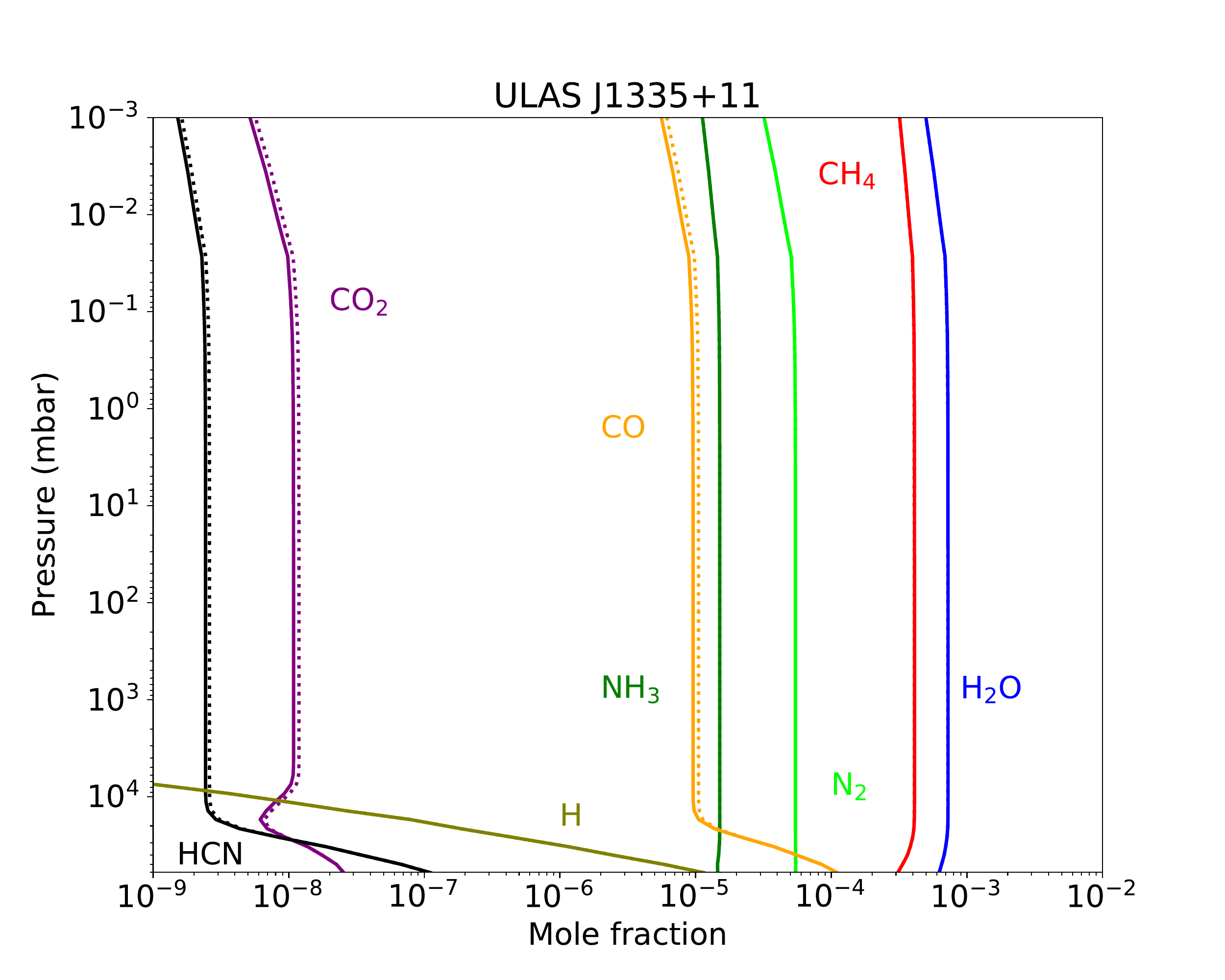}
  \caption{Vertical abundances profiles of the main species in GJ 436b with a solar metallicity (top), with a metallicity of 100$\Sun$ (middle), and in ULAS J1335+11 (bottom). The abundances obtained with the updated chemical scheme (solid lines) are compared to the ones obtained with the reduced scheme (dotted lines).}
  \label{fig:ULAS_red}
  \end{figure}

\begin{figure}[!h]
  \includegraphics[width=\columnwidth]{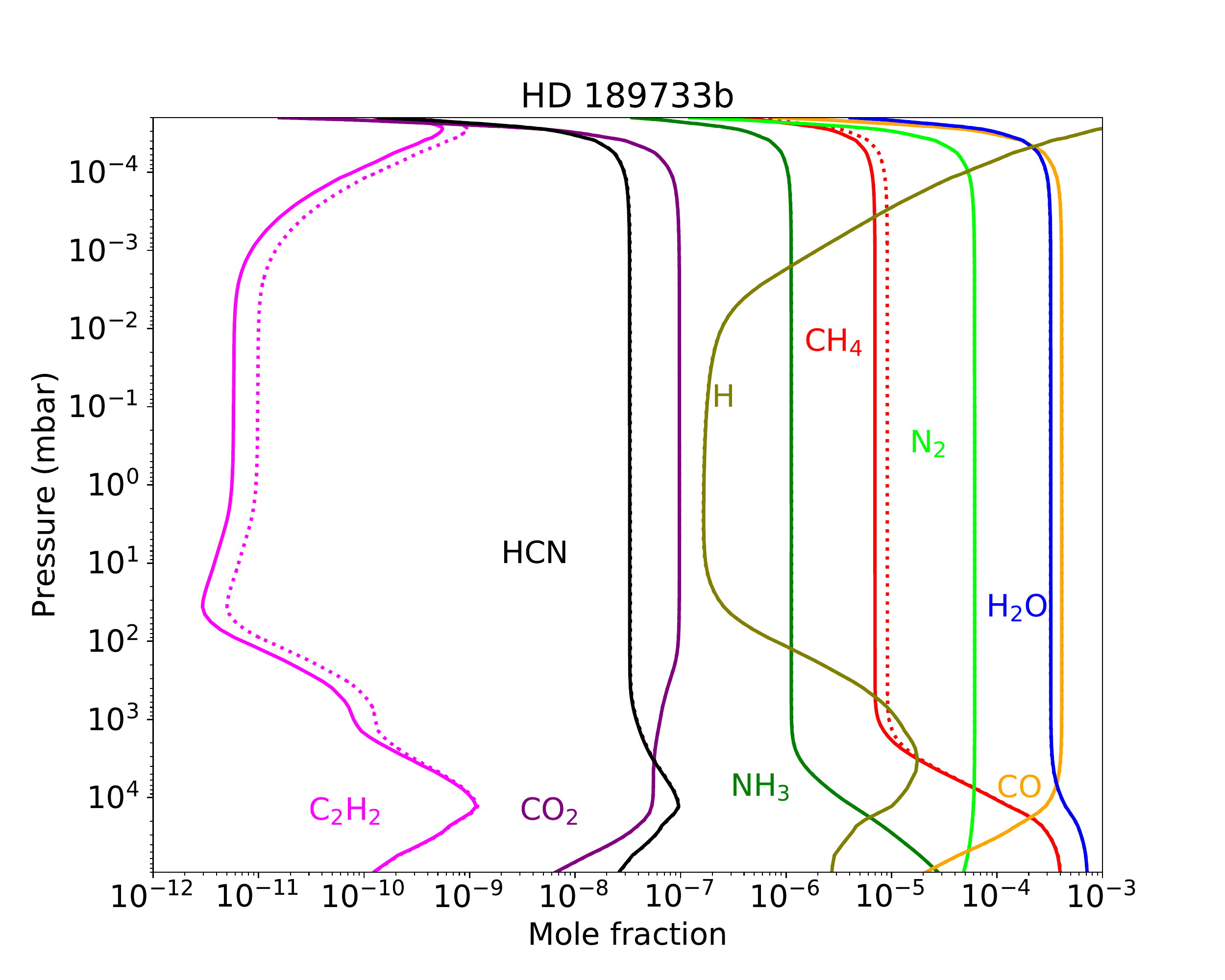}
   \includegraphics[width=\columnwidth]{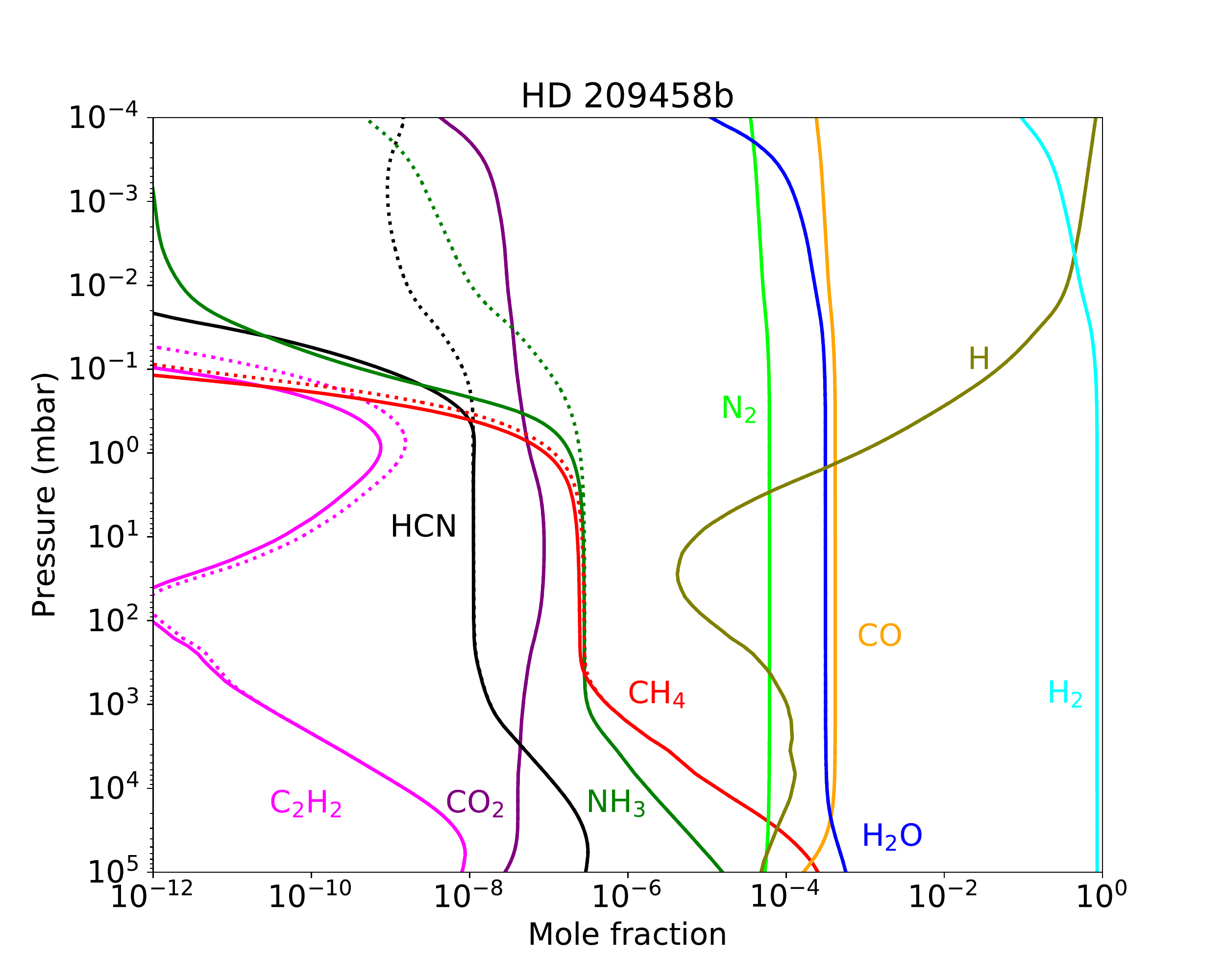}
  \includegraphics[width=\columnwidth]{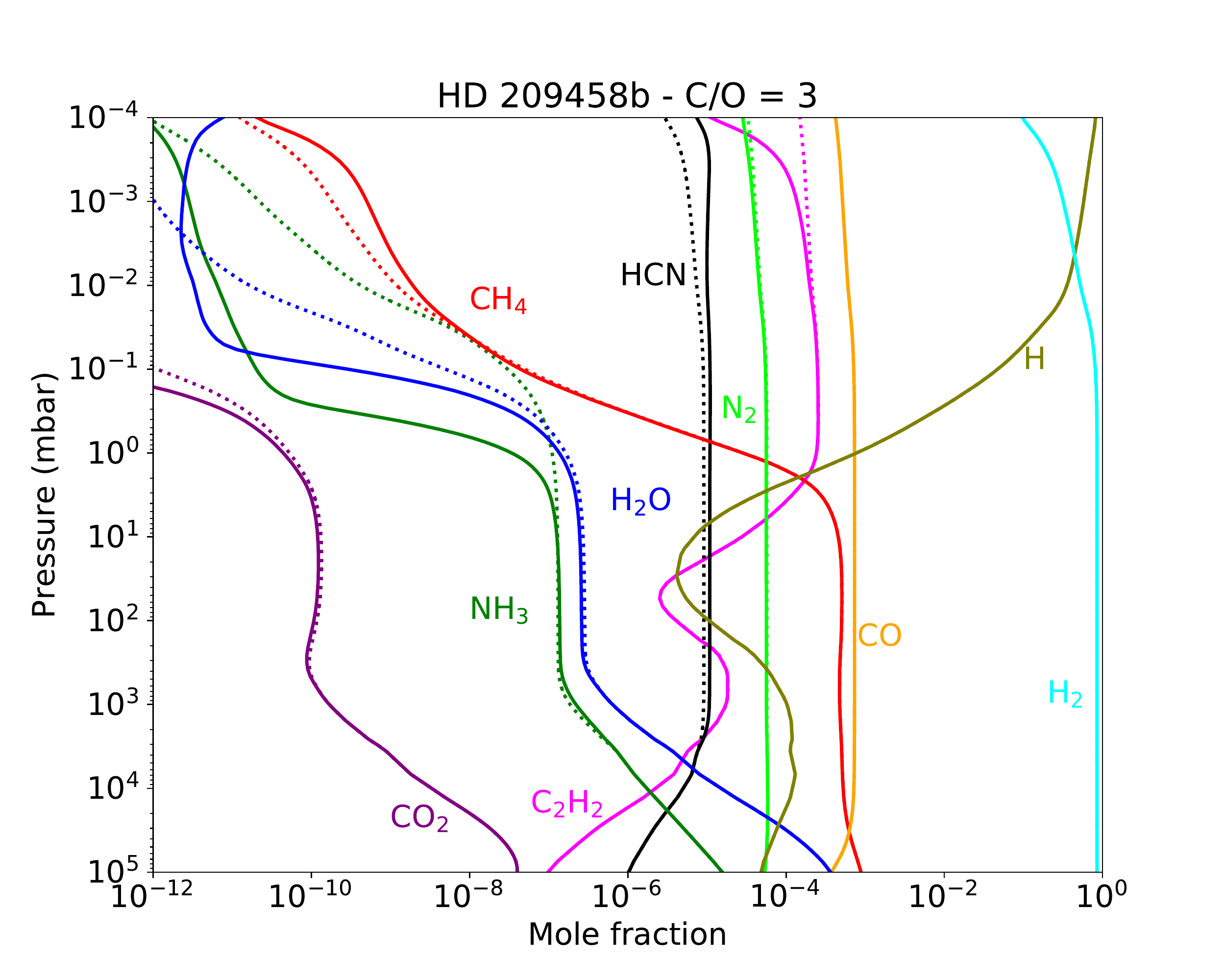}
  \caption{Vertical abundances profiles of the main species in HD 189733b (top), HD 209458b (middle), and HD 209458b with a C/O ratio of 3$\Sun$ (bottom). The abundances obtained with the updated chemical scheme (solid lines) are compared to the ones obtained with the reduced scheme (dotted lines).}
  \label{fig:hotjup_red}
  \end{figure}

\begin{figure}[!h]
  \includegraphics[width=\columnwidth]{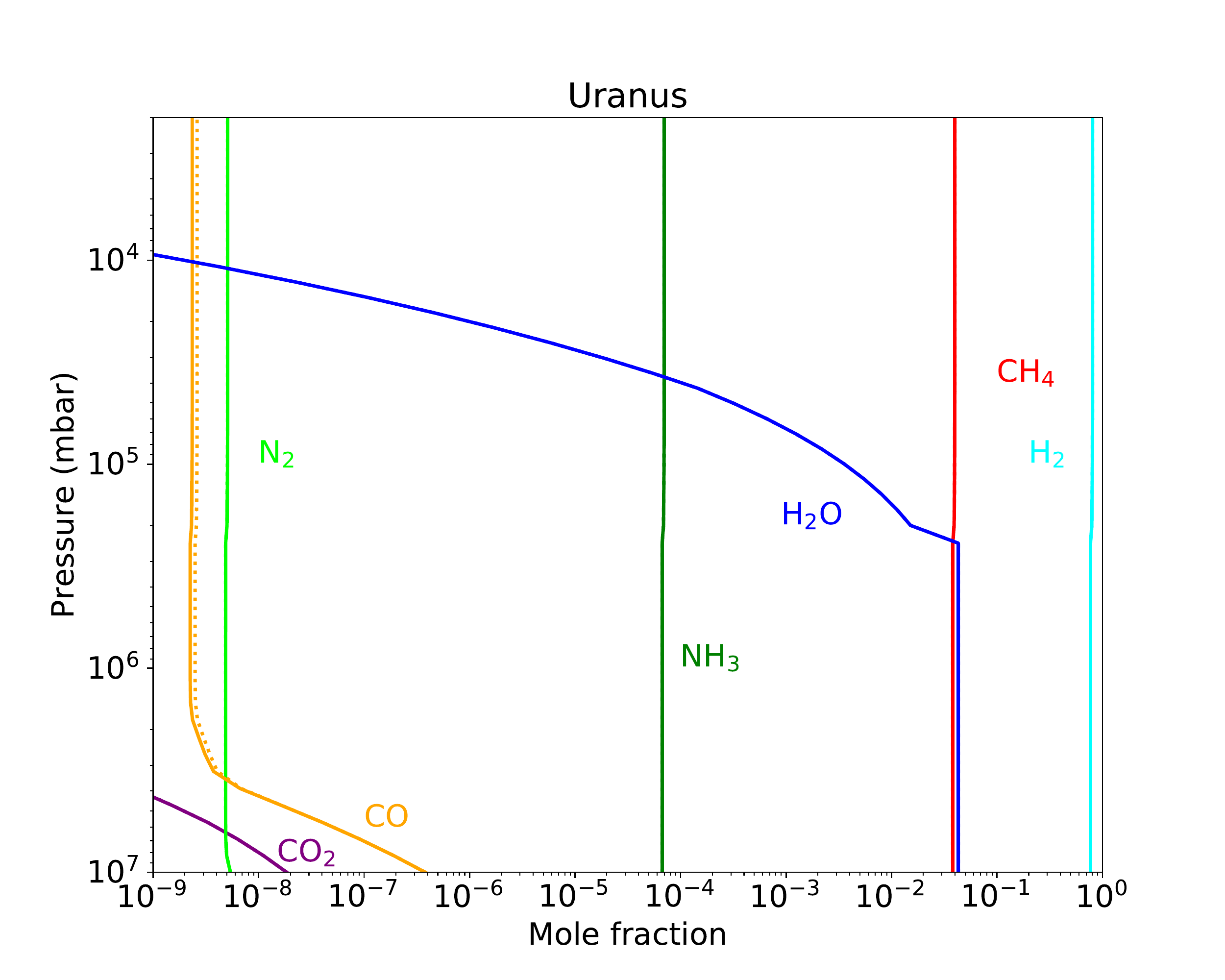}
  \includegraphics[width=\columnwidth]{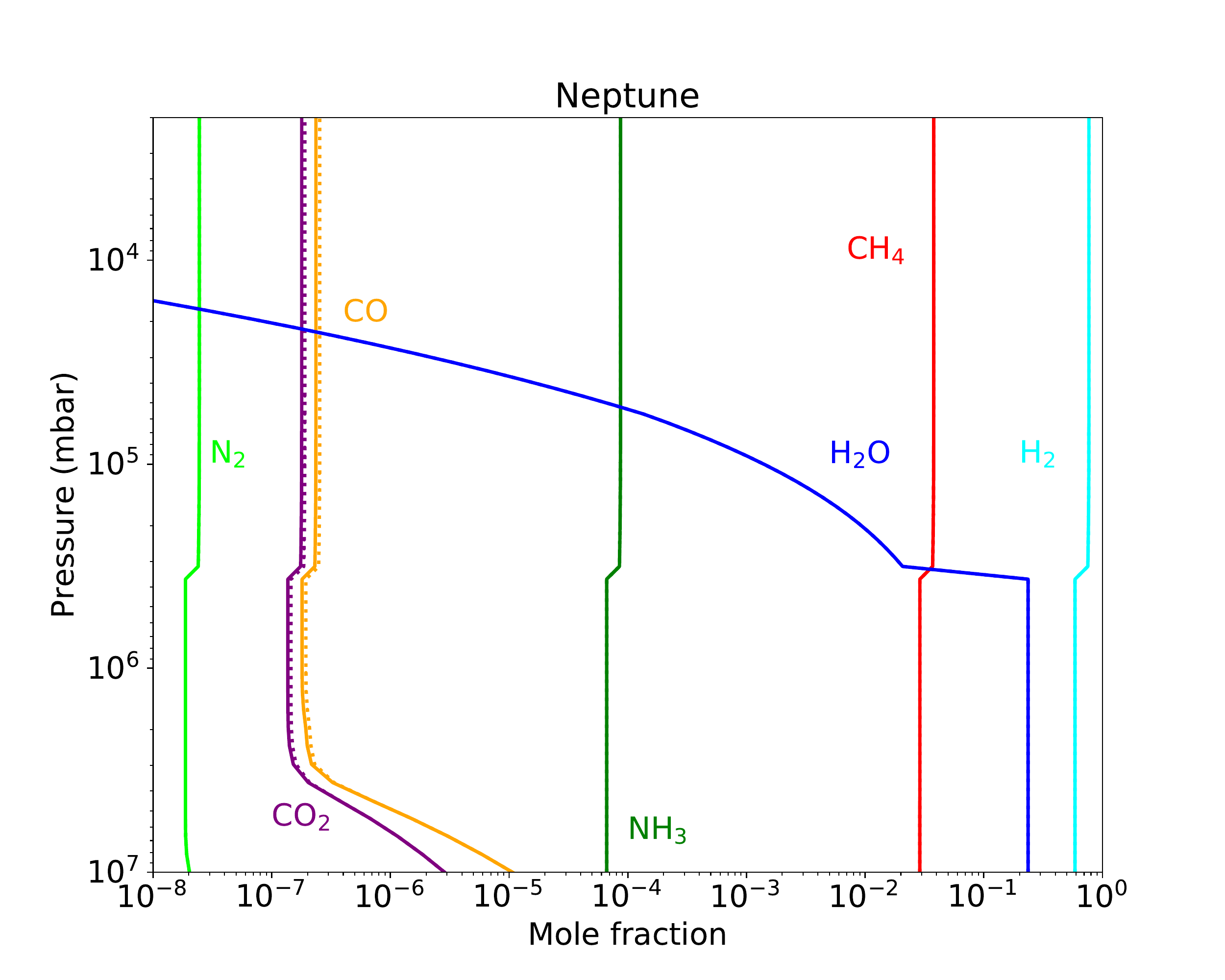}
  \caption{Vertical abundances profiles of the main species in Uranus (top) and Neptune (bottom). The abundances obtained with the updated chemical scheme (solid lines) are compared to the ones obtained with the reduced scheme (dotted lines).}
  \label{fig:Ura_Nept_red}
  \end{figure}

The updated reduced scheme gives very good results for the planets modeled in this study (see Figs.\ref{fig:ULAS_red}, \ref{fig:hotjup_red}, and \ref{fig:Ura_Nept_red}). In order to show the validity of the reduced scheme for hot C-rich atmospheres, we have modeled HD 209458b with a high C/O ratio (3$\Sun$), like in \cite{Venot2019}. While CH$_4$ was clearly overestimated in the upper atmosphere with the reduced scheme of \cite{Venot2019} (see their Fig.11), our new reduced scheme provides a better agreement for CH$_4$ thanks to the addition of C$_2$H$_2$ in the scheme.

  \begin{table*}[]
\caption{For GJ 436b models, with different metallicities, and the ULAS J1335+11 model, maximum variations of abundances (in \%) for each species for which the reduced scheme is designed. The pressure level (@level in mbar) where the maximum variation is reached is indicated within parentheses. These values are calculated within the regions probed by infrared observations: [0.1--1000] mbar and [10--10$^4$] mbar for the warm Neptune and the T Dwarf, respectively.}
\label{tab:red_var_1}
\centering
\begin{tabular}{l|l|l|l}
\hline \hline
Species &  GJ 436b & GJ 436b, $Z=100\Sun$ & ULAS J1335+11 \\
\hline
H$_2$O  &   7$\times$10$^{-2}$ (@6$\times$10$^{2}$) & 1 (@6$\times$10$^{-1}$) &  1$\times$10$^{-1}$ (@6$\times$10$^{3}$)\\
 \hline
CH$_4$ &   2$\times$10$^{-1}$ (@6$\times$10$^{2}$)    & 3 (@3$\times$10$^{2}$)  & 2$\times$10$^{-1}$ (@1$\times$10$^{1}$) \\
  \hline
CO            &   1$\times$10$^{1}$ (@1$\times$10$^{-1}$)  &  2 (@1$\times$10$^{-1}$)   &  1$\times$10$^{1}$ (@6$\times$10$^{3}$)  \\
 \hline
CO$_2$   &    1$\times$10$^{1}$ (@1$\times$10$^{-1}$)     &  1 (@1$\times$10$^{-1}$)  & 1$\times$10$^{1}$ (@5$\times$10$^{3}$)\\
 \hline
NH$_3$   &    2$\times$10$^{-2}$ (@8$\times$10$^{2}$)   & 2$\times$10$^{-1}$ (@9$\times$10$^{2}$) &  3$\times$10$^{-3}$ (@5$\times$10$^{4}$)  \\
 \hline
HCN       &  1$\times$10$^{1}$ (@1$\times$10$^{-1}$) & 3 (@1$\times$10$^{-1}$) & 7 (@7$\times$10$^{3}$)  \\ 
\hline
C$_2$H$_2$       &  4$\times$10$^{-1}$ (@6$\times$10$^{-1}$) & 8 (@1) & 5$\times$10$^{-1}$ (@1$\times$10$^{3}$)\\ 
\hline
\end{tabular}
\end{table*}

 \begin{table*}[]
\caption{For HD 209458b models, with different C/O ratios, and HD 189733b models, with different eddy diffusion coefficients, maximum variations of abundances (in \%) for each species for which the reduced scheme is designed. The pressure level (@level in mbar) where the maximum variation is reached is indicated in parentheses. These values are calculated within the region probed by infrared observations ([0.1--1000] mbar).}
\label{tab:red_var_2}
\centering
\begin{tabular}{l|l|l|l}
\hline \hline
Species &  HD 209458b & HD 209458b, C/O $=$ 3$\Sun$ & HD 189733b \\
\hline
H$_2$O  &   1$\times$10$^{-1}$ (@9$\times$10$^{2}$) & 1$\times$10$^{3}$ (@1$\times$10$^{-1}$) &  5$\times$10$^{-1}$ (@1$\times$10$^{3}$)\\
 \hline
CH$_4$ &   1$\times$10$^{3}$ (@1$\times$10$^{-1}$)    & 1$\times$10$^{1}$ (@1$\times$10$^{-1}$)  & 3$\times$10$^{1}$ (@3$\times$10$^{1}$)\\
  \hline
CO            &   9$\times$10$^{-2}$ (@9$\times$10$^{2}$)  &  8$\times$10$^{-3}$ (@6$\times$10$^{-1}$)   &  4$\times$10$^{-1}$ (@1$\times$10$^{3}$)  \\
 \hline
CO$_2$   &    4$\times$10$^{-2}$ (@9$\times$10$^{2}$)     &  1$\times$10$^{3}$ (@1$\times$10$^{-1}$)   & 8$\times$10$^{-2}$ (@1$\times$10$^{-1}$)   \\
 \hline
NH$_3$   &    2$\times$10$^{4}$ (@1$\times$10$^{-1}$)   & 2$\times$10$^{5}$ (@1$\times$10$^{-1}$) &  4$\times$10$^{-1}$ (@1$\times$10$^{-1}$)  \\
 \hline
HCN       &  8$\times$10$^{2}$ (@1$\times$10$^{-1}$) & 2$\times$10$^{1}$ (@1$\times$10$^{-1}$) & 1 (@1$\times$10$^{3}$)  \\ 
\hline
C$_2$H$_2$       &  2$\times$10$^{3}$ (@1$\times$10$^{-1}$) & 9$\times$10$^{-1}$ (@7$\times$10$^{1}$) & 7$\times$10$^{1}$ (@2$\times$10$^{2}$)\\ 
\hline
\end{tabular}
\end{table*}

  \begin{table}[]
\caption{For Uranus and Neptune models, maximum differences of abundances (in \%) for each species for which the reduced scheme is designed. The pressure level (@level in mbar) where the maximum variation is reached is indicated in parentheses. These values are calculated within the region where quenching occurs ([10$^{6}$--10$^{7}$] mbar).}
\label{tab:red_var_3}
\centering
\begin{tabular}{l|l|l}
\hline \hline
Species &  Uranus & Neptune \\
\hline
H$_2$O  &   9$\times$10$^{-4}$ (@8$\times$10$^{6}$) & 2$\times$10$^{-4}$ (@2$\times$10$^{6}$) \\
 \hline
CH$_4$ &   2$\times$10$^{-2}$ (@8$\times$10$^{6}$)    & 4$\times$10$^{-3}$ (@8$\times$10$^{6}$)  \\
  \hline
CO            &   1$\times$10$^{1}$ (@8$\times$10$^{6}$)  &  8 (@1$\times$10$^{6}$)    \\
 \hline
CO$_2$   &    8 (@1$\times$10$^{6}$)     &  6 (@1$\times$10$^{6}$)   \\
 \hline
NH$_3$   &    8$\times$10$^{-4}$ (@8$\times$10$^{6}$)   & 2$\times$10$^{-4}$ (@8$\times$10$^{6}$) \\
 \hline
HCN       &  4$\times$10$^{-1}$ (@3$\times$10$^{6}$) & 3 (@2$\times$10$^{6}$)  \\ 
\hline
C$_2$H$_2$       &  4$\times$10$^{-2}$ (@8$\times$10$^{6}$) & 1$\times$10$^{-2}$ (@1$\times$10$^{6}$) \\ 
\hline
\end{tabular}
\end{table}

  For each species of interest, the maximum difference of abundances obtained using the two chemical networks (with the corresponding pressure level) are gathered in Tables \ref{tab:red_var_1}, \ref{tab:red_var_2}, and \ref{tab:red_var_3}. For the exoplanets and the brown dwarf, we restricted our comparison to the pressure range probed by observations. For the solar system giant planets, we focused on the quenching area, as this level governs the abundances observed at 2 bar and is thus decisive for the conclusions drawn on their elemental composition \citep{Cavalie2009,Cavalie2014,Cavalie2017}. For GJ 436b, ULAS J1335+11, Uranus, and Neptune, variations are below 10\%.  We notice that a lower agreement is found for hot Jupiter atmospheres, especially in the upper atmosphere of HD 209458b, where the pressure is low and the temperature high. However, we have checked using the \texttt{TauRex} code in forward mode that these variations occur high enough in the atmosphere and do not impact the synthetic spectra computed with these two chemical compositions.

Note that the comparison is made for models without photodissociation, as the reduced scheme does not contain photolysis reactions.

\end{appendix}

\end{document}